\documentclass[apj]{emulateapj}

\usepackage{apjfonts}
\usepackage{times}

\usepackage{amsmath}
\usepackage{graphicx}
\usepackage{natbib}
\usepackage{lineno,hyperref}
\usepackage{amssymb}

\usepackage{color}

 \font\sevenrm=cmr7 scaled 1000

\shorttitle{Received 2020 May 16; revised 2020 July 27; accepted 2020 August 06}
\shortauthors{Le et al.}

\newcommand{\Hb}{H{$\beta$}}

\newcommand{\FeII}{\ion{Fe}{2}}

\newcommand{\MgII}{\ion{Mg}{2}}

\newcommand{\OIII}{[\ion{O}{3}]}

\newcommand{\mbh}{M$_{\rm BH}$}

\newcommand{\kms}{km~s$^{\rm -1}$}
\newcommand{\ergs}{erg s$^{-1}$}

\def\gsim{\mathrel{\rlap{\lower4pt\hbox{\hskip1pt$\sim$}}
    \raise1pt\hbox{$>$}}}         

\def\lsim{\mathrel{\rlap{\lower4pt\hbox{\hskip1pt$\sim$}}
    \raise1pt\hbox{$<$}}}         

\begin{document}

\title{Calibrating \ion{Mg}{2}-based black-hole mass estimators using Low-to-High-Luminosity Active Galactic Nuclei} 

\author{Huynh Anh N. Le$^{1,2,3,*}$}
\author{Jong-Hak Woo$^{1}$}
\author{Yongquan Xue$^{2,3}$}
\affil{$^{1}$ Astronomy Program, Department of Physics and Astronomy, Seoul National University, Seoul, 08826; woo@astro.snu.ac.kr \\
$^{2}$ CAS Key Laboratory for Research in Galaxies and Cosmology, Department of Astronomy, University of Science and Technology of China, Hefei 230026, China; lha@ustc.edu.cn; xuey@ustc.edu.cn \\
$^{3}$ School of Astronomy and Space Science, University of Science and Technology of China, Hefei 230026, China
}
\altaffiltext{*}{PIFI Fellow.}

\begin{abstract}

We present single-epoch black-hole mass (\mbh) estimators based on the rest-frame ultraviolet (UV) \MgII\ 2798\AA\ and optical \Hb\ 4861\AA\ emission lines. To enlarge the luminosity range of active galactic nuclei (AGNs), we combine the 31 reverberation-mapped AGNs with relatively low luminosities from \citet{Bahk+19}, 47 moderate-luminosity AGNs from \citet{Woo+18}, and 425 high-luminosity AGNs from the Sloan Digital Sky Survey (SDSS). The combined sample has the monochromatic luminosity at 5100\AA\ ranging $\mathrm{\log \lambda L_{5100} \sim 41.3-46.5}$ \ergs, over the range of 5.5 $<$ $\log$\mbh\ $<$ 9.5. Based on the fiducial mass from the line dispersion or full width half maximum (FWHM) of \Hb\ paired with continuum luminosity at 5100\AA, we calibrate the best-fit parameters in the black hole mass estimators using the \MgII\ line. We find that the differences in the line profiles between \MgII\ and \Hb\ have significant effects on calibrating the UV \mbh\ estimators. By exploring the systematic discrepancy between the UV and optical \mbh\ estimators as a function of AGN properties, we suggest to add correction term $\Delta$M = -1.14 $\rm \log(FWHM_{MgII}/\sigma_{MgII}$) + 0.33 in the UV mass estimator equation. We also find a $\sim$0.1 dex bias in the \mbh\ estimation due to the difference of the spectral slope in the 2800-5200 \AA\ range. Depending on the selection of \mbh\ estimator based on either line dispersion or FWHM and either continuum or line luminosity, the derived UV mass estimators show $\gsim 0.1$ dex intrinsic scatter with respect to the fiducial \Hb\ based \mbh.  

\end{abstract}
\keywords{galaxies: active -- galaxies: nuclei -- galaxies: Seyfert}

\section{INTRODUCTION} \label{section:intro}

Black-hole mass (\mbh) is a fundamental parameter for understanding the physical nature of active galactic nuclei (AGNs). 
The correlations between \mbh\ and galaxy properties have been extensively investigated over the last two decades, in order to constrain the nature of coevolution between black holes and their host galaxies \citep[e.g.,][]{Ferrarese00, Gebhardt00, MJ02, Tremaine+02, Woo06, Woo+10, Xue+10, Kormendy&Ho13, Le+14, Sun+15, Shankar+16, Xue+17, Shankar+19}. 

The reverberation-mapping (RM) technique is to measure the time lag between flux variabilities of AGN continuum and broad emission lines, providing \mbh\ measurements \citep[e.g.,][]{BM82, Peterson93}. Assuming that the gas in the broad-line region (BLR) is governed by the gravitational potential of the central BH, one can use the virial theorem to determine \mbh\ as:
\begin{equation}
	\rm M_{\rm BH} = f\frac{\rm R_{BLR} V^{2}}{\rm G}
	\label{eq:mbh1}
\end{equation}
where, G is the gravitational constant, V is the gas velocity (line dispersion or FWHM), and $\rm R_{BLR}$ is the size of the BLR (i.e., speed of light $\times$ time lag), while f is a scale factor, representing the unknown geometry and kinematic distribution of the BLR gas, which is mainly calibrated from the \mbh $-$ $\rm \sigma_*$ relation based on the assumption that active and quiescent galaxies follow the same relation between \mbh\ and stellar velocity dispersion ($\sigma_*$) {\citep[e.g.,][]{Onken+04, Woo+10, Woo15}.

While the RM method is powerful for measuring \mbh, an extensive monitoring is required \citep[e.g.,][]{Wandel1999, Kaspi+00, Peterson+04, Bentz+09, Lira+18, Barth2011, Grier2013, Barth2015, Du+16, Du+17, Park+17, Grier+17, Grier+19, Rakshit+19, Woo+19a, Woo+19b, Cho+20}. Therefore, a simple recipe from the measured BLR size-luminosity relation is popularly used because only a single spectroscopic observation is required to estimate \mbh\ \citep[e.g.,][]{Woo&Urry02, VW02, MD04, VP06, Shen+11}. 
By combining the size-luminosity relation and the virial theorem, \mbh\ can be expressed as: 
\begin{equation}
	\log \rm M_{\rm BH}= \alpha + \beta \log V + \gamma \log L
	\label{eq:mbh2}
\end{equation}
where, $\rm V$ is gas velocity measured from the width of broad emission lines, and L is either continuum or emission line luminosity. $\beta$ is close to 2 based on the virial assumption while $\gamma$ is empirically determined as 0.533$^{+0.035}_{-0.033}$ based on the \Hb\ reverberation mapping results \citep{Bentz+13}.  


While the size-luminosity relation is calibrated with the \Hb\ lag, the \Hb-based \mbh\ estimator is typically applied to low-redshift AGNs at z $\lsim$ 0.8 since many ground-based AGN surveys have been performed in the optical wavelength range. 
At intermediate z (0.8 $\lsim$ z $\lsim$ 2.5) \MgII\ substitutes \Hb\ to estimate \mbh\ \citep[e.g.,][]{MJ02, MD04, McGill+08, Woo+18}, while at higher z (3 $\lsim$ z $\lsim$ 5) \ion{C}{4} is used for \mbh\ estimation \citep[e.g.,][]{VW02, VP06, Assef+11, Denney12, Shen+12, Park+13, Runnoe+13, Brotherton+15, Park+17, Coatman+17, Sun+18}. 
Since the UV lines play an essential role in determining \mbh\ of higher-z AGNs, it is crucial to validate and improve the UV line based mass estimators. 


To utilize the \MgII\ line, we need to determine the size-luminosity relation based on the \MgII\ lag measurements. However, there are only a few \MgII\ reverberation mapping measurements \citep[e.g.,][]{Clavel+91, Reichert+94, Dietrich+95, Metzroth+06, Shen+16, Wang+19}. Some studies failed to measure the time lag of \MgII\ {\citep[e.g.,][]{Woo+08, Cackett+15}. Thus, the \Hb\ BLR size measurements are instead used to calibrate the \MgII-based \mbh\ estimator. \citet{MJ02} performed the first calibration study based on the \MgII\ line. Using a sample of 34 objects (17 Seyferts from \citealp{Wandel1999} and 17 Palomar-Green (PG) quasars from \citealp{Kaspi+00}), they determined the relation between the \Hb-BLR size and the UV continuum luminosity at 3000 \AA\ ($\rm L_{3000}$). Later on, \citet{MD04} updated the \Hb\ BLR size-$\rm L_{3000}$ relation for high luminosity AGNs. Based on the enlarged sample of the reverberation-mapped AGNs and UV data, other authors re-calibrated the relation between the BLR size and UV luminosity
\citep[e.g.,][]{Kong+06, VP06, Wang+09, Rafiee+11}. 
The \MgII-based \mbh\ estimator derived from the \Hb\ reverberation-mapped AGNs provided consistent \mbh, albeit with additional large uncertainties. 

These early studies have two main limitations. First, they used a small sample of the reverberation-mapped AGNs, which are relatively low-luminosity 
and low-redshift AGNs. Second, the rest-frame UV and optical spectra were not obtained simultaneously, suffering from the variability issues. Thus, proper comparisons of the line widths as well as luminosities between UV (\MgII\ and L$_{\rm 3000}$) and optical (\Hb\ and L$_{5100}$) were not available. 
Later studies utilized higher luminosity AGNs along with simultaneous observations of the rest-frame UV and optical, providing better calibrated \mbh\ estimators
\citep{McGill+08, Shen+11, Shen+12, Trakhtenbrot+12, Tilton+13, Mejia-Restrepo+16}.

So far various \mbh\ estimators based on \MgII\ have been reported. However, there are still considerable discrepancies among those estimators \citep[see][]{Woo+18}. Most previous studies focused on the relatively limited luminosity range \citep[e.g.,][]{MD04, McGill+08, Wang+09, Shen+11, Shen+12, Bahk+19}. Thus, a
calibration study over a broad dynamic range of luminosity using simultaneous observations of \MgII\ and \Hb\ is needed to improve the \MgII-based \mbh\ estimator.

In our previous studies, \citet{Woo+18} used a sample of 52 moderate-luminosity AGNs to calibrate the \mbh\ estimator, reporting that the \MgII-based masses typically show $\gsim$0.2 dex intrinsic scatter with respect to \Hb-based masses, while \citet{Bahk+19} utilized a sample of 31 \Hb\ reverberation-mapped AGNs,
reporting that the \MgII-based masses are consistent with the \Hb\ reverberation masses within a factor of 2. 
In this study we calibrate the \MgII-based mass estimator over a large dynamic range of luminosity (i.e., $\mathrm{\log \lambda L_{5100} \sim41.3 - 46.5}$ \ergs), by combining the low-to-moderate luminosity AGNs from \citet{Woo+18} and \citet{Bahk+19} with the high-luminosity AGNs at z $\sim$ 0.4$-$0.8 selected from the Sloan Digital Sky Survey (SDSS) archive.
We also select the sample with simultaneous observations of both \MgII\ and \Hb\ emission lines. This combined sample with the high-quality UV and optical spectra provide an unique opportunity to minimize intrinsic scatter and biases, leading to proper calibration of the \MgII-based mass estimator. 
In Section \ref{section:obs}, we describe the sample selection. Section \ref{section:meas} presents the measurements. Section \ref{section:scaling} reports the scaling of line widths and luminosities. Section \ref{section:mbh} presents the \mbh\ calibrations. The discussion and summary are presented in Sections \ref{section:discuss} and \ref{section:sum}, respectively. The following cosmological parameters are used throughout the paper: $H_0 = 70$~km~s$^{-1}$~Mpc$^{-1}$, $\Omega_{\rm m} = 0.30$, and $\Omega_{\Lambda} = 0.70$.

\section{Sample Selection}\label{section:obs}

To increase the dynamic range of AGN luminosity, we combined three different subsamples. 
Firstly, we selected 31 AGNs with $\mathrm{\lambda L_{5100} \sim10^{41.3} - 10^{44.3}}$ \ergs\ from the study of \citet{Bahk+19}, who used the \Hb\ reverberation-mapped AGNs with high-quality UV and optical spectra. Among these 31 AGNs, high quality UV and optical 
spectra were obtained simultaneously for 6 objects using the Space Telescope Imaging Spectrograph (STIS) on the Hubble Space Telescope (HST). Thus, we included these 6 objects (hereafter HST targets), when comparing the line width and luminosity of the \MgII\ and \Hb\ lines.
For the other 25 AGNs, we obtained the high-quality optical spectra with high signal-to-noise ratio (S/N $>$20) from various single-epoch observations from the literature (see Table \ref{table:31sample}). In the case of UV spectra, we used the measurement of \citet{Bahk+19}. The difference in time between the UV and optical observations of these 25 objects are from 1 month to 14 years. Since the line width and the luminosity of continuum change due to time variability, it is not proper to compare line widths or luminosities obtained from different epochs. Thus, we excluded these 25 objects when we compared line widths or luminosities, respectively. In contrast, when we calibrated mass estimators, we included these objects, because the virial product (i.e., $\rm L^{0.5} \times V^{2}$) is constant independent of time variability. For the \MgII\ mass estimators, we found consistent results, including these 25 objects or not.


Secondly, we adopted 52 moderate-luminosity AGNs (i.e., $\mathrm{\lambda L_{5100} \sim10^{43.8} - 10^{44.4}}$ \ergs) at 0.4 $<$ z $<$ 0.6 from \citep{Woo+18}. 
These objects were observed using the Low-resolution Imaging Spectrometer (LRIS) at the Keck telescope and the \MgII\ and \Hb\ lines were obtained at the same time. \cite{Woo+18} presented a detailed study of UV and optical comparison, and we included those measurements in this study. Among the 52 targets, we removed 5 targets with strong internal extinction (see details in \citealp{Woo+18}).

Thirdly, we selected the high-luminosity AGNs from the SDSS archive. Initially we selected 14,367 quasars at a limited redshift range, i.e., 0.4 < z < 0.8 where the SDSS spectral range covered both \Hb\ and \MgII\ lines in the rest frame. Then we focused on only 487 AGNs , which have high-quality spectra, i.e., S/N $\geqslant$ 20 at both 3000 \AA\ and 5100 \AA. Among them, we removed 62 targets which showed strong absorption features in the \MgII\ line profile or have poor results in the emission line fitting. Thus, the SDSS subsample included 425 targets with $\mathrm{\lambda L_{5100} \sim10^{44.5} - 10^{46.5}}$ \ergs. 

Finally, we combined these three subsamples. Note that we corrected the spectral resolution of each instrument when modeling the data for each subsample. The total sample is composed of 503 AGNs, which expands over five orders 
of magnitude in the optical luminosity, i.e., $\mathrm{\log \lambda L_{5100} \sim41.3 - 46.5}$ \ergs, providing a sufficient dynamic range to calibrate
\mbh\ estimators. 

\section{Measurements}\label{section:meas}

As performed in our previous works (e.g., \citealp{McGill+08}; \citealp{Park+15}; and \citealp{Woo+18}), we applied the same procedure of the multi-component spectral analysis for measuring the line widths and luminosities of the \MgII\ and \Hb\ emission lines. In this section, we briefly describe the fitting process.

\subsection{\ion{\rm Mg}{2} and \Hb} \label{section:measure}

In our analysis, for moderate-luminosity AGNs, we adopted the measurements from our previous study in \citet{Woo+18}. In the case of RM sample, we used the measurements of UV spectra from \citet{Bahk+19}, while we measured the optical spectra by our fitting models. For the SDSS sample, the detailed estimation was presented in our previous study in \citet{Le+19}.

The UV spectra were fitted in the range of 2600\AA\ $-$ 3090\AA, where the \MgII\ emission line region (2750\AA\ $-$ 2850\AA) was masked out. The pseudo-continuum was modeled simultaneously with a combination of three components including: a single power-law, a Balmer continuum, and an \FeII\ template based on the I Zw 1 by \citet{Tsuzuki06} (see Figure \ref{fig:allspec}). After subtracting the pseudo-continuum from the observed spectra, we fitted the \MgII\ line by using a sixth-order Gauss-Hermite series (see more details in Section 3.2 in \citealp{Woo+18}). By using the best-fit models which were determined by $\chi^{2}$ minimization using the nonlinear Levenberg-Marquardt least-squares fitting routine technique, MPFIT \citep{markwardt09}, we determined the line width ($\rm FWHM_{MgII}$), line dispersion (the second moment of the line profile, $\rm \sigma_{MgII}$), the luminosity of the \MgII\ line ($\rm L_{MgII}$), and the monochromatic luminosity at 3000\AA\ ($\rm L_{3000}$). The \MgII\ line profile of our sample does not show a clear signature of narrow component. Therefore, similar to other works in the literature (except for \citealp{Wang+09}), we did not subtract the narrow component in measuring $\rm FWHM_{MgII}$. Note that subtracting the narrow component of \MgII\ should be performed with caution since it is difficult to determine how much the narrow component contributes to the line profile. The measurement errors of line width and luminosity were determined based on the Monte Carlo simulations. We generated 100 mock spectra, for which the flux at each wavelength was added randomly based on the flux errors, then we applied the same fitting method for each spectrum. We adopted 1$\sigma$ dispersion of the measured distributions as the error value.

In the case of \Hb, the observed optical spectral range was modeled with a combination of the pseudo-continuum including a single power law, an \FeII\ component based on the I Zw 1 \FeII\ template \citep{BG92}, and a host-galaxy component which was adopted from the stellar template from the Indo-US spectral library in \citet{Valdes04} (see Figure \ref{fig:allspec}). The spectra were fitted in the wavelength ranges of 4430\AA\ $-$ 4770\AA\ and 5080\AA\ $-$ 5450\AA. After subtracting the pseudo-continuum from the observed spectra, we fitted the broad component of the \Hb\ line using a sixth order Gauss-Hermite series (see Section 3.1 in \citealp{Woo+18}). The narrow component of \Hb\ was modeled separately by using the \OIII\ 5007\AA\ best-fit model. The best-fit model was determined using the $\chi^{2}$ minimization from MPFIT. From the best-fit model, we measured the line width ($\rm FWHM_{H\beta}$), line dispersion ($\rm \sigma_{H\beta}$), the luminosity of the \Hb\ line ($\rm L_{H\beta}$), and the monochromatic luminosity at 5100\AA\ ($\rm L_{5100}$). The best-fit models of the optical spectra of the 31 sources from \citet{Bahk+19} are shown in Figures \ref{fig:HST1}$-$\ref{fig:HST3}. Similar to the error measurements of the \MgII\ line, we used the Monte Carlo simulations for determining the errors of the line width and luminosity of \Hb.

Figures \ref{fig:hist_L} presents the luminosity distributions of the sample. The continuum luminosity at 5100\AA\ ($\rm L_{5100}$) or at 3000\AA\ ($\rm L_{3000}$) has a broad dynamic range from $10^{41}$ to $\rm \sim10^{47}\ erg\ s^{-1}$, while the line luminosity $\rm L_{H\beta}$ or $\rm L_{MgII}$ is around 10$^{39}$ to $\rm \sim10^{45}\ erg\ s^{-1}$. The luminosity range of the sample is typically a factor of 2 broader than that of the previous studies, which mainly focused on either high-luminosity objects from SDSS or low-luminosity AGNs. For example, the samples in \citet{Shen+11} and \citet{Trakhtenbrot+12} have continuum luminosity around $\mathrm{\sim 10^{45}-10^{47} erg\ s^{-1}}$.
Figure \ref{fig:hist_W} shows the line width distributions of the sample. The line width $\rm FWHM_{H\beta}$ extends in a range of $\mathrm{\sim 2000-14000\ km\ s^{-1}}$, while the line dispersion $\rm \sigma_{H\beta}$ is around $\mathrm{\sim 1000-5000\ km\ s^{-1}}$. In the case of \MgII, the line width shows smaller ranges at $\mathrm{\sim 2000-9000\ km\ s^{-1}}$ and $\mathrm{\sim 1000-3500\ km\ s^{-1}}$ for $\rm FWHM_{MgII}$ and $\rm \sigma_{MgII}$, respectively.

\begin{figure*}
\centering
	\includegraphics[width = 0.399\textwidth]{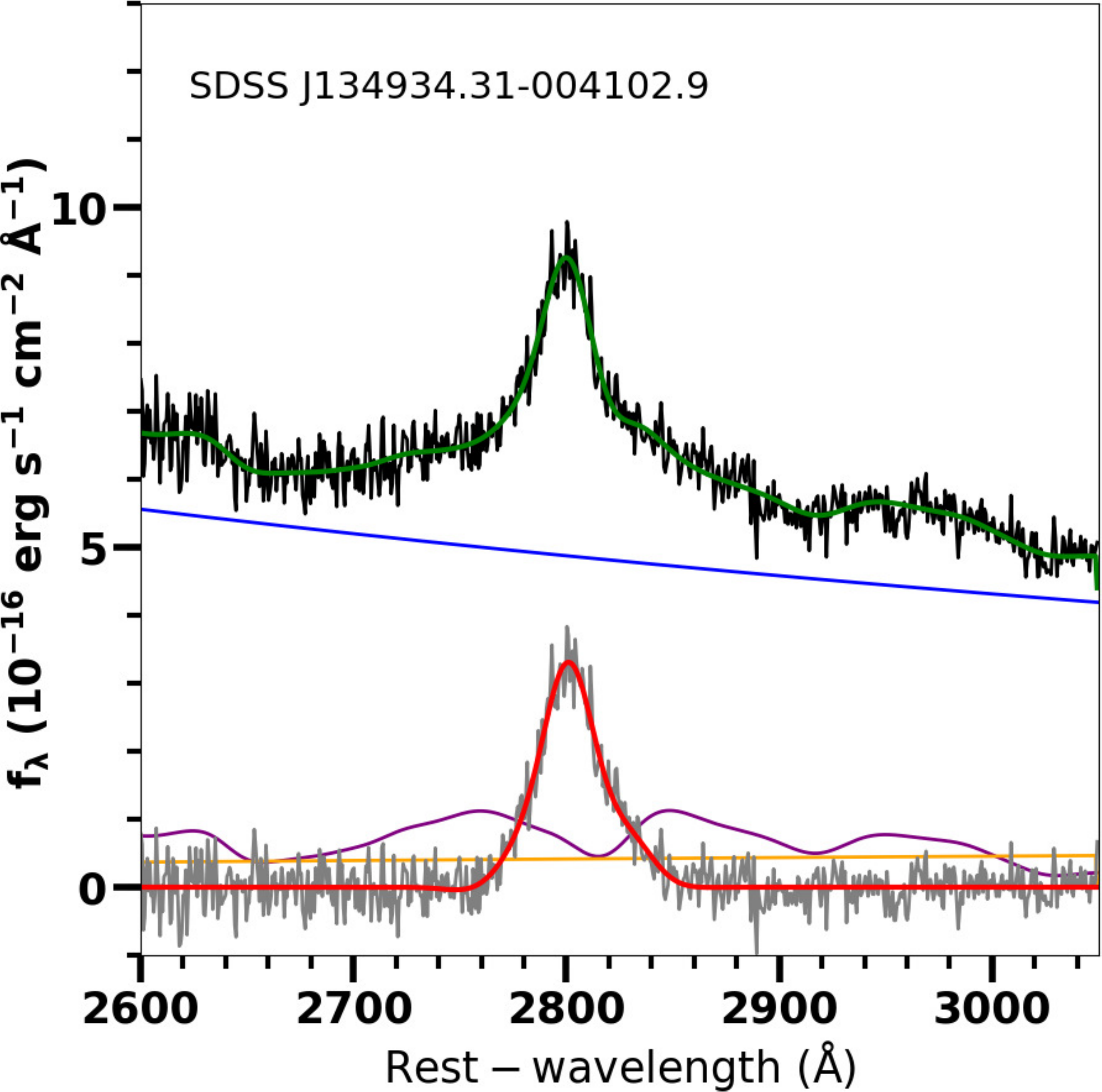}
	\includegraphics[width = 0.59\textwidth]{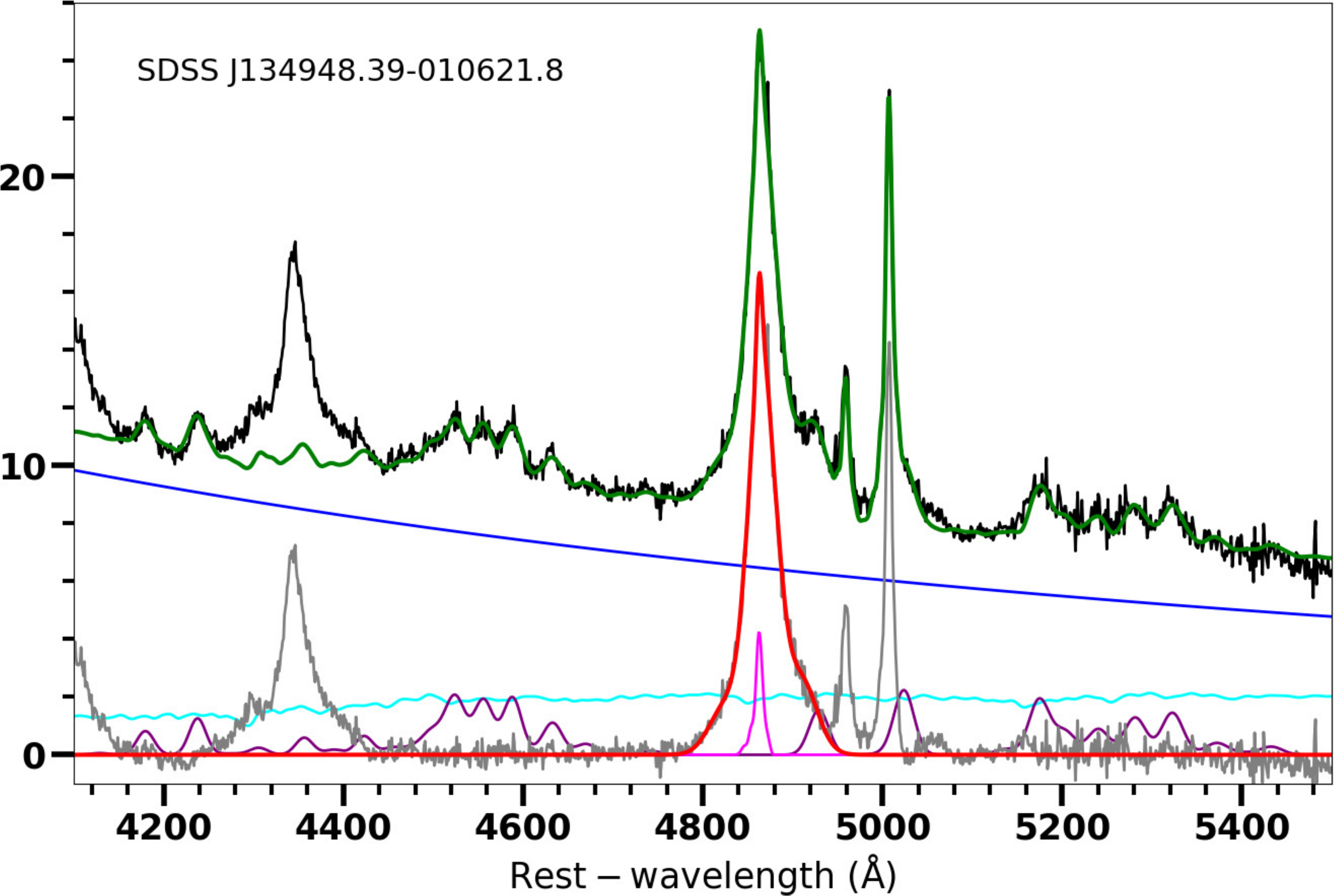}
	\caption{Multi-component fitting results for the \ion{Mg}{2} and \Hb\ emission line regions. Left panel: example of \ion{Mg}{2} fitting for the SDSS spectrum, SDSS J134934.31$-$004102.9. The rest-frame SDSS spectrum is in thick black. The total model (green) includes power-law continuum (blue), \ion{Fe}{2} model (purple), and Balmer continuum model (orange). The continuum subtracted emission line is displayed in gray and the \MgII\ line model is presented in red. Right panel: example of \Hb\ fitting for the SDSS spectrum, SDSS J134948.39$-$010621.8. The color schemes are the same as in the left panel, except that the stellar model is shown in cyan and narrow component of \Hb\ is plotted in pink.}
	\label{fig:allspec}
\end{figure*}

\begin{figure*}
\centering
	\includegraphics[width = 0.8\textwidth]{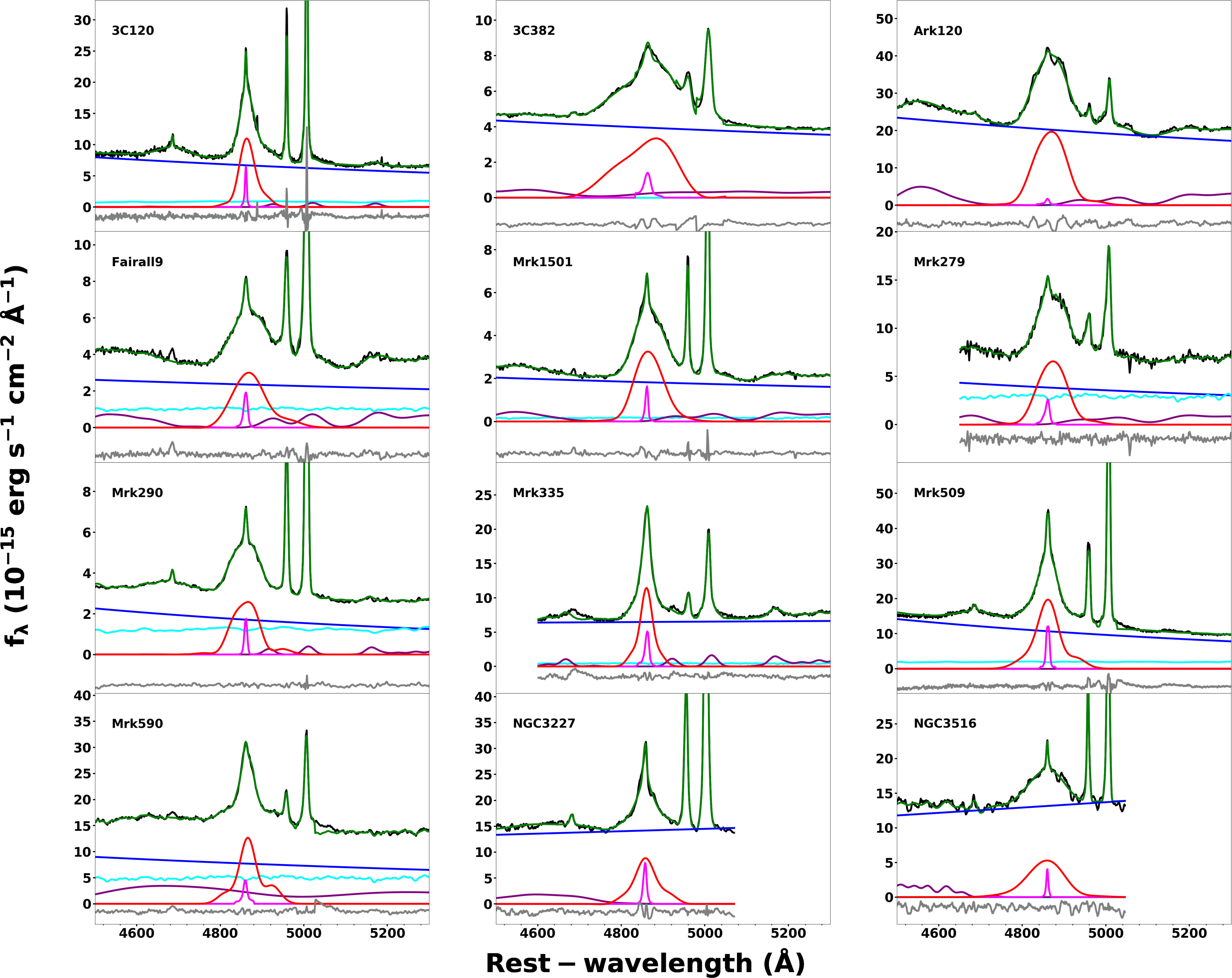}
	\caption{Multi-component fitting results for the \Hb\ emission line regions of 31 sources from \citet{Bahk+19}. The total model (green) includes power-law continuum (blue), \ion{Fe}{2} model (magenta), and Balmer continuum model (orange). The stellar model is shown in cyan and narrow component of \Hb\ is plotted in pink. The continuum subtracted emission line is displayed in gray.}
	\label{fig:HST1}
\end{figure*}

\begin{figure*}
\centering
	\includegraphics[width = 0.8\textwidth]{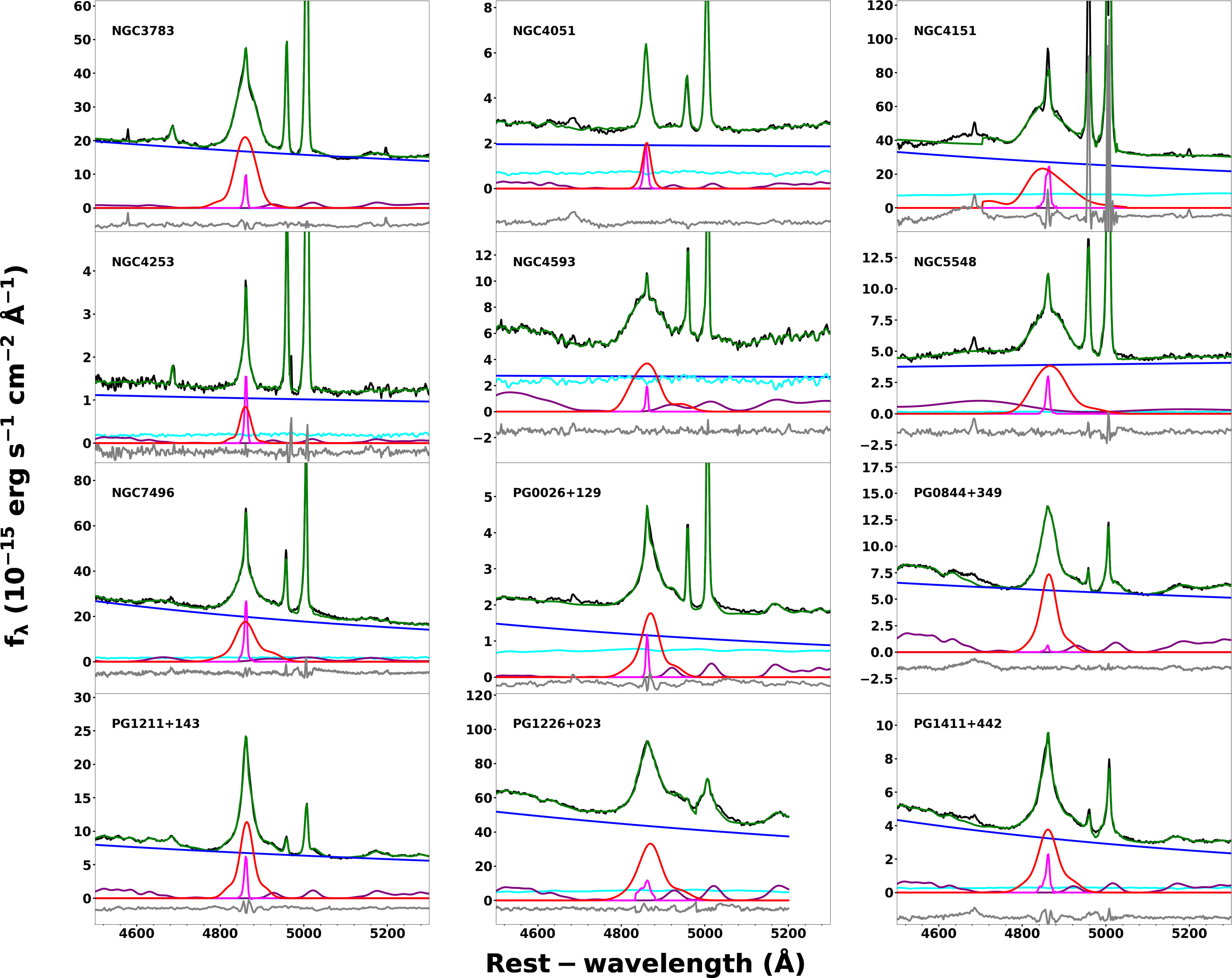}
	\caption{Continue of Figure \ref{fig:HST1}.}
	\label{fig:HST2}
\end{figure*}

\begin{figure*}
\centering
	\includegraphics[width = 0.8\textwidth]{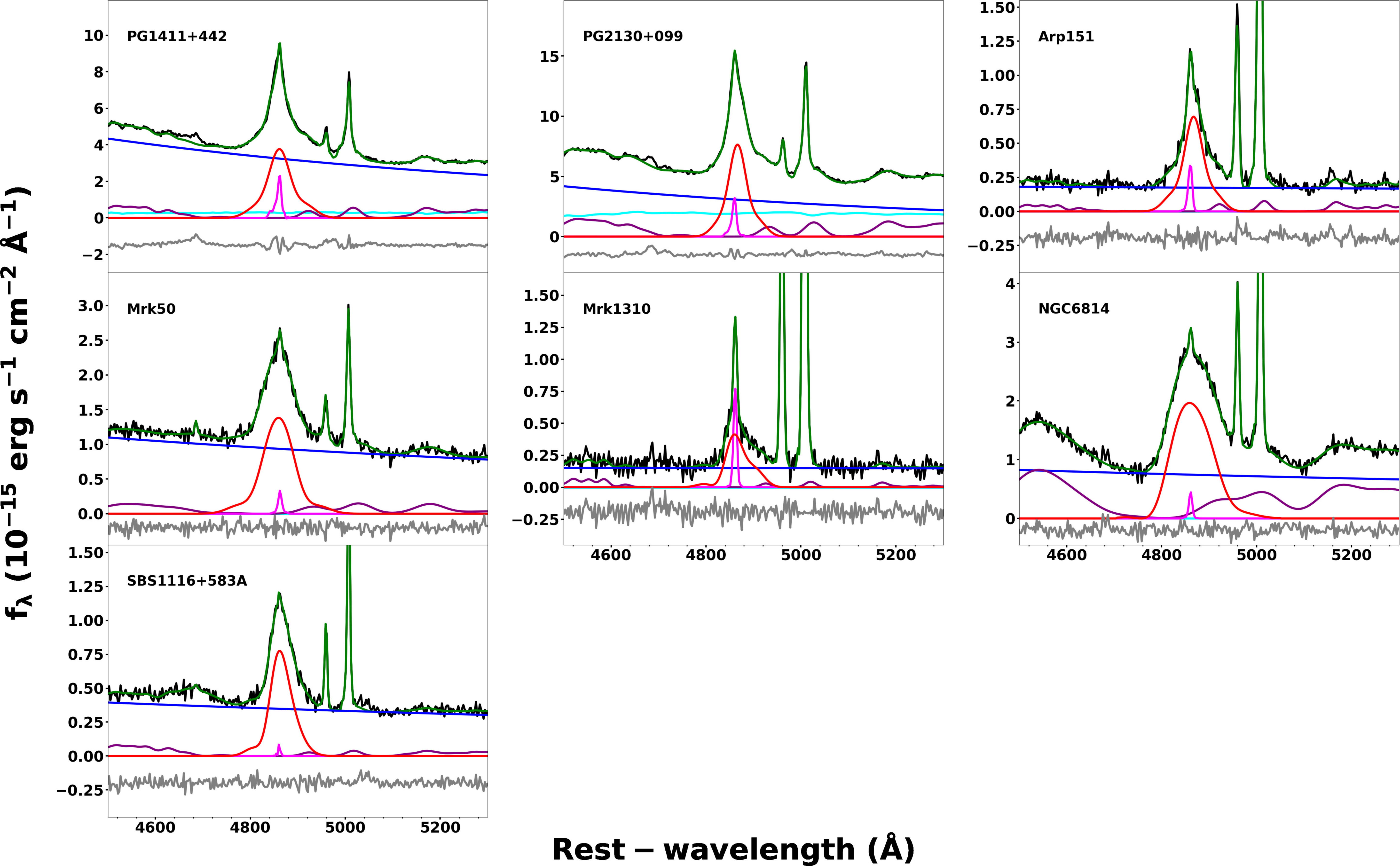}
	\caption{Continue of Figure \ref{fig:HST2}.}
	\label{fig:HST3}
\end{figure*}

\subsection{Line Profiles of \ion{\rm Mg}{2} and \Hb}\label{section:lineprofiles}

In this section, we compare line width (FWHM) and line dispersion ($\sigma$) of the \MgII\ and \Hb\ emission lines to investigate the characteristics of their line profiles (see Figure \ref{fig:compare}). As mentioned in \citet{Bahk+19}, the \MgII\ emission line in the UV spectrum of NGC~4051 showed strong contamination by absorption features while this target has the smallest \mbh. Nonetheless, excluding or including this target in their analysis had no significant effect on the final results. Therefore, we included NGC~4051 in our analysis, but marked it with a different color for clarification in all the figures throughout the paper.

In the case of \MgII, the linear regression between FWHM and $\sigma$ shows a slope of 0.97 $\pm$ 0.03 with an intrinsic scatter $\sigma_{\mathrm{inst}}$ = 0.05 dex, indicating a linear relationship. The ratio of FWHM and $\sigma$ is in a range 1.23$-$3.91, with an average of 1.98 $\pm$ 0.34, which is smaller than the case of a Gaussian profile (i.e., 2.35). 
While the FWHM and $\sigma$ of \MgII\ show a linear relationship in general, we separated the sample into two groups for understanding the line profile of \MgII\ in more detail. We divided the sample at FWHM $=$ 3200 \kms, which is the mean value of the sample, and perform a linear regression. 
Separately, we found that the AGNs with narrower \MgII\ show a slope of 0.70 $\pm$ 0.04 ($\sigma_{\mathrm{inst}}$ = 0.03 dex), while the AGNs with broader \MgII\ have a slope of 0.91 $\pm$ 0.07 ($\sigma_{\mathrm{inst}}$ = 0.06 dex). This difference shows that there is significant change in the line profile between the narrow and broad \MgII\ lines. Narrower \MgII\ lines tend to have broader wings and a narrow core than broader Mg II lines.

In the case of \Hb, FWHM and $\sigma$ show a sub-linear relationship with a slope of 1.48 $\pm$ 0.03 ($\sigma_{\mathrm{inst}}$ = 0.06 dex). The ratio of FWHM and $\sigma$ is in a range of 1.03$-$3.53 with an average of 1.93 $\pm$ 0.40, which is similar to the case of \MgII, albeit with a larger scatter, 0.40 dex. We also separated the sample into two groups for understanding the line profile of \Hb\ in more detail. We divided the sample at FWHM $=$ 4000 \kms, which is the mean value of the sample, and performed a linear regression. We found that the AGNs with narrower \Hb\ show a slope of 1.16 $\pm$ 0.07 ($\sigma_{\mathrm{inst}}$ = 0.05 dex), while the AGNs with broader \Hb\ have a slope of 1.25 $\pm$ 0.06 ($\sigma_{\mathrm{inst}}$ = 0.05 dex). In contrast with the \MgII\ profile, the two groups show consistent slopes. However, as a function of line width, the ratio of FWHM and $\sigma$ increases as line width increases. This result suggests that there is a systematic trend between the narrower and broader \Hb\ lines.

In addition, we compared the difference of line profiles between \Hb\ and \MgII\ as a function of line width. In the case of \Hb, we found that as the line width increases, FWHM-to-line dispersion ratio increases. We found a similar trend for \MgII, but with larger scatter. This result is consistent with those from our previous study with a limited luminosity range by \citet{Woo+18}.

\begin{figure}
 \centering
	\includegraphics[width = 0.23\textwidth]{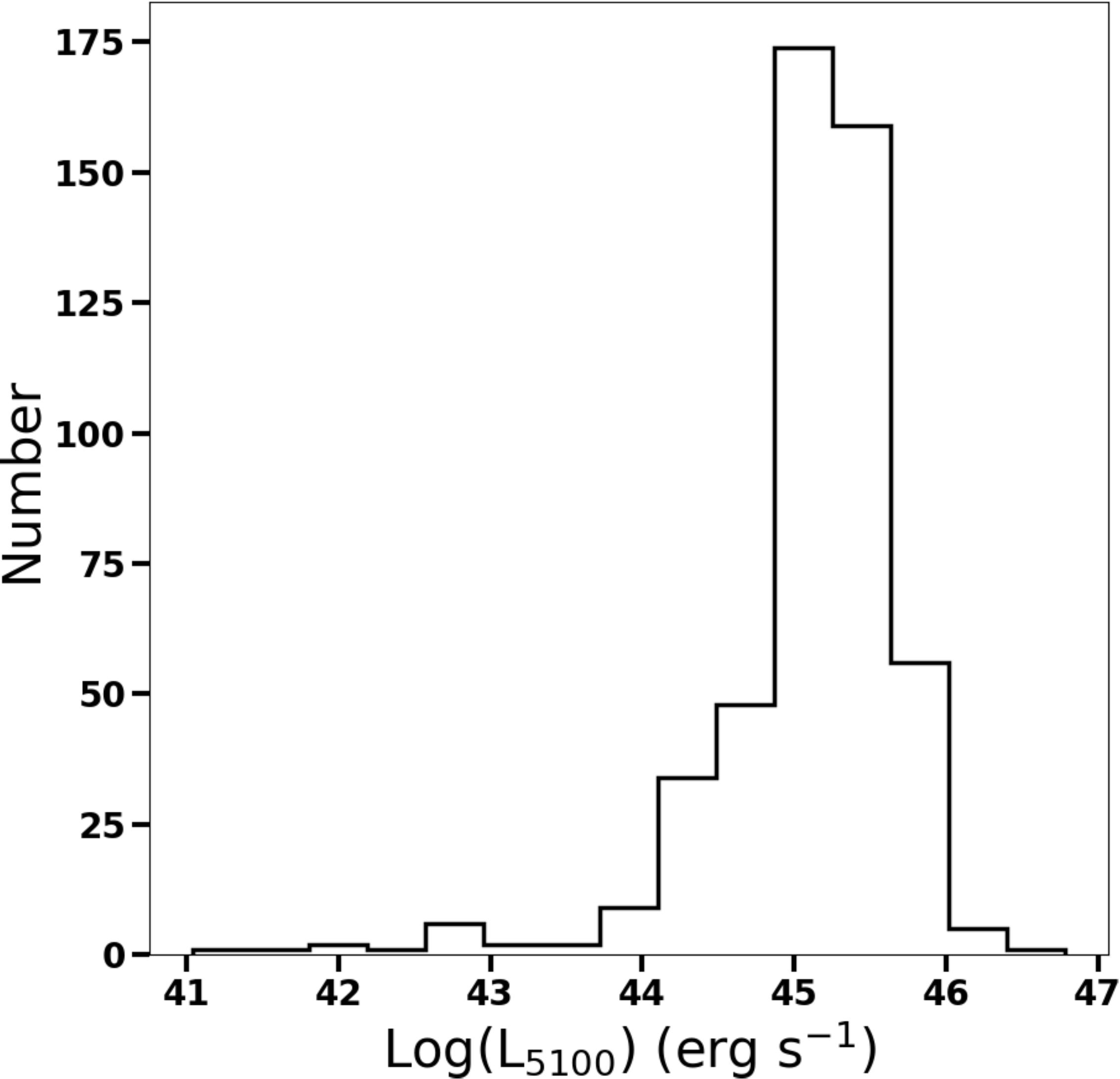}
	\includegraphics[width = 0.23\textwidth]{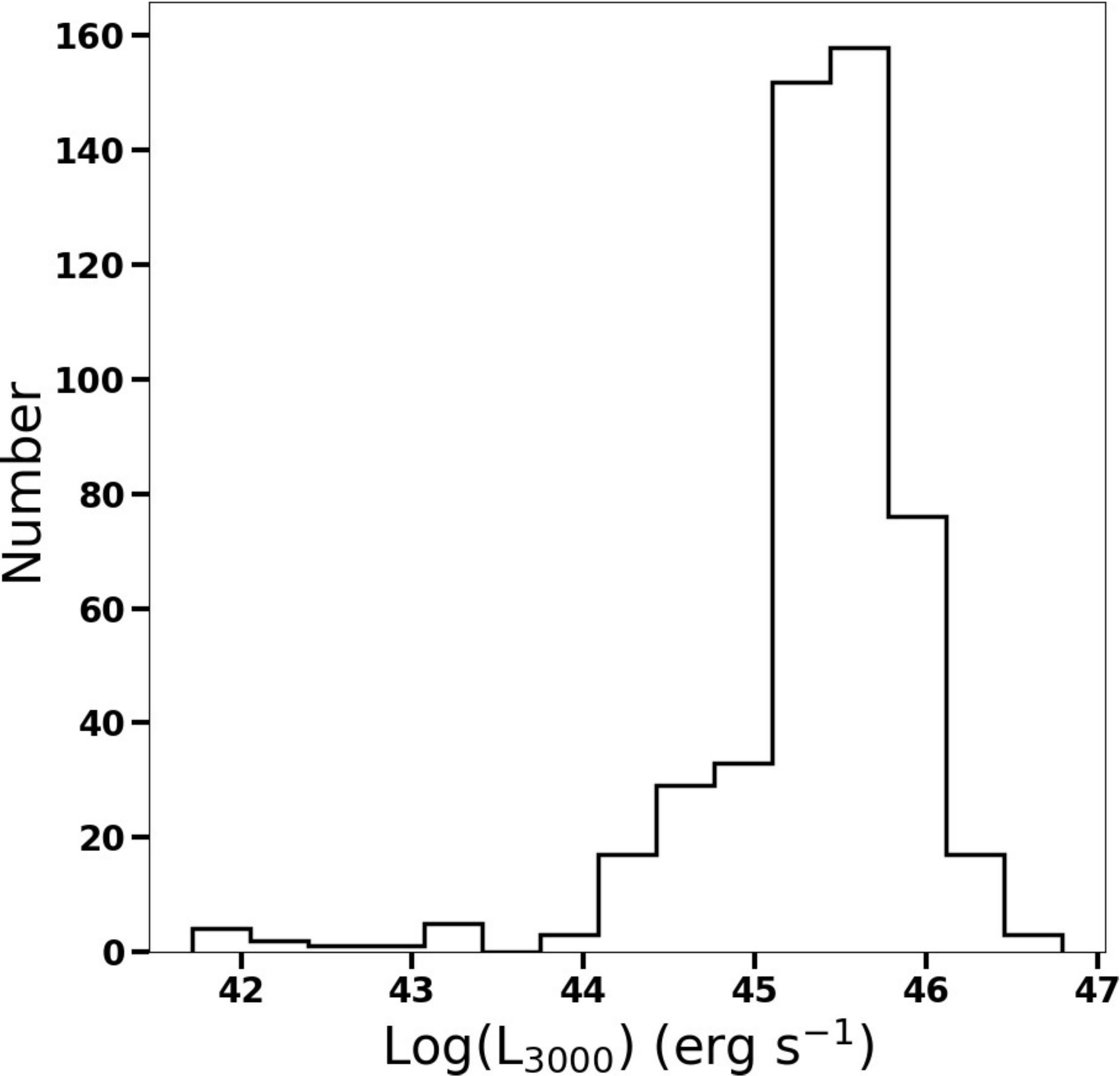} \\
	\includegraphics[width = 0.23\textwidth]{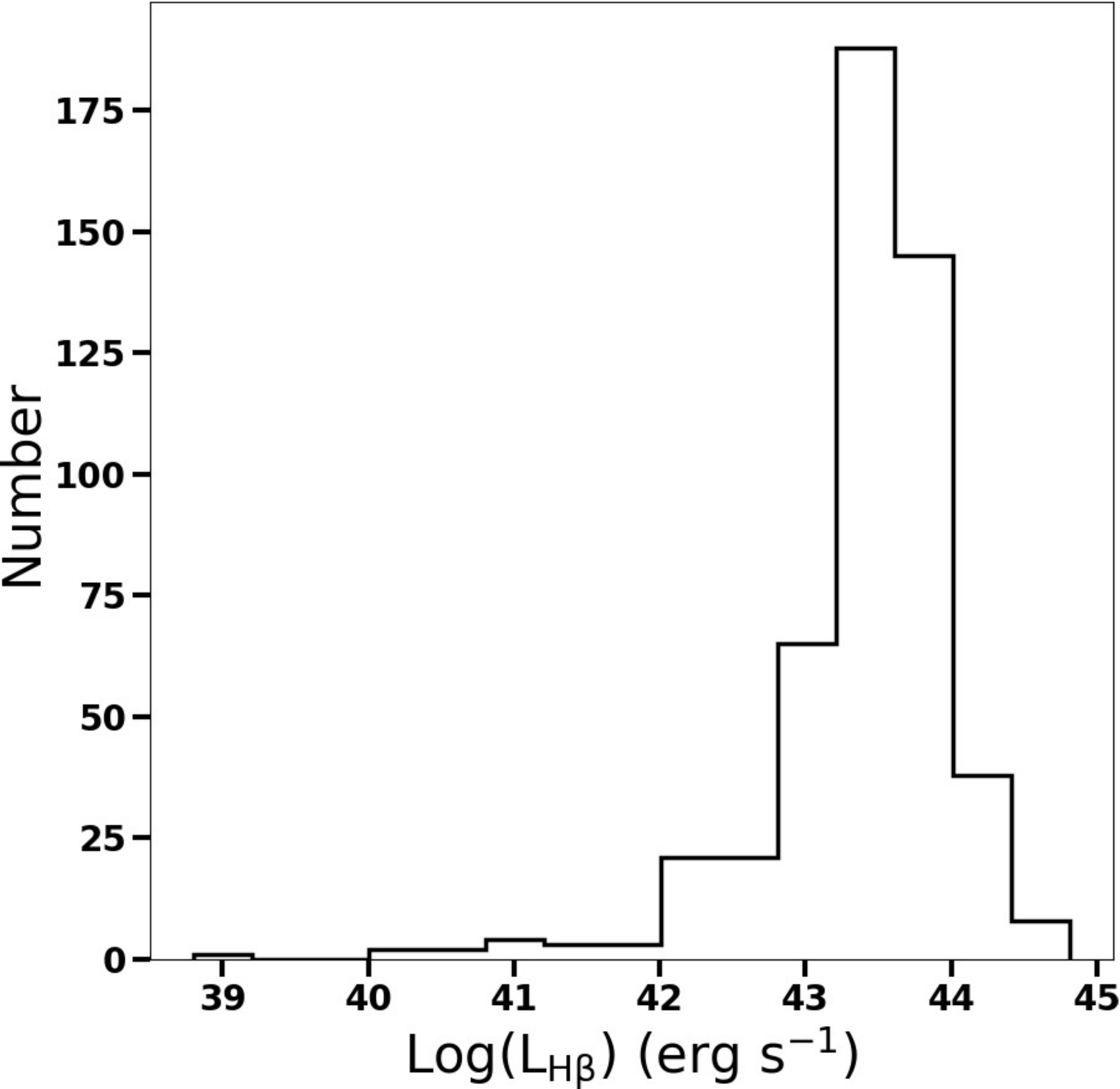}
	\includegraphics[width = 0.23\textwidth]{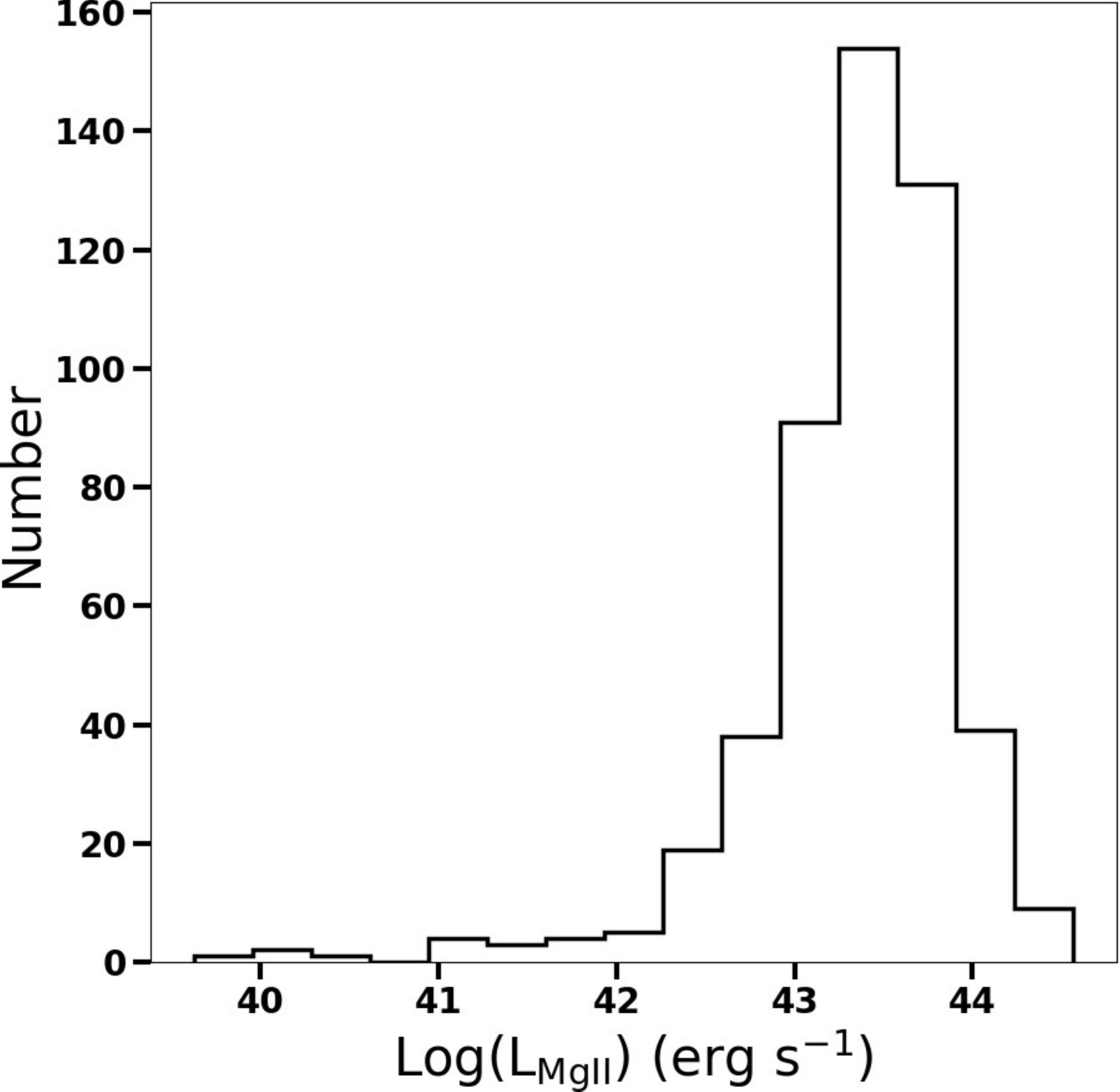}
	\caption{ Upper panels: Distributions of the $\mathrm{L_{5100}}$ (left panel) and $\mathrm{L_{3000}}$ (right panel). Bottom panels: Distributions of the $\mathrm{L_{H\beta}}$ (left panel) and $\mathrm{L_{MgII}}$ (right panel).}
\label{fig:hist_L}
\end{figure}

\begin{figure}
 \centering
	\includegraphics[width = 0.23\textwidth]{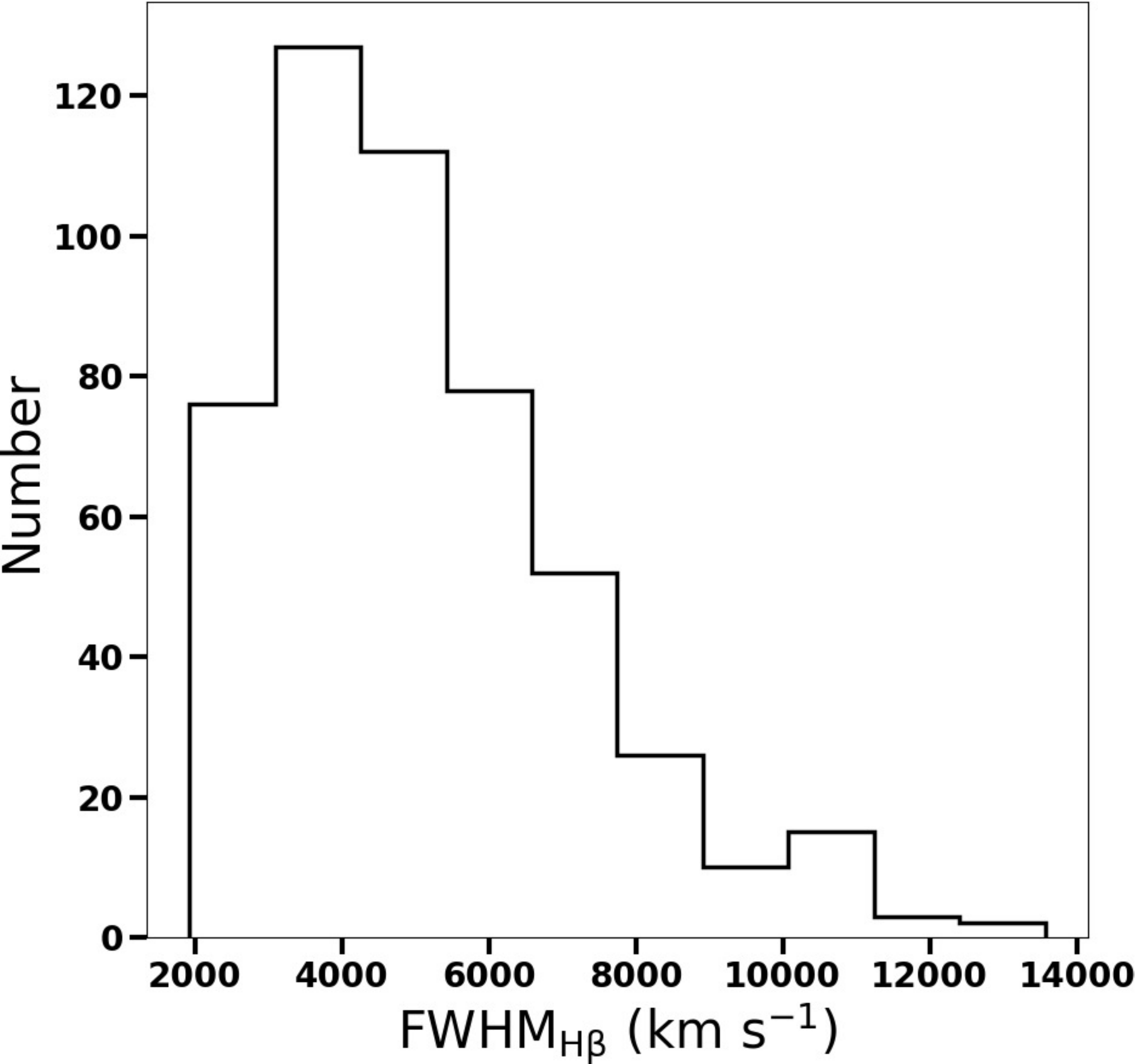}
	\includegraphics[width = 0.23\textwidth]{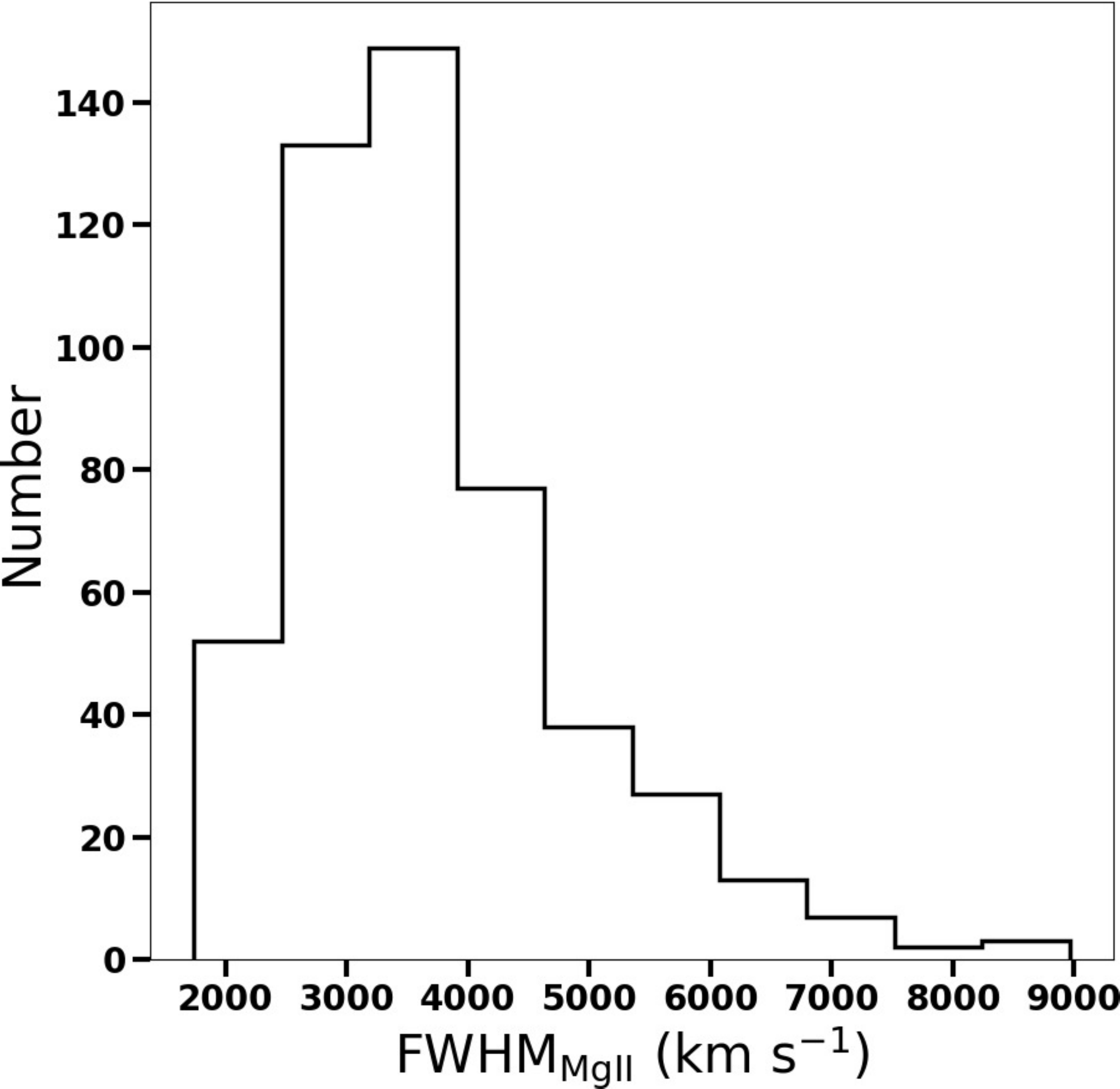} \\
	\includegraphics[width = 0.23\textwidth]{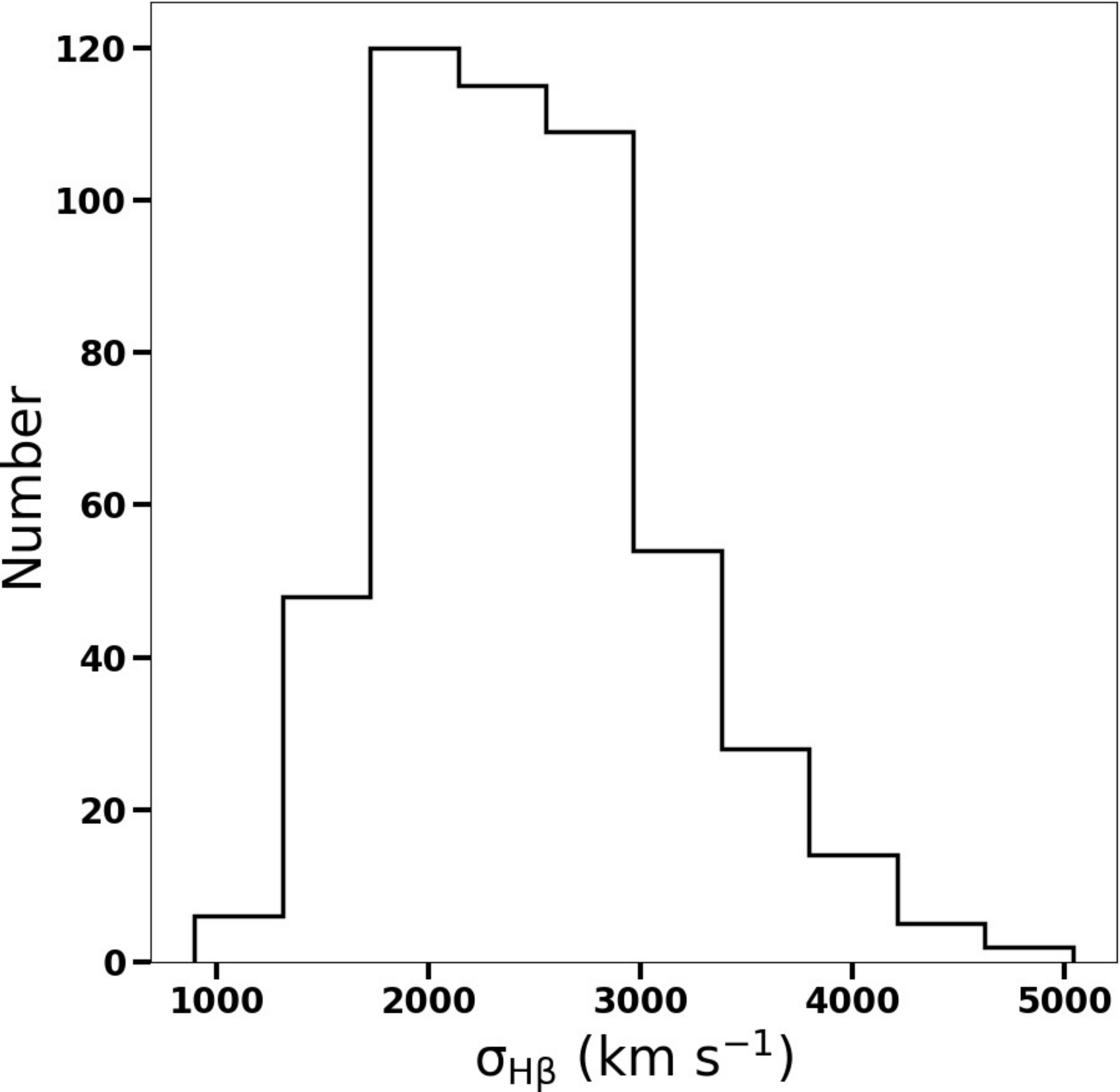}
	\includegraphics[width = 0.23\textwidth]{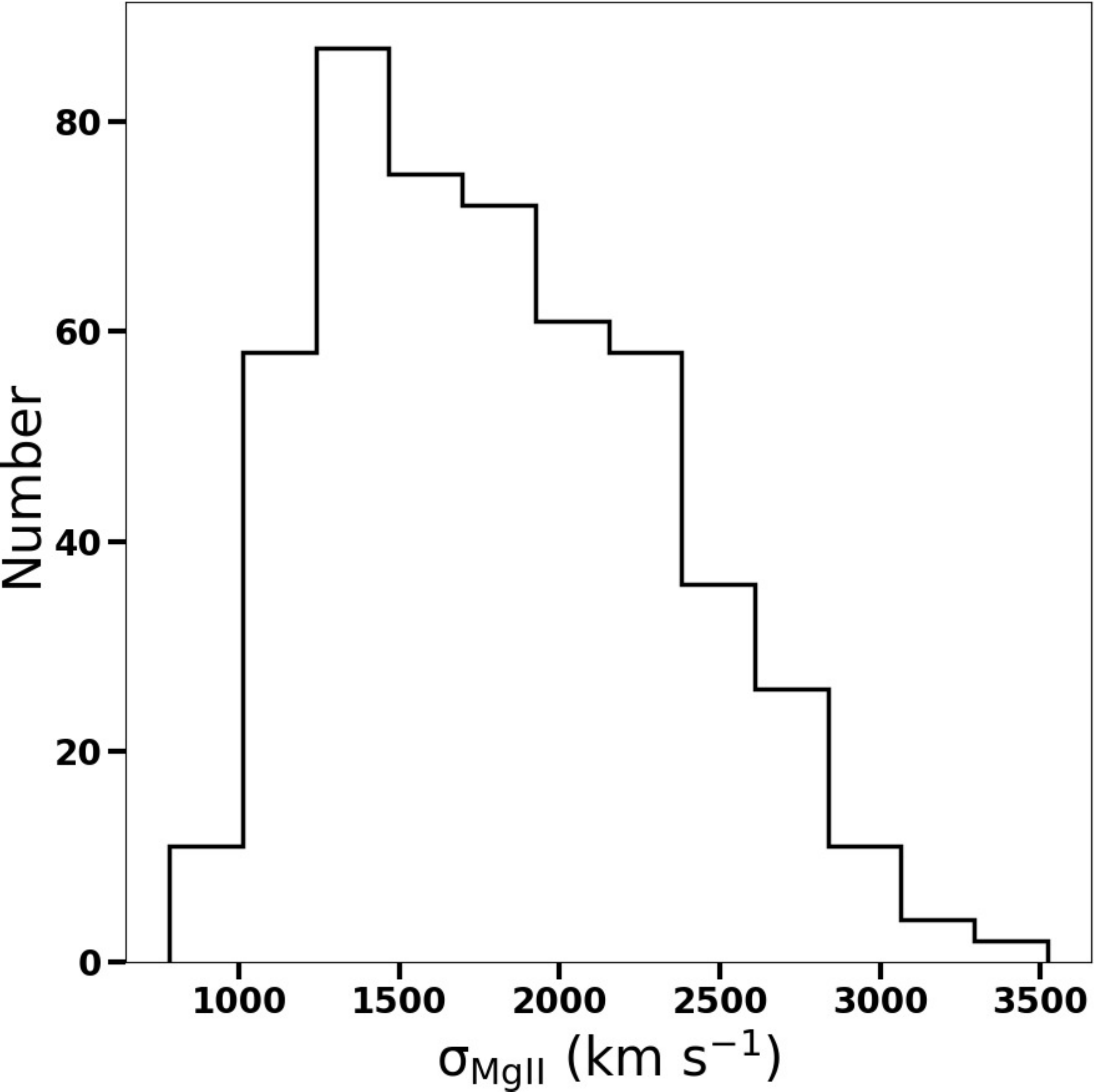}
	\caption{Upper panels: Distributions of the line width $\rm FWHM_{H\beta}$ (left panel) and $\rm FWHM_{MgII}$ (right panel). Bottom panels: Distributions of the $\rm \sigma_{H\beta}$ (left panel) and $\rm \sigma_{MgII}$ (right panel).}
\label{fig:hist_W}
\end{figure}

\section{Line Width and Luminosity Relations} \label{section:scaling}

We applied the cross correlation analysis for the line widths and luminosities. We used the FITEXY method \citep{Park12, Woo+18} to find the best-fit results, including slope, intercept, and intrinsic scatter $\rm \sigma_{inst}$. 

\subsection{Line width comparison}\label{section:wcompare}

In Figure \ref{fig:scale_w}, we compare the line widths between \MgII\ and \Hb\ emission lines. In the case of $\sigma$, we found that \MgII\ $\sigma$ is narrower than that of \Hb\ by $\sim$0.1 dex. The best-fit result is 
\begin{equation}
\begin{aligned}
	\log \left(\frac{\sigma_{\rm Mg \sevenrm II}}{\rm 1000\ km\ s^{-1}}\right) =  (-0.10 \pm 0.01) \ \ \ \ \ \ \ \ \ \ \ \ \ \ \ \ \ \ \ \ \ \ \ \ \ \ \ \             \\                
	                  +\ (0.94 \pm0.03) \times \log \left(\frac{\sigma_{\rm H\beta}}{\rm 1000\ km\ s^{-1}}\right), 	                  
\label{eq:sigcom}
\end{aligned}
\end{equation}
with an intrinsic scatter $\sigma_{\mathrm{inst}}$ = 0.06, indicating a linear relationship of $\sigma$ between \MgII\ and \Hb. The best-fit slope is consistent with our previous study of using 47 intermediate luminosity AGNs by \citet{Woo+18}, who reported the best-fit slope 0.84 $\pm$ 0.07. This result is also consistent with that of \citet{Bahk+19}, who obtained the slope of 0.89 $\pm$ 0.20 using low-luminosity reverberation-mapped AGNs. 

In the case of FWHM, we found a shallower slope than that of line dispersion as:  
\begin{equation}
\begin{aligned}
	\log \left(\frac{\rm FWHM_{\rm Mg \sevenrm II}}{\rm 1000\ km\ s^{-1}}\right) =  (0.11 \pm 0.01) \ \ \ \ \ \ \ \ \ \ \ \ \ \ \ \ \ \ \ \ \ \ \ \ \ \ \ \              \\                
	                  +\ (0.63 \pm0.02) \times \log \left(\frac{\rm FWHM_{\rm H\beta}}{\rm 1000\ km\ s^{-1}}\right), 	                  
	\label{eq:wcom}
\end{aligned}
\end{equation}
with an intrinsic scatter $\sigma_{\mathrm{inst}}$ = 0.07. The best-fit slope is consistent with that of our previous work using moderate-luminosity AGNs \citep{Woo+18}, who reported a slope of 0.60 $\pm$ 0.07, while \citet{Wang+09} obtained a steeper slope of 0.81 $\pm$ 0.02. Note that \citet{Wang+09} subtracted the narrow component of \MgII\ in measuring of $\rm FWHM_{MgII}$. Thus, we expect the slope of \citet{Wang+09} is systematically steeper than that of ours since their $\rm FWHM_{MgII}$ could be overestimated.  

To test the systematic difference between AGNs with broader and narrower lines, we divided the sample into two groups at FWHM $=$ 4000 \kms\
\citep[see][]{Marziani+13}. For the sources with narrower lines, the FWHMs of \MgII\ and \Hb\ show comparable values to each other, while the best-fit slope is 0.59 $\pm$ 0.04 ($\sigma_{\mathrm{inst}}$ = 0.05). In contrast, AGNs with broader lines show a steeper slope of 0.70 $\pm$ 0.05 ($\sigma_{\mathrm{inst}}$ = 0.07) and \MgII\ line is typically narrower than \Hb. Our result is consistent with that of \citet{Marziani+13}, who reported that $\rm FWHM_{H\beta}$ is broader than that of \MgII\ by $\sim$ 20$\%$. These results suggest systematic difference of the line profiles depending on the width of the line.
In addition, we investigated the systematic effect on the slope due to the luminosity or Eddington ratio range. By dividing the sample into two subsamples using the median $\rm L_{3000}$ or the median Eddington ratio, we obtained the best fit for each subsample. However, we found no significant difference of the slope between these subsamples.  

Since we found a sub-linear relationship between the FWHMs of \Hb\ and \MgII, $\beta$ in Equation \ref{eq:mbh2} cannot be the same for \Hb\
and \MgII. In other words, if we use $\beta$ = 2 for \Hb\ based on the virial assumption, we need to use $\beta$ > 2 for \MgII, breaking the virial assumption. The nonlinear relationship between the FWHMs of \Hb\ and \MgII\ raises large uncertainties in \mbh\ estimators. We also note that in the case of AGNs with a broad \Hb\ line, the \Hb\ line profile is complex, while the \MgII\ line profile shows no strong complexity \citep[see Figure 4 in][]{Woo+18}. The asymmetry in the \Hb\ line profile increases with increasing FWHM when FWHM is larger than 4000 \kms\ \citep{Wolf+19}. Thus, the FWHM measurements and the \mbh\ estimates suffer from significantly large uncertainty when the line width is large. In contrast,
we found no such trend in the case of line dispersion, which may suggest that \mbh\ estimators based on the line dispersion of \Hb\ and \MgII\ provide better mass estimates. 

\begin{figure*}
\center
	\includegraphics[width = 0.4\textwidth]{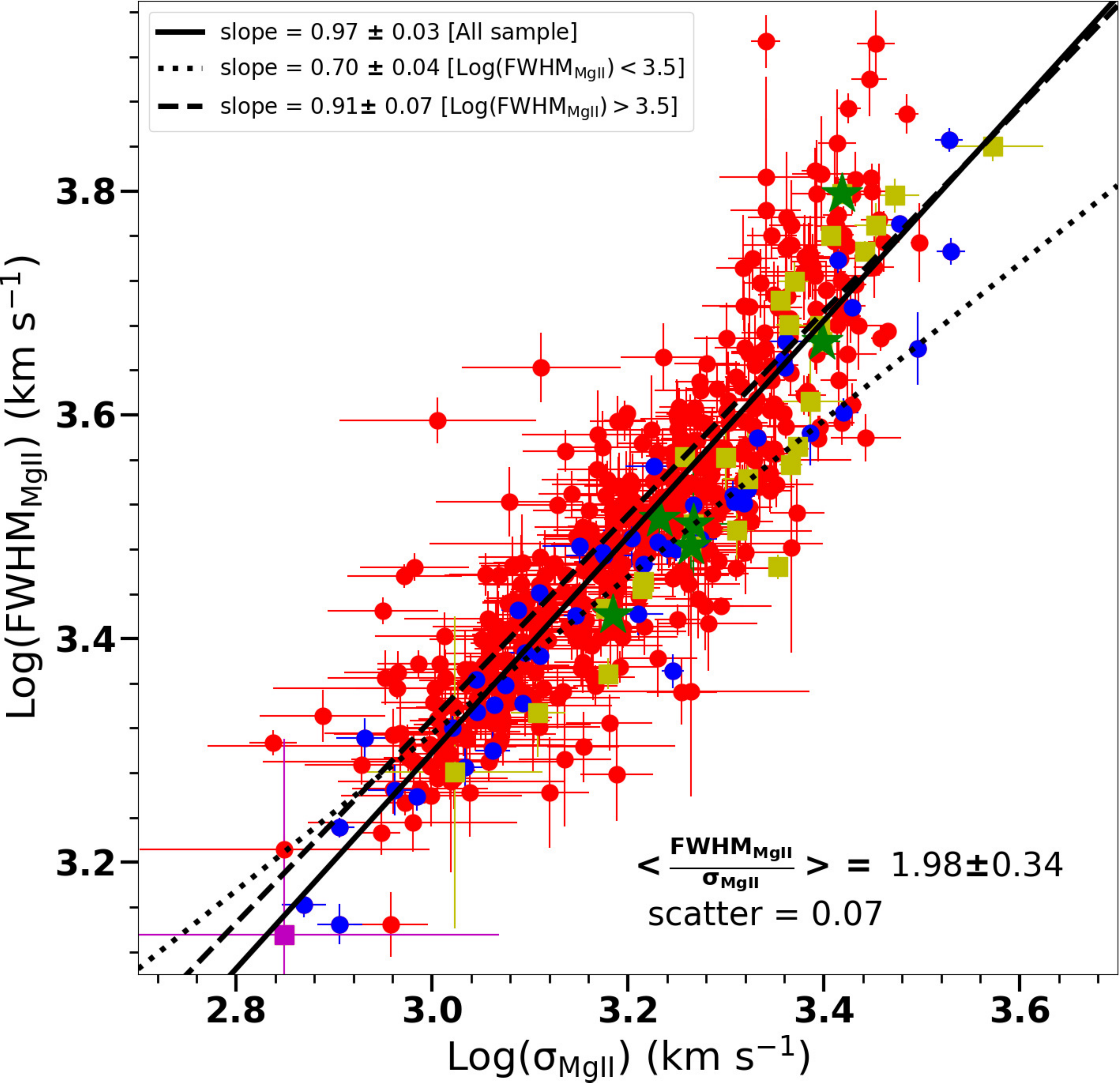}
	\includegraphics[width = 0.4\textwidth]{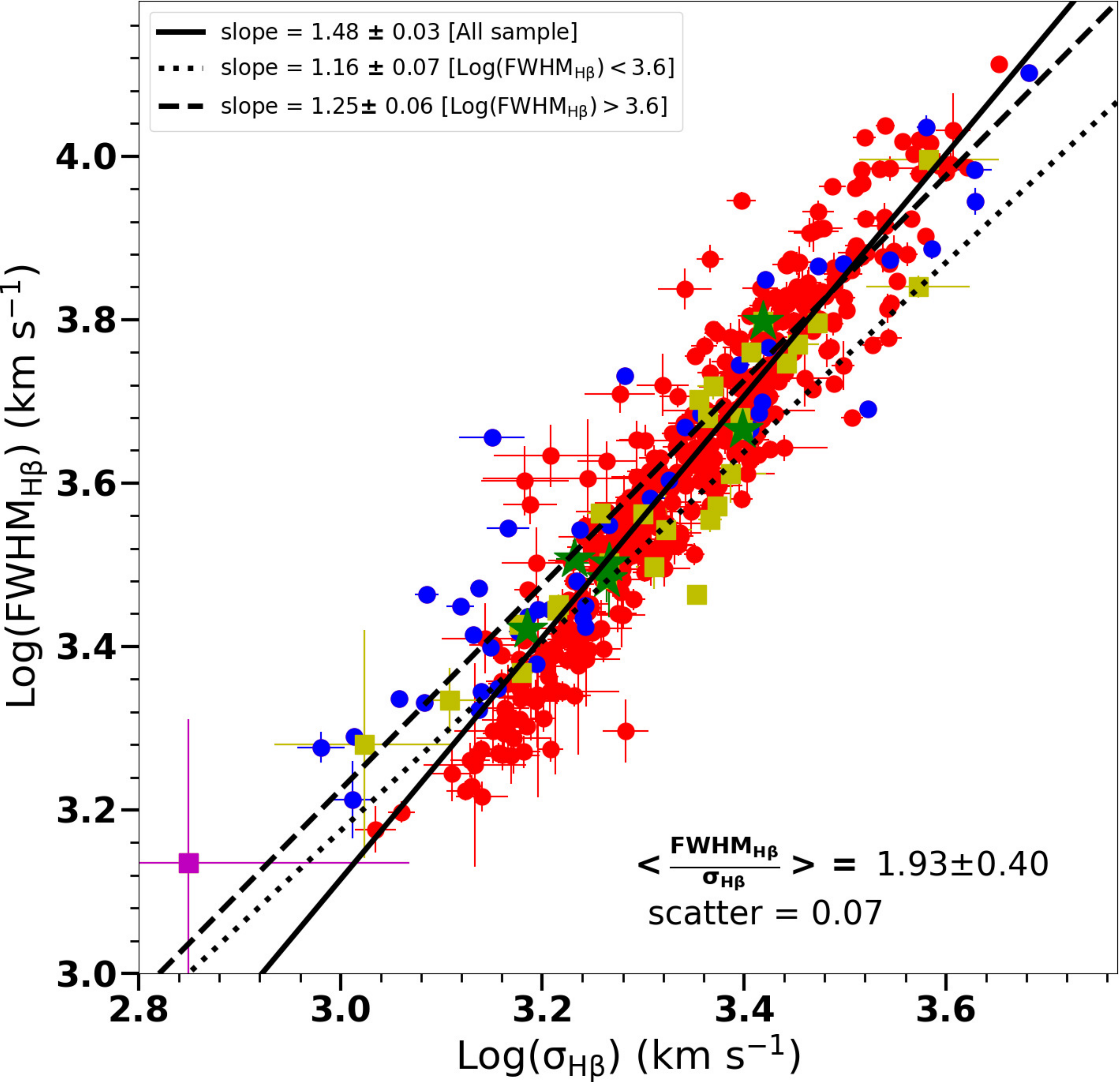}
	\includegraphics[width = 0.41\textwidth]{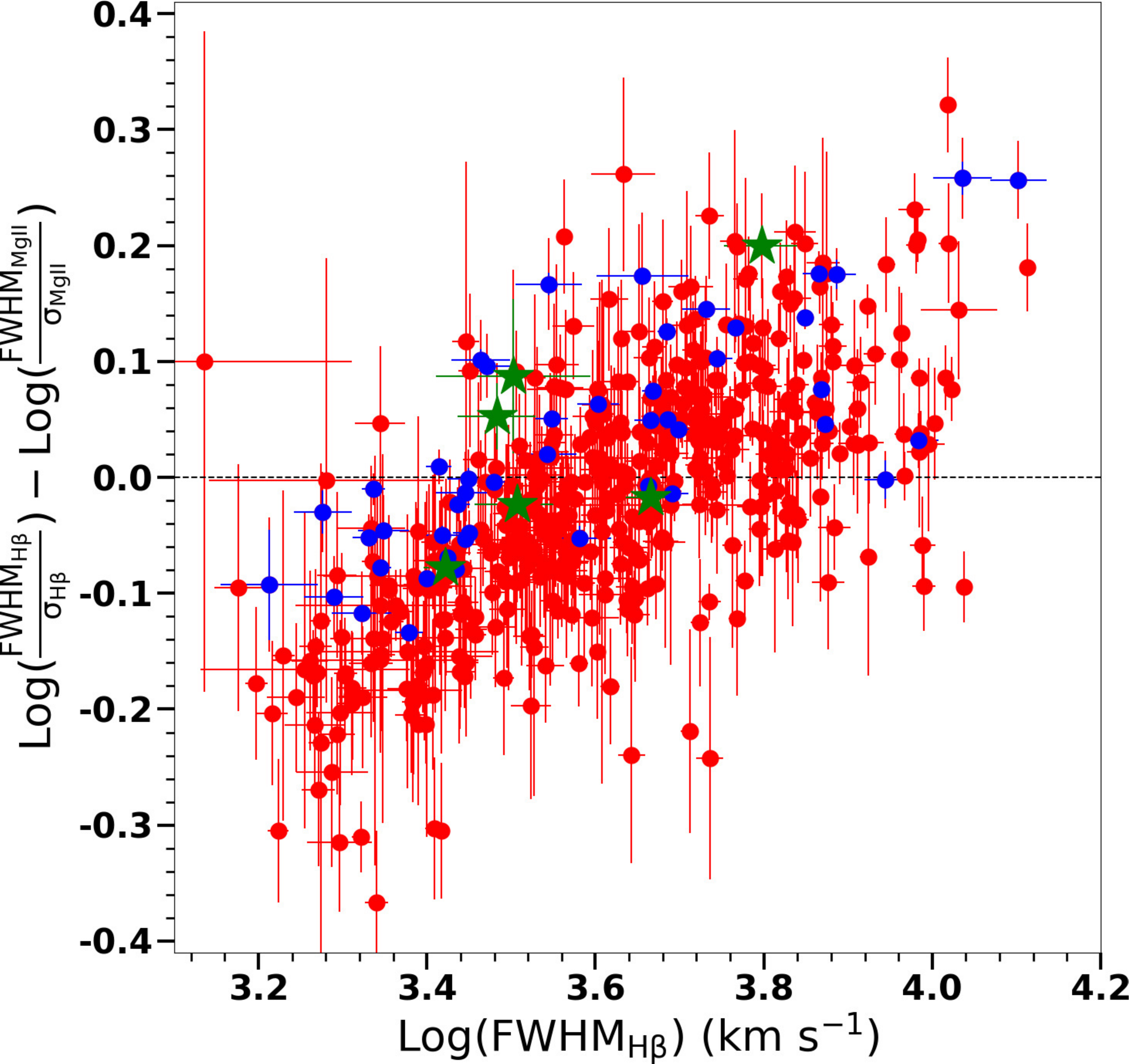}
	\includegraphics[width = 0.41\textwidth]{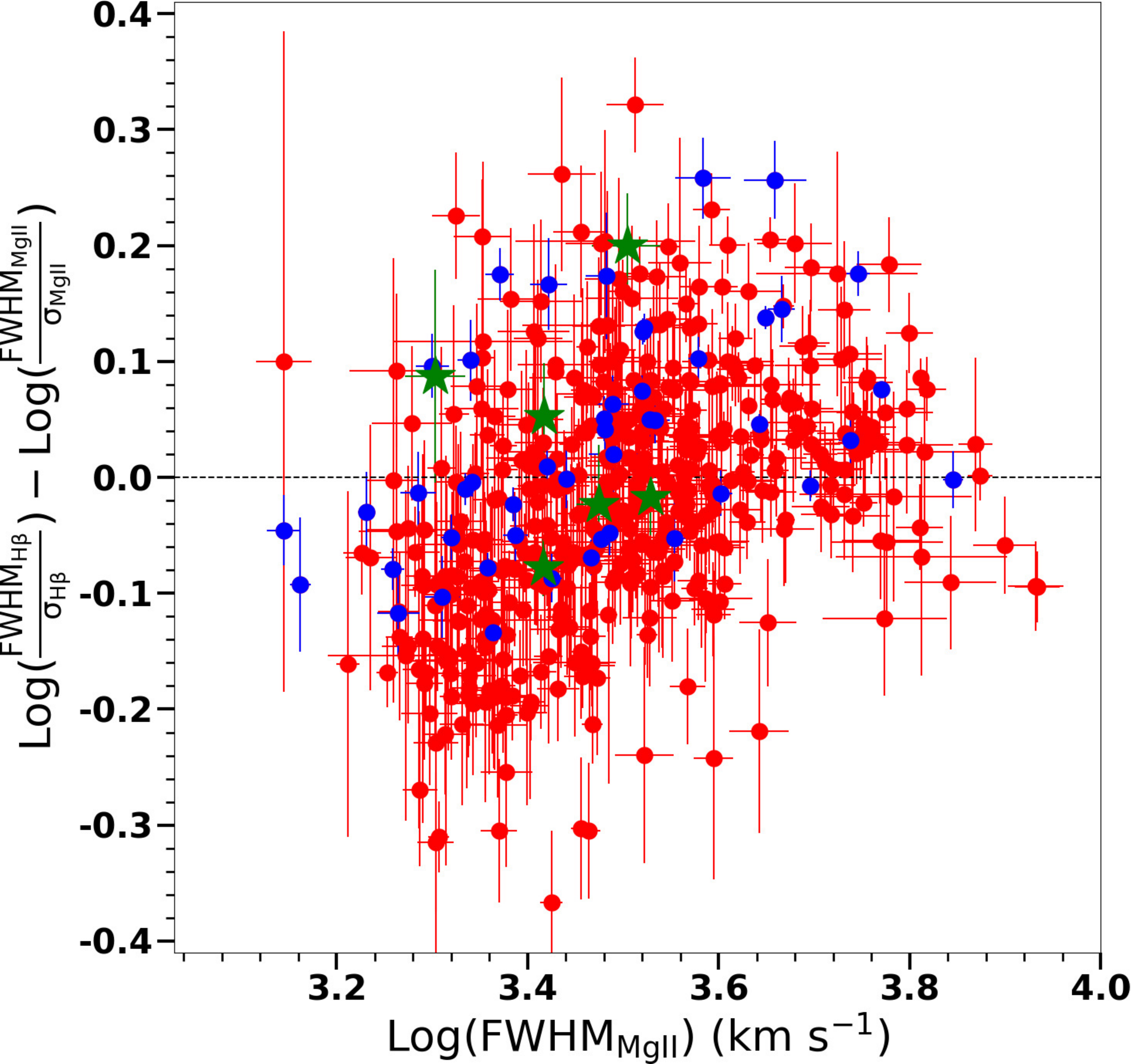}
	\caption{Top left panel: comparison of the line width $\rm FWHM_{MgII}$ and line dispersion $\rm \sigma_{MgII}$. The best-fit slopes are presented for the total sample (solid line),  AGNs with a broader \MgII, i.e., FWHM $>$ 3200 \kms (dashed line), and AGNs with a narrower \MgII, i.e., FWHM $<$ 3200 \kms (dotted line). The symbols represent the moderate-luminosity AGNs (blue), the SDSS sample (red) and the reverberation-mapped AGNs (yellow), the six HST targets (green), and NGC~4051 (pink). Top right panel: same as the top left panel, but for \Hb\ emission line. Bottom panels: comparing the difference of line profile (FWHM and $\sigma$) between \ion{Mg}{2} and \Hb\ emission lines as a function of FWHM of \Hb\ (left panel) and \MgII\ (right panel).}
	\label{fig:compare}
\endcenter
\end{figure*}

\begin{figure}
 \center
	\includegraphics[width = 0.48\textwidth]{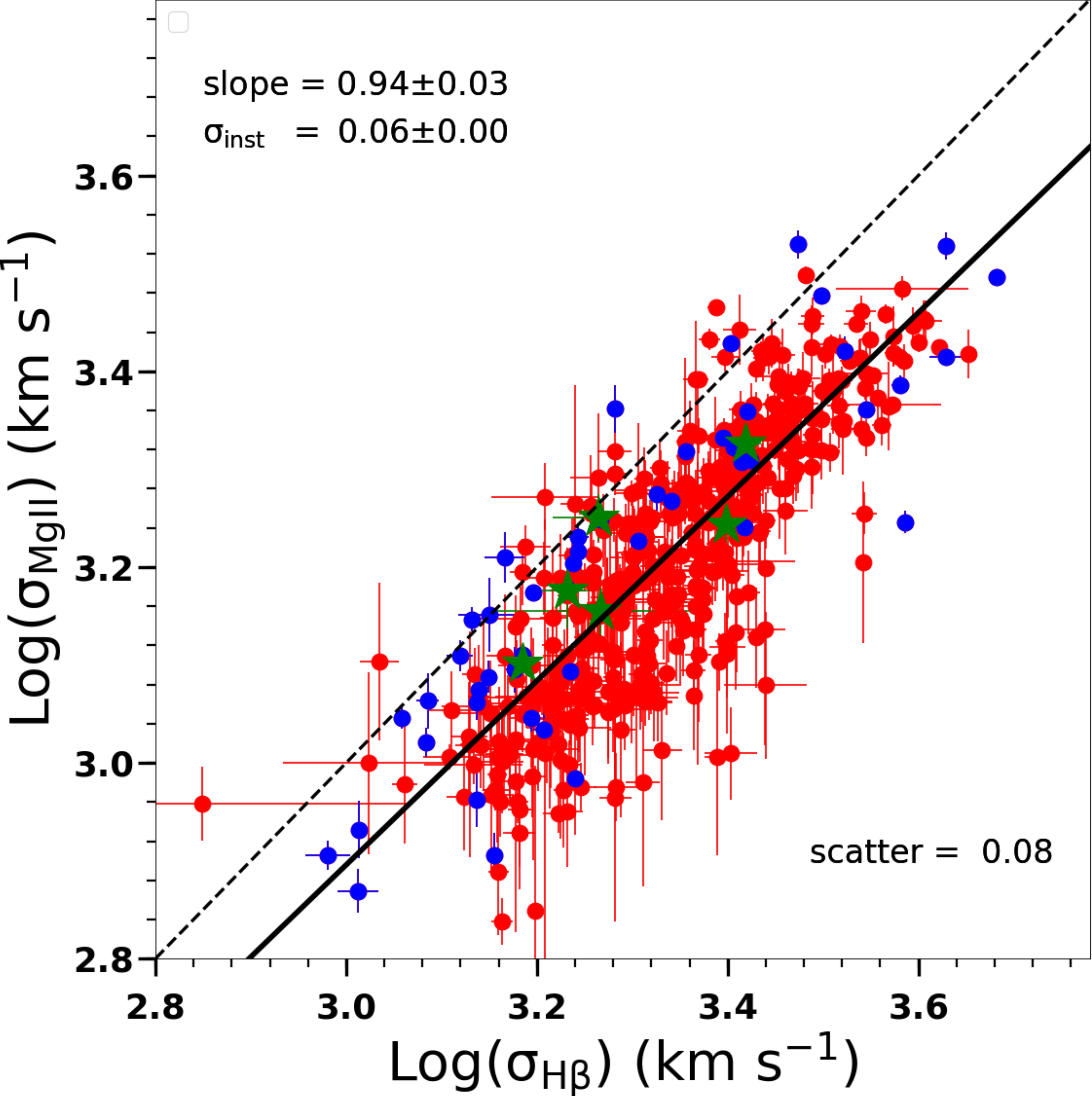}
	\includegraphics[width = 0.49\textwidth]{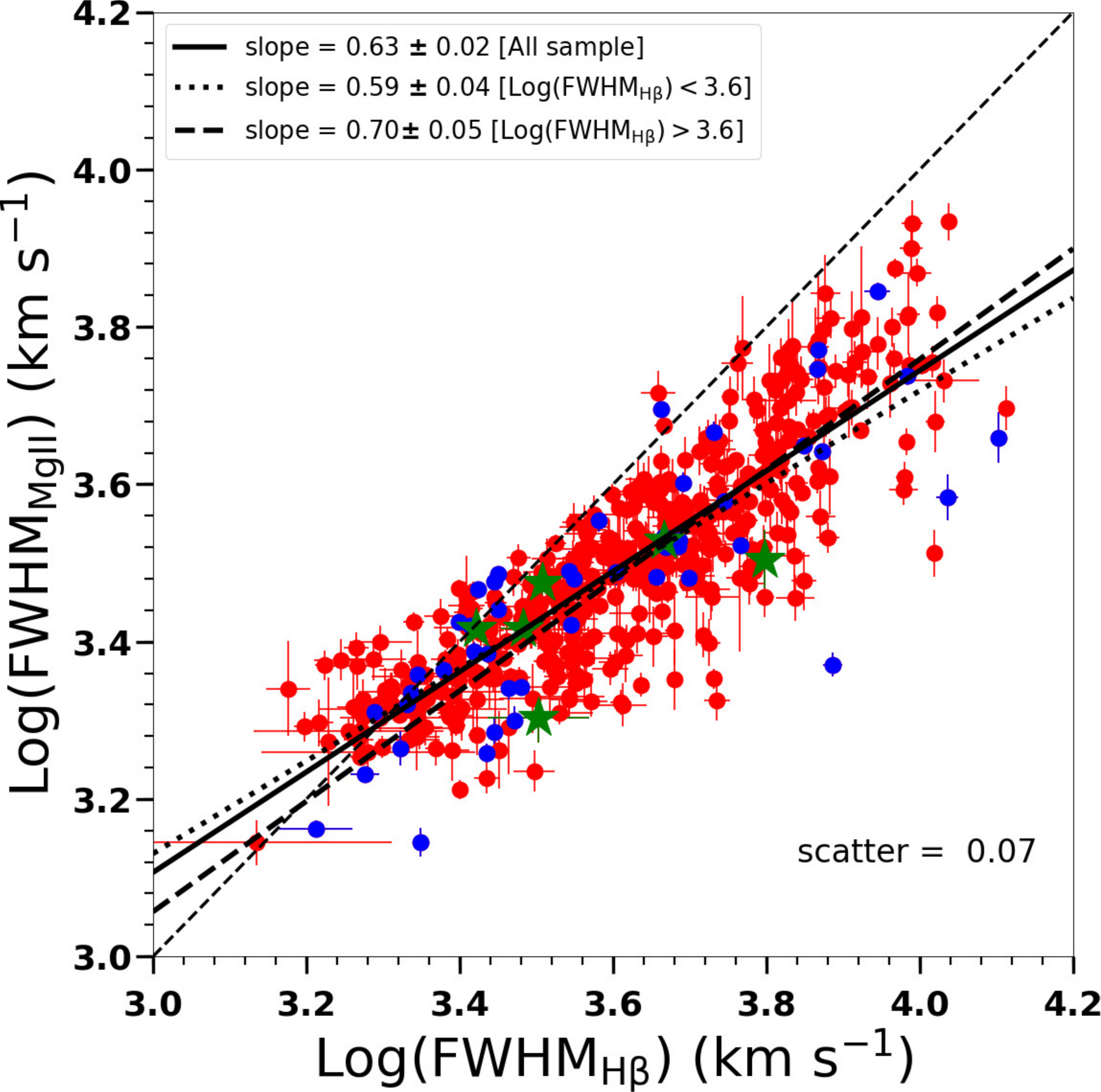}
	\caption{Top panel: comparison of line dispersion $\rm \sigma_{MgII}$ and $\rm \sigma_{H\beta}$. Bottom panel: same as the top panel, but for the comparison of line width $\rm FWHM_{MgII}$ and $\rm FWHM_{H\beta}$. The best-fit slopes are presented for the total sample (solid line), AGNs with a broader \Hb, i.e., FWHM $>$ 4000 \kms (dashed line), and AGNs with a narrower \MgII, i.e., FWHM $<$ 4000 \kms (dotted line). The color schemes of sample are similar to those in Figure \ref{fig:compare}. The symbols are shown for the moderate-luminosity AGNs (blue), the SDSS sample (red), and the six HST targets (green). \label{fig:scale_w}}
\end{figure}

\subsection{Luminosity comparison}\label{section:lcompare}

In Figure \ref{fig:scalelum}, we compare various continuum and emission line luminosities. First, we measured the best-fit slope between $\mathrm{L_{H\beta}}$ and $\mathrm{L_{5100}}$ as,
\begin{equation}
\begin{aligned}
	\log \left(\frac{\rm L_{\rm H\beta}}{\rm 10^{42}\ erg\ s^{-1}}\right) =  (0.31 \pm 0.02) \ \ \ \ \ \ \ \ \ \ \ \ \ \ \ \ \ \ \ \ \ \ \ \ \ \ \ \              \\                
	                  +\ (0.98 \pm0.02) \times \log \left(\frac{\rm L_{\rm 5100}}{\rm 10^{44}\ erg\ s^{-1}}\right), 	                  
	 \label{eq:hbline}
\end{aligned}
\end{equation}
with $\sigma_{\mathrm{inst}}$ = 0.19 $\pm$ 0.01. This slope indicates a linear relationship between $\mathrm{L_{H\beta}}$ and $\mathrm{L_{5100}}$, and which is consistent with our previous work by \citet{Woo+18} while the slope is shallower than 1.13 $\pm$ 0.01 reported by \citet{Greene05}. Using only high luminosity sample, $\rm \log \lambda L_{5100}\ $>$\ 45.4$ \ergs, in contrast, \citet{Shen+12} presented a much steeper slope of 1.25 $\pm$ 0.07. The large dynamic range of our sample may overcome any systematic trend implemented in a limited luminosity range. 

Second, we compared $\mathrm{L_{3000}}$ with $\mathrm{L_{5100}}$, and obtained the best-fit result as, 
\begin{equation}
\begin{aligned}
	\log \left(\frac{\rm L_{3000}}{\rm 10^{44}\ erg\ s^{-1}}\right) =  (0.29 \pm 0.01) \ \ \ \ \ \ \ \ \ \ \ \ \ \ \ \ \ \ \ \ \ \ \ \ \ \ \ \              \\                
	                  +\ (0.98 \pm0.02) \times \log \left(\frac{\rm L_{\rm 5100}}{\rm 10^{44}\ erg\ s^{-1}}\right), 	                  
	\label{eq:lum3000}
\end{aligned}
\end{equation} 
with $\sigma_{\mathrm{inst}}$ = 0.13 $\pm$ 0.01, being consistent with the result of our previous work \citep{Woo+18}. Our result is also consistent with that of \citet{Shen+12}, who presented a slope of 0.98 $\pm$ 0.01. 

Third, we compared the \MgII\ line luminosity with \Hb\ and continuum luminosities.
The best-fit slopes show somewhat sub-linear relationships between those luminosities as 
\begin{equation}
\begin{aligned}
	\log \left(\frac{\rm L_{\rm Mg \sevenrm II}}{\rm 10^{42}\ erg\ s^{-1}}\right) =  (0.43 \pm 0.02) \ \ \ \ \ \ \ \ \ \ \ \ \ \ \ \ \ \ \ \ \ \ \ \ \ \ \ \              \\                
	                  +\ (0.82 \pm0.02) \times \log \left(\frac{\rm L_{\rm 5100}}{\rm 10^{44}\ erg\ s^{-1}}\right), 	                  
	\label{eq:lummgii_5100}
\end{aligned}
\end{equation}
with $\sigma_{\mathrm{inst}}$ = 0.23 $\pm$ 0.01,
\begin{equation}
\begin{aligned}
	\log \left(\frac{\rm L_{\rm Mg \sevenrm II}}{\rm 10^{42}\ erg\ s^{-1}}\right) =  (0.14 \pm 0.03) \ \ \ \ \ \ \ \ \ \ \ \ \ \ \ \ \ \ \ \ \ \ \ \ \ \ \ \              \\                
	                  +\ (0.87 \pm0.02) \times \log \left(\frac{\rm L_{\rm 3000}}{\rm 10^{44}\ erg\ s^{-1}}\right), 	                  
	\label{eq:lummgii_3000}
\end{aligned}
\end{equation}
with $\sigma_{\mathrm{inst}}$ = 0.18 $\pm$ 0.01,
\begin{equation}
\begin{aligned}
	\log \left(\frac{\rm L_{\rm Mg \sevenrm II}}{\rm 10^{42}\ erg\ s^{-1}}\right) =  (0.14 \pm 0.02) \ \ \ \ \ \ \ \ \ \ \ \ \ \ \ \ \ \ \ \ \ \ \ \ \ \ \ \              \\                
	                  +\ (0.87 \pm0.02) \times \log \left(\frac{\rm L_{\rm H\beta}}{\rm 10^{42}\ erg\ s^{-1}}\right), 	                  
	\label{eq:mgii_hbline}
\end{aligned}
\end{equation}
with $\sigma_{\mathrm{inst}}$ = 0.17 $\pm$ 0.01, respectively.
These results are consistent with our previous study with moderate-luminosity AGNs \citep{Woo+18}. The relationship between $\rm L_{Mg \sevenrm II}$ and $\mathrm{L_{5100}}$ is also consistent with that of \citet{Shen+12}, who reported a slope of 0.86 $\pm$ 0.07. In the case of $\rm L_{Mg \sevenrm II}$ with $\mathrm{L_{3000}}$, we obtained a shallower slope than that of \citet{Shen+11}, who showed a slope of 0.98. We expect that the discrepancy is from the difference of luminosity range in the sample. 
\citet{Shen+11} used the high-luminosity SDSS sample, while our sample has a much broader luminosity range. This indicates that there is significant change in the line profile of \MgII. This significant change is also shown in the comparison between FWHM and line dispersion $\sigma$ of \MgII\ in Figure~\ref{fig:compare}.

The comparison between line and continuum luminosity is consistent with that of \citet{Dong+09} who reported a sub-linear relationship between $\rm L_{Mg \sevenrm II}$ and $\mathrm{L_{3000}}$ with a slope of 0.91$\pm$ 0.01. \citet{Dong+09} explained that the sub-linear relationship indicates the Baldwin effect \citep{Baldwin77} in the UV range since for higher luminosity AGNs, the continuum luminosity near the Big Blue Bump will be higher because of the increase of the thermal component in the UV continuum (e.g., \citealp{Malkan&Sargent82}; \citealp{Zheng&Malkan93}). 


\begin{figure*}
 \centering
	\includegraphics[width = 0.42\textwidth]{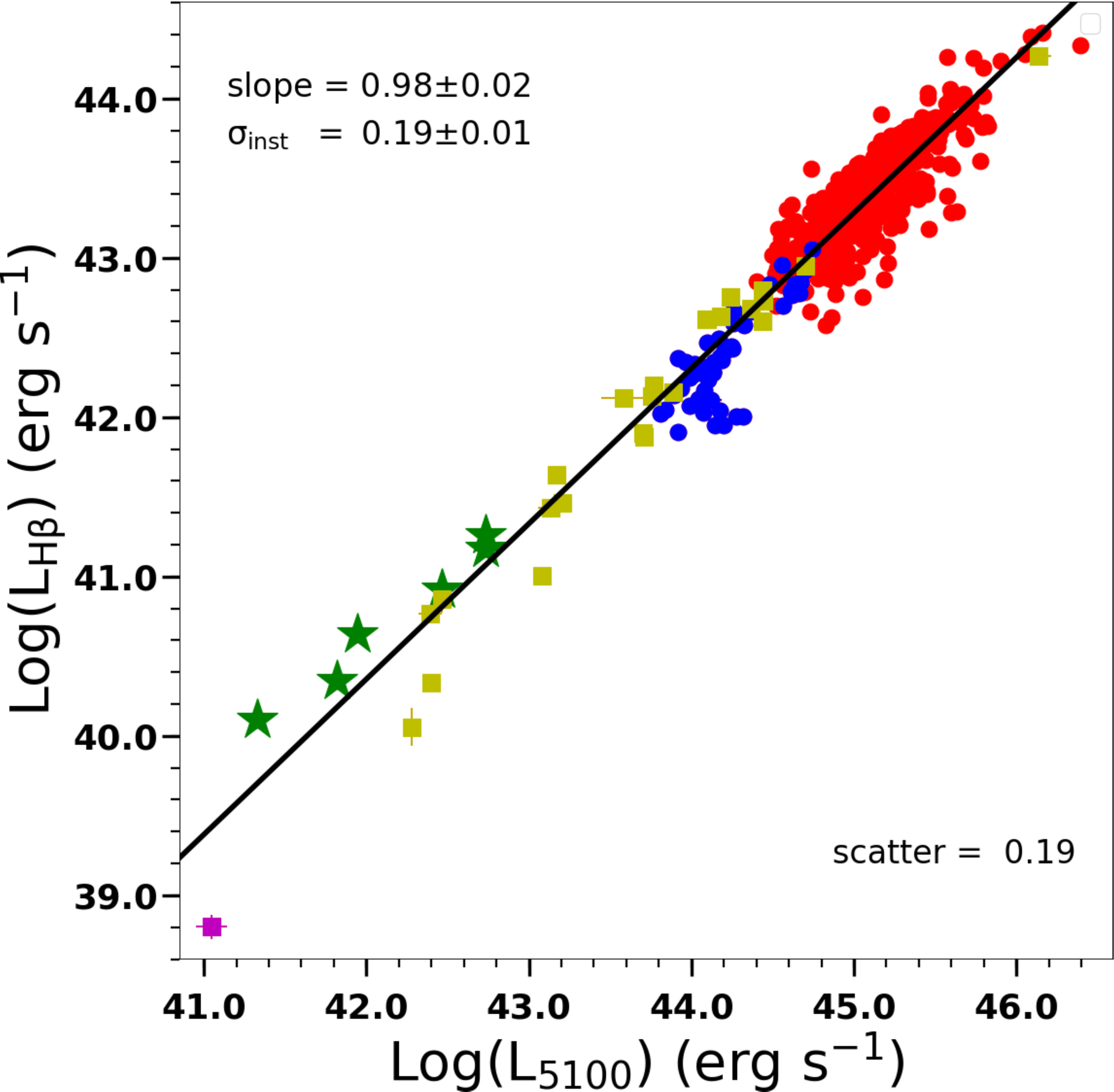} 
	\includegraphics[width = 0.42\textwidth]{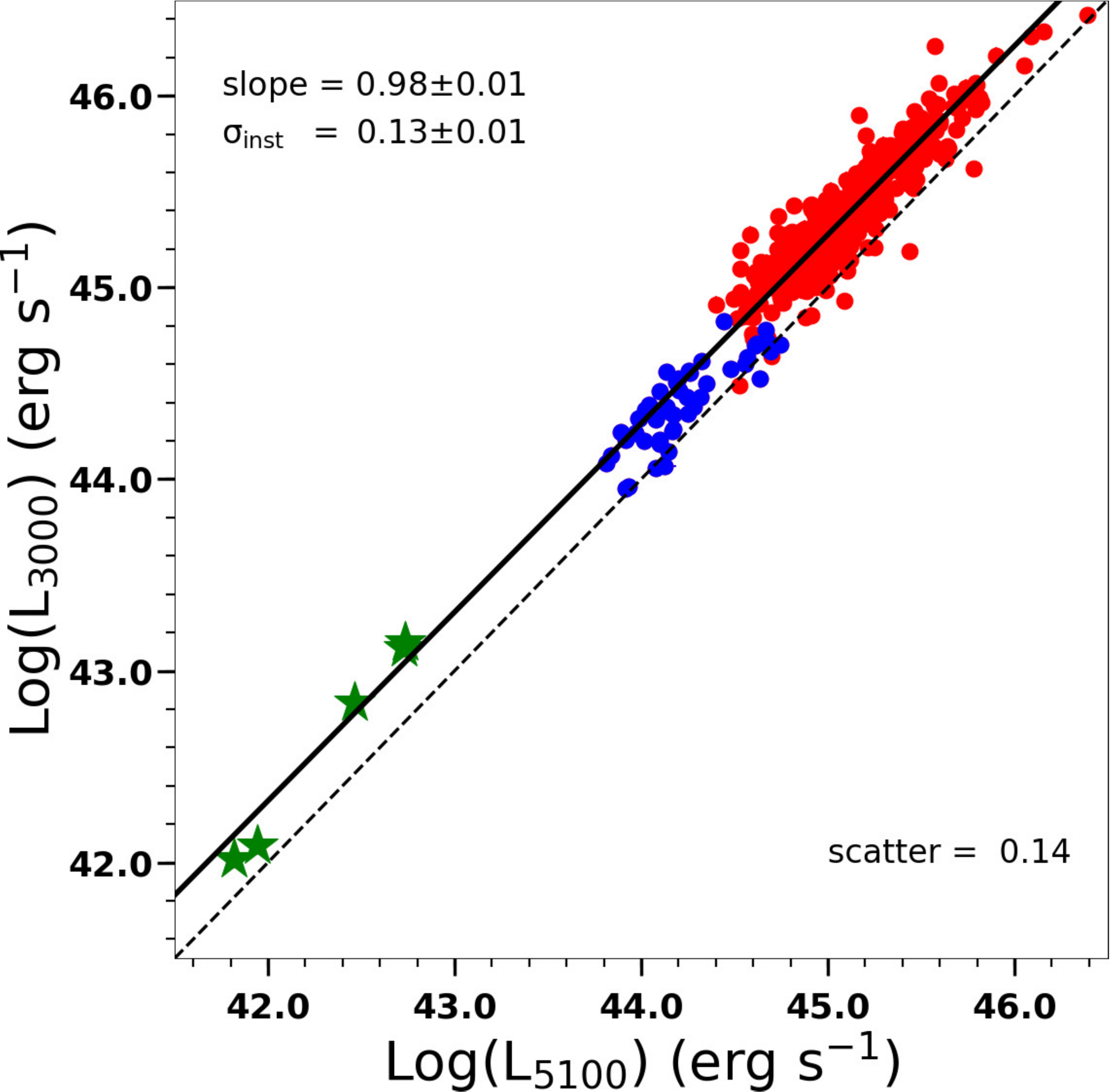}\\
	\includegraphics[width = 0.42\textwidth]{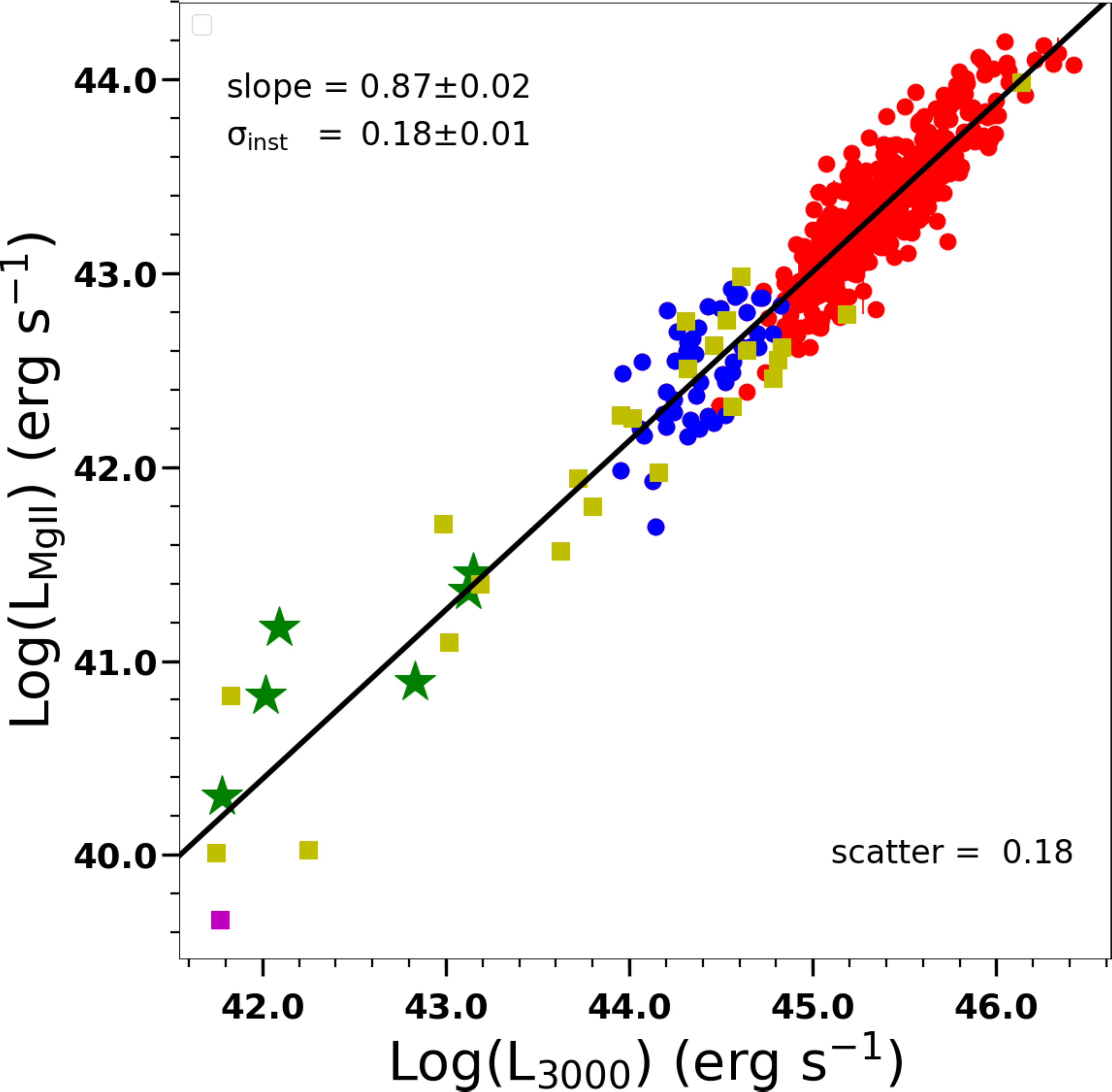}
	\includegraphics[width = 0.42\textwidth]{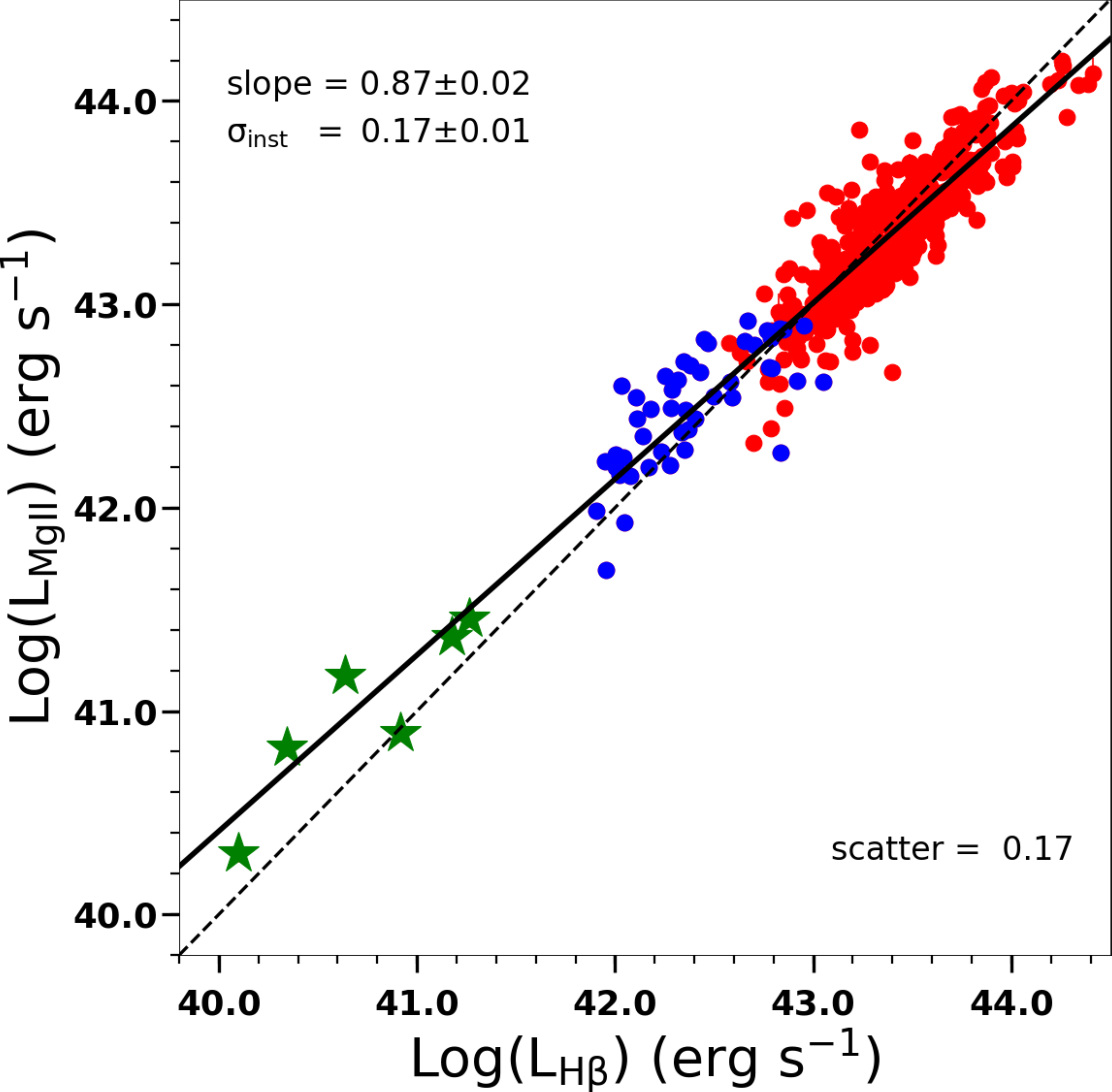}
	\caption{Comparison of continuum and emission line luminosities. Top panels: comparing $\rm L_{5100}$ with $\rm L_{H\beta}$ (left) and $\rm L_{3000}$ (right). Bottom panels: Same as those top panels, but for the comparison of for $\rm L_{3000}$ vs. $\rm L_{MgII}$ (left) and $\rm L_{H\beta}$ vs. $\rm L_{MgII}$ (right). The symbols are shown for the moderate-luminosity AGNs (blue), the SDSS sample (red), the 25 RM sources (yellow), the six HST targets (green), and NGC~4051 (pink). The best-fit slope is shown in thick black solid line. The dash-lines represent one to one relation.}
\label{fig:scalelum}
\end{figure*}

%
%
\begin{table*}
\begin{center}
\tablewidth{1\textwidth}
\fontsize{7}{5}\selectfont
\caption{\label{table:31sample} Optical spectral properties of 31 sample}
\begin{tabular}{cccccccccc}

\tableline
\tableline\\

Target  & z  & Ref  & Date-observation & Gap & S/N & $\rm FWHM_{H\beta}$  & $\rm \sigma_{H\beta}$ & $\rm \log \lambda L_{5100}$ & $\rm \log L_{H\beta}$ \\  
(1) & (2) & (3) & (4) & (5) & (6) & (7) & (8) & (9) & 10 \\

\tableline
\tableline

~ \\

3C120	&	0.033	&	M03	&	24 Sep 1995	&	11	&	39	&	2673	$\pm$	45	&	1505	$\pm$	21	&	43.886	$\pm$	0.001	&	42.157	$\pm$	0.004	\\
3C382	&	0.058	&	B09	&	10 Aug 2007	&	4	&	30	&	9906	$\pm$	425	&	3831	$\pm$	612	&	44.182	$\pm$	0.005	&	42.635	$\pm$	0.026	\\
Ark120	&	0.033	&	M03	&	03 Apr 1990	&	5	&	45	&	5754	$\pm$	66	&	2554	$\pm$	43	&	44.367	$\pm$	0.002	&	42.684	$\pm$	0.005	\\
Fairall9	&	0.047	&	M03	&	20 Dec 1993	&	0.1	&	40	&	5575	$\pm$	120	&	2769	$\pm$	41	&	43.768	$\pm$	0.044	&	42.203	$\pm$	0.006	\\
Mrk1501	&	0.089	&	M03	&	08 Oct 1994	&	2	&	43	&	5037	$\pm$	66	&	2266	$\pm$	47	&	44.238	$\pm$	0.016	&	42.757	$\pm$	0.004	\\
Mrk279	&	0.031	&	M03	&	26 Mar 1989	&	11	&	25	&	5236	$\pm$	208	&	2343	$\pm$	85	&	43.581	$\pm$	0.133	&	42.121	$\pm$	0.021	\\
Mrk290	&	0.030	&	M03	&	16 Feb 1990	&	5	&	45	&	4789	$\pm$	65	&	2314	$\pm$	68	&	43.168	$\pm$	0.024	&	41.640	$\pm$	0.007	\\
Mrk335	&	0.026	&	M03	&	13 Oct 1996	&	11	&	39	&	2158	$\pm$	199	&	1284	$\pm$	83	&	43.706	$\pm$	0.035	&	41.877	$\pm$	0.015	\\
Mrk509	&	0.034	&	M03	&	12 Oct 1996	&	4	&	51	&	3733	$\pm$	42	&	2364	$\pm$	28	&	44.097	$\pm$	0.011	&	42.613	$\pm$	0.002	\\
Mrk590	&	0.026	&	M03	&	13 Oct 1996	&	5	&	48	&	2911	$\pm$	74	&	2256	$\pm$	45	&	43.756	$\pm$	0.013	&	42.136	$\pm$	0.005	\\
NGC3227	&	0.004	&	Ho95	&	29 Mar 1986	&	14	&	20	&	3647	$\pm$	267	&	1995	$\pm$	105	&	42.396	$\pm$	0.003	&	40.332	$\pm$	0.013	\\
NGC3516	&	0.009	&	Ho95	&	29 Mar 1986	&	10	&	20	&	6253	$\pm$	221	&	2969	$\pm$	172	&	43.083	$\pm$	0.003	&	41.006	$\pm$	0.015	\\
NGC3783	&	0.010	&	M03	&	23 May 1993	&	1	&	42	&	3654	$\pm$	70	&	1811	$\pm$	58	&	43.209	$\pm$	0.010	&	41.464	$\pm$	0.006	\\
NGC4051	&	0.002	&	M06	&	\nodata	&	\nodata	&	20	&	1366	$\pm$	551	&	707	$\pm$	357	&	41.050	$\pm$	0.096	&	38.802	$\pm$	0.074	\\
NGC4151	&	0.003	&	M03	&	01 Jul 1995	&	5	&	28	&	6922	$\pm$	218	&	3738	$\pm$	442	&	42.467	$\pm$	0.002	&	40.858	$\pm$	0.082	\\
NGC4253	&	0.013	&	M03	&	25 June 2001	&	1	&	19	&	1908	$\pm$	613	&	1056	$\pm$	217	&	42.279	$\pm$	0.032	&	40.057	$\pm$	0.120	\\
NGC4593	&	0.009	&	M03	&	04 Apr 1990	&	3	&	26	&	4785	$\pm$	135	&	2489	$\pm$	121	&	42.391	$\pm$	0.071	&	40.766	$\pm$	0.016	\\
NGC5548	&	0.017	&	M03	&	21 May 1993	&	1	&	35	&	5884	$\pm$	262	&	2839	$\pm$	139	&	43.133	$\pm$	0.078	&	41.432	$\pm$	0.014	\\
NGC7496	&	0.016	&	M03	&	12 Oct 1996	&	0.3	&	37	&	3595	$\pm$	127	&	2324	$\pm$	63	&	43.700	$\pm$	0.009	&	41.900	$\pm$	0.006	\\
PG0026+129	&	0.142	&	M03	&	11 Oct 1990 	&	4	&	47	&	3141	$\pm$	192	&	2046	$\pm$	86	&	44.439	$\pm$	0.039	&	42.798	$\pm$	0.007	\\
PG0844+349	&	0.064	&	M03	&	22 Feb 1991	&	1	&	44	&	2783	$\pm$	48	&	1638	$\pm$	29	&	44.438	$\pm$	0.001	&	42.603	$\pm$	0.005	\\
PG1211+143	&	0.081	&	M03	&	01 May 1995 	&	4	&	42	&	2336	$\pm$	74	&	1512	$\pm$	41	&	44.701	$\pm$	0.002	&	42.950	$\pm$	0.006	\\
PG1226+023	&	0.158	&	M03	&	04 Apr 1990 	&	9	&	30	&	4091	$\pm$	344	&	2435	$\pm$	199	&	46.135	$\pm$	0.080	&	44.268	$\pm$	0.020	\\
PG1411+442	&	0.090	&	M03	&	23 Jun 2001	&	0.3	&	44	&	3491	$\pm$	372	&	2103	$\pm$	125	&	44.443	$\pm$	0.017	&	42.728	$\pm$	0.011	\\
PG2130+099	&	0.063	&	M03	&	18 Sep 1990	&	6	&	40	&	2824	$\pm$	86	&	1645	$\pm$	42	&	44.093	$\pm$	0.042	&	42.616	$\pm$	0.008	\\
\cline{1-10}\\
Arp151	&	0.021	&	B19	&	29 Apr 2013	&	0	&	17	&	3039	$\pm$	199	&	1837	$\pm$	101	&	41.943	$\pm$	0.010	&	40.637	$\pm$	0.011	\\
Mrk50	&	0.023	&	B19	&	12 Dec 2012	&	0	&	19	&	4633	$\pm$	122	&	2502	$\pm$	107	&	42.731	$\pm$	0.005	&	41.174	$\pm$	0.009	\\
Mrk1310	&	0.020	&	B19	&	07 Jan 2013	&	0	&	14	&	3179	$\pm$	487	&	1847	$\pm$	171	&	41.818	$\pm$	0.013	&	40.345	$\pm$	0.025	\\
NGC6814	&	0.005	&	B19	&	07 Jan 2013	&	0	&	20	&	6274	$\pm$	95	&	2622	$\pm$	52	&	41.331	$\pm$	0.007	&	40.100	$\pm$	0.005	\\
SBS1116+583A	&	0.028	&	B19	&	12 Jul 2013	&	0	&	13	&	3215	$\pm$	103	&	1706	$\pm$	71	&	42.466	$\pm$	0.006	&	40.917	$\pm$	0.008	\\
Zw229-015	&	0.028	&	B19	&	23 Jul 2013	&	0	&	18	&	2638	$\pm$	49	&	1529	$\pm$	37	&	42.734	$\pm$	0.005	&	41.262	$\pm$	0.006	\\
~\\
\tableline
\tableline
\end{tabular}
\tablecomments{Col. (1): Name of target. Col. (2): Redshift.  Col. (3): Reference of observed spectra e.g., M03: \citet{Marziani+13}; M06: \citet{M06}; B09: \citet{B09}; Ho95: \citet{Ho+95}; B19: \citet{Bahk+19}. Col. (4): Date of observed spectra. Col. (5): Difference in time between the UV and optical observations in units of year.
Col (6): Signal-to-noise of spectra at 5100\AA\ continuum. Col. (7): $\rm FWHM_{H\beta}$ in units of $\rm km s^{-1}$. Col. (8): Line dispersion $\rm \sigma_{H\beta}$ in units of $\rm km s^{-1}$. Col. (9): Continuum luminosity $\rm L_{5100}$ at 5100\AA\ in units of $\rm erg s^{-1}$. Col. (10): Emission line luminosity $\rm L_{H\beta}$ in units of $\rm erg s^{-1}$.} 
\end{center}
\end{table*}

\section{Calibrating \mbh \ estimators} \label{section:mbh}

In this section, we calibrate \mbh\ estimators for each pair of velocity and luminosity from \MgII, \Hb, $\rm L_{3000}$ and $\rm L_{5100}$, using the best fits from Section \ref{section:scaling}. We determined the parameters in Equation \ref{eq:mbh2} by comparing with the fiducial \mbh. As a reference, we used two fiducial masses. The first fiducial mass is determined from $\rm \sigma_{H\beta}$ and $\rm L_{5100}$, and the second fiducial mass is obtained from $\rm FWHM_{H\beta}$ and $\rm L_{5100}$. As in our previous study in \citet{Woo+18}, we adopted the virial theorem and \Hb\ size-luminosity relation ($\beta = 2.0$ and $\gamma = 0.533$) for calculating fiducial masses. 
For the virial factor, we used the best-fit value f = 4.47 ($\alpha$ = 7.47) and f = 1.12 ($\alpha$ = 6.87) from \cite{Woo15}, respectively, for the fiducial masses based on $\rm \sigma_{H\beta}$ and $\rm FWHM_{H\beta}$.

\subsection{\Hb-based mass estimators}\label{section:mbh:Hb}

In Figure \ref{fig:hbcross}, we present the \mbh\ estimator based on the \Hb\ emission line. Firstly, in the case of the fiducial mass based on the \Hb\ line dispersion $\rm \sigma_{H\beta}$ and $\rm L_{5100}$, we fixed $\beta = 2.0$ for $\rm \sigma_{H\beta}$, and for $\rm FWHM_{H\beta}$, we fixed $\beta = 2.0/1.48 = 1.35$ (based on the obtained slopes of $\rm FWHM_{H\beta}$ and $\rm \sigma_{H\beta}$ in Figure \ref{fig:compare}). Secondly, when the fiducial mass is based on $\rm FWHM_{H\beta}$ and $\rm L_{5100}$, we fixed $\beta = 2.0$ for $\rm FWHM_{H\beta}$, and for $\rm \sigma_{H\beta}$, we fixed $\beta = 2.0/(1.0/1.48) = 2.96$. For both fiducial masses, we used $\gamma$ = 0.533/0.98 = 0.55 when adopting luminosity from $\rm L_{H\beta}$ (Figure \ref{fig:scalelum}). Using those $\beta$ and $\gamma$ values, we determined $\alpha$ based on the $\chi^2$ minimization with the FITEXY method in \citet{Park+12b}. The root-mean-square (rms) scatters of both \mbh\ estimators are 0.10$-$0.13 dex and 0.10$-$0.19 dex for the fiducial masses from $\rm \sigma_{H\beta}$ and $\rm FWHM_{H\beta}$, respectively. When luminosity is adopted from the \Hb\ emission line, the rms scatter becomes larger compared to that of using continuum 5100\AA\ luminosity. Similarly, the choice of velocity by using $\rm FWHM_{H\beta}$ has larger rms scatter than when $\rm \sigma_{H\beta}$ is used as velocity. By enlarging the sample using SDSS, our estimators have slightly smaller rms scatter ($\sim$0.05 dex) compared to that of our previous study in \citet{Woo+18}.
    
\begin{figure*}
\center
	\includegraphics[width = 0.49\textwidth]{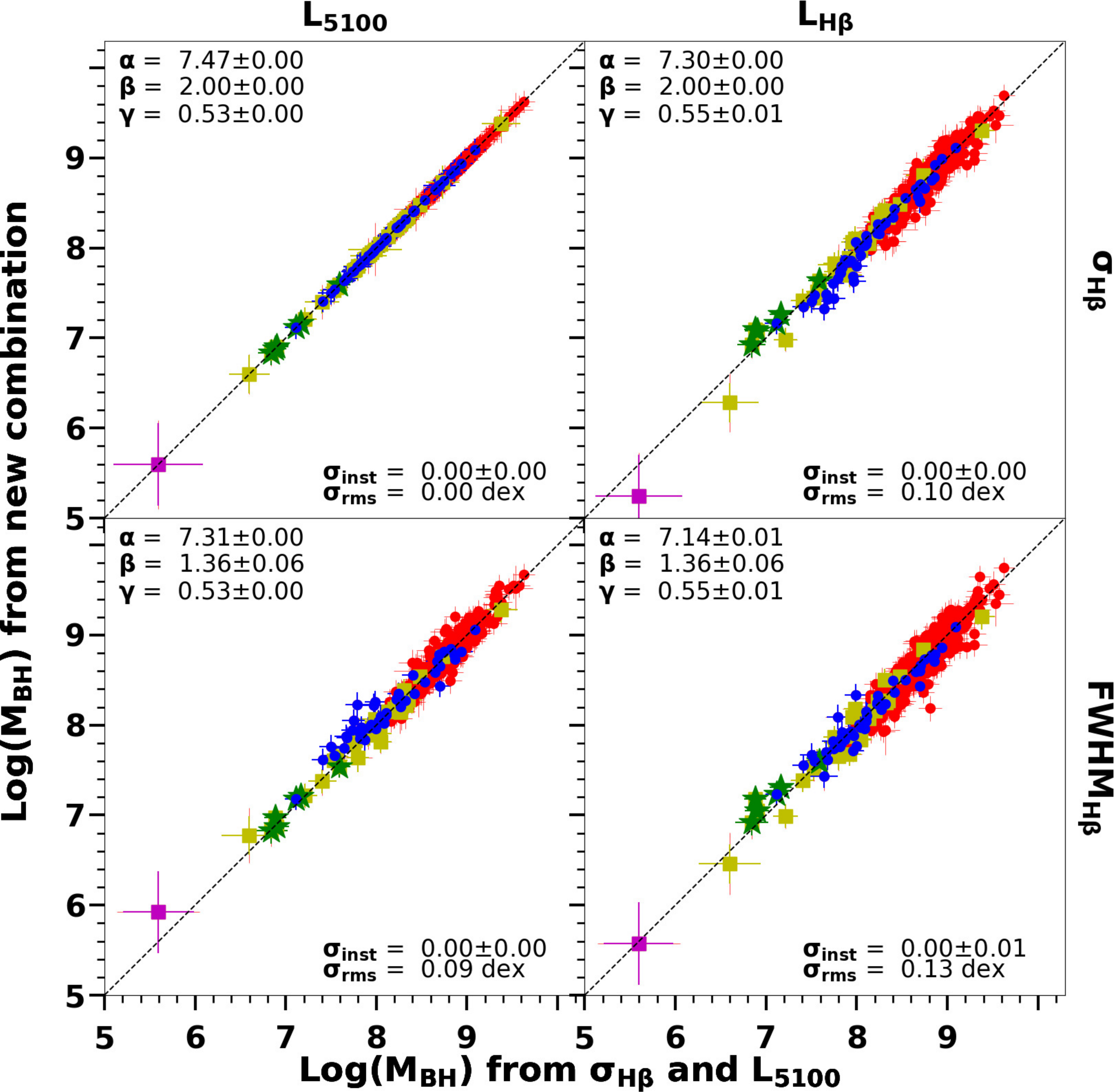}
	\includegraphics[width = 0.49\textwidth]{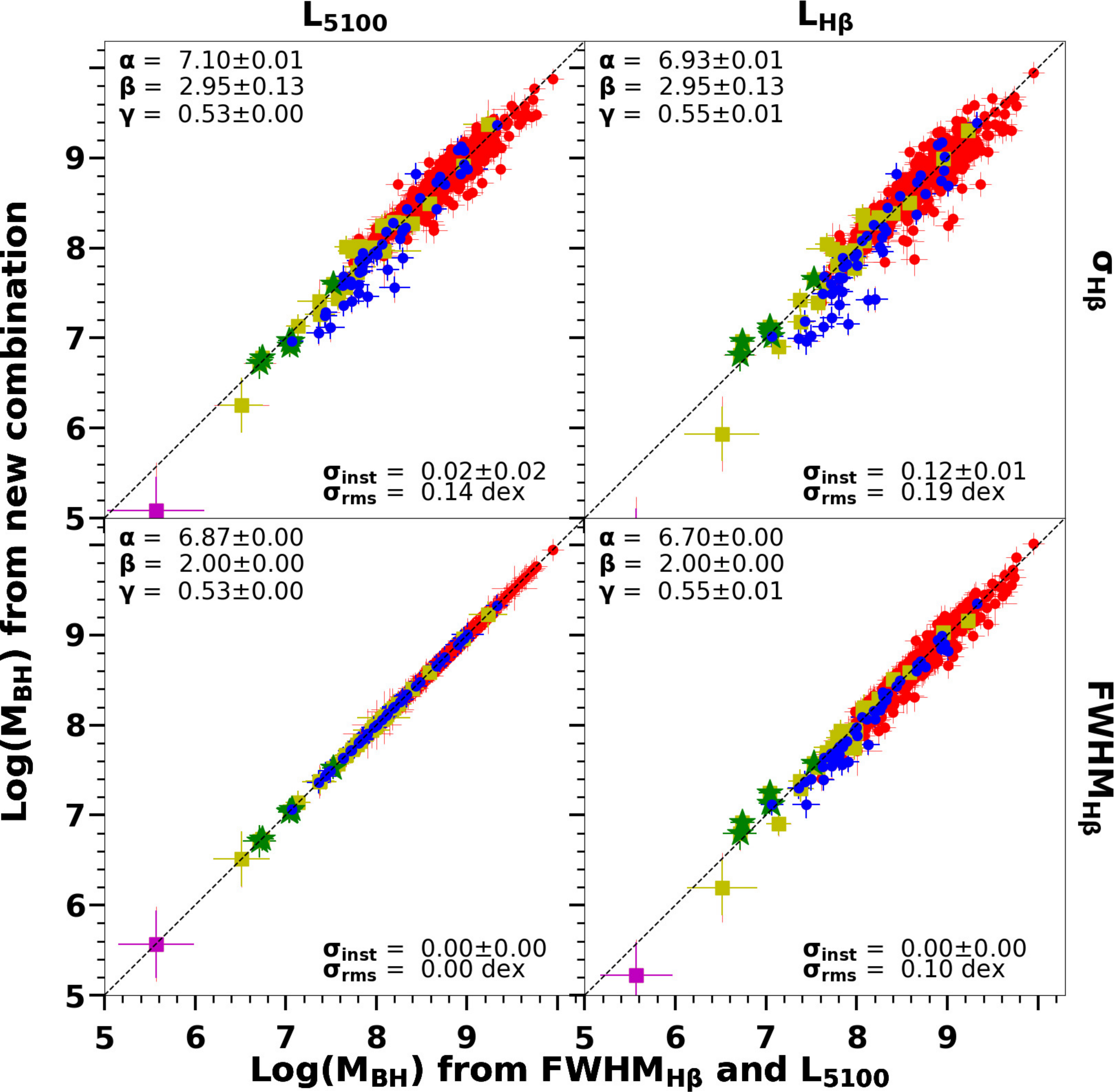}		
	\caption{Newly derived \mbh\ with estimator from \Hb\ emission line. Left panel: cross-calibration fitting between newly derived \mbh\ and the fiducial mass with estimator based on 
	$\sigma_{\rm H\beta}$ and $\rm L_{5100}$. Newly derived \mbh\ is from $\rm \alpha + \beta \log V_{1000} + \gamma \log L$. $\rm V_{1000}$ is velocity estimator in units of $\rm 1000~ km~ s^{-1}$, L is luminosity estimator in units of $\rm 10^{44}~erg~s^{-1}$ for continuum or $\rm 10^{42}~erg~s^{-1}$ for emission line. $\alpha$ is estimated by $\chi^{2}$ minimization fitting. $\beta$ and $\gamma$ in each panel depend on different estimators, shown in the top and upper right part of each figure. Right panel: same as the left panel, but for the fiducial mass with estimator based on $\rm FWHM_ {\rm H\beta}$ and $\rm L_{5100}$. The symbols are shown for the moderate-luminosity AGNs (blue), the SDSS sample (red), the 25 RM sources (yellow), the six HST targets (green), and NGC~4051 (pink). The dash-lines represent one to one relation. 
\label{fig:hbcross}}
\endcenter
\end{figure*}

\subsection{\ion{\rm Mg}{2}-based mass estimators}\label{section:mbh:MgII}

We calibrated the \mbh\ estimators based on the \MgII\ emission line by determining $\alpha$, $\beta$, and $\gamma$ in Equation \ref{eq:mbh2}.
As we performed in our previous study \citep{Woo+18}, we used five schemes in the fitting process: \\ 
\smallskip
\hspace{4.1mm} $\bullet$ Scheme 1: $\beta$ and $\gamma$ are adopted from scaling relations in Section \ref{section:scaling}.\\
\smallskip
\hspace{5mm}  $\bullet$ Scheme 2: $\beta = 2.0$ and $\gamma = 0.5$.\\
\smallskip
\hspace{5mm}  $\bullet$ Scheme 3: $\beta = 2.0$ and $\gamma$ is a free parameter.\\
\smallskip
\hspace{5mm}  $\bullet$ Scheme 4: $\gamma = 0.5$ and $\beta$ is a free parameter.\\
\smallskip
\hspace{5mm}  $\bullet$ Scheme 5: both $\beta$ and $\gamma$ are free parameters.

We present all calibrated parameters for these five Schemes in Tables \ref{table:mbh_sigma} and \ref{table:mbh_fwhm} based on the fiducial masses from $\rm \sigma_{H\beta}$ and $\rm FWHM_{H\beta}$, respectively. In Figure \ref{fig:mgcross_2}, we show 3 cases (Schemes 1, 2, and 5). 
In total, we have 25 AGNs, for which UV and optical spectra were not observed simultaneously. 
By excluding these 25 AGNs, we performed the calibration of \mbh\ estimators. However, we found consistent results with/without these 25 AGNs. Therefore, we presented the calibration results for the total sample. Note that we presented the results based on the FITEXY method to be consistent with our previous studies. However, we also used the Bayesian method using PYMC (Python Markov chain Monte Carlo), and obtained consistent results.

In the case of Scheme 1, $\beta$ and $\gamma$ were fixed as determined from the scaling relations in 
Sections \ref{section:wcompare} and \ref{section:lcompare}. With respect to the fiducial mass based on the \Hb\ line dispersion, 
we obtained $\beta = 2/0.94 = 2.13$ for $\rm \sigma_{MgII}$ because of log $\sigma_{\rm MgII}$ $\propto$ 0.94 log $\rm \sigma_{H\beta}$ in Equation \ref{eq:sigcom}. For $\rm FWHM_{MgII}$, we adopted $\beta = 2.0/0.93 = 2.14$  since log FWHM$_{\rm MgII}$ $\propto$ 0.93 log $\rm \sigma_{H\beta}$}. When we used the fiducial mass based on $\rm FWHM_{H\beta}$, we fixed $\beta = 2.0/0.63 = 3.17$ for $\rm FWHM_{MgII}$ because of Equation \ref{eq:wcom}, and $\beta = 2.0/0.625 = 3.20$ for $\rm \sigma_{MgII}$ because of log $\sigma_{\rm MgII}$ $\propto$ 0.625 log $\rm FWHM_{H\beta}$. For $\gamma$, we also used the scaling relation. For example, we obtained $\gamma$ = 0.533/0.98 = 0.54 for  $\rm L_{3000}$,
and $\gamma$ = 0.533/0.82 = 0.65 for $\rm L_{MgII}$, using the best-fit slopes in Equations \ref{eq:lum3000} and \ref{eq:lummgii_5100}, respectively. 

The results of Scheme 1 show that the \mbh\ estimators based on $\sigma_{\rm MgII}$ have a rms scatter of 0.19$-$0.25 dex, which is smaller than the case of \mbh\ estimators based on $\rm FWHM_{MgII}$, 0.25$-$0.31 dex (see Figure \ref{fig:mgcross_2}). In general, the rms scatter becomes larger when we adopted $\rm L_{MgII}$ and $\rm FWHM_{MgII}$, indicating that the pair of continuum luminosity from 3000\AA\ and $\rm \sigma_{MgII}$ is the best choice for the UV \mbh\ estimator. By enlarging the sample size and the dynamic range of AGN luminosity, 
the calibration is improved as the rms scatters become smaller than that of our previous study \citep{Woo+18} by $\sim$ 0.05$-$0.1 dex.  

In the cases of Schemes 2 and 3, we fixed $\beta = 2.0$ and set $\gamma$ as a free parameter (Scheme 3) or fixed both $\beta = 2.0$ and $\gamma = 0.5$ (Scheme 2), following the virial theorem and expected size-luminosity relation. With respect to the fiducial mass based on the \Hb\ line dispersion, we obtained a smaller rms scatter, 0.18$-$0.23 dex than that of the Scheme 1. Compared to the previous study by \citet{Woo+18}, the rms scatter is reduced by 0.03$-$0.06 dex. When we adopted the fiducial mass based on $\rm FWHM_{H\beta}$, the rms scatter is 0.22$-$0.24 dex, which is also smaller than that of \citet{Woo+18} by 0.01$-$0.08 dex. 

Turning to the cases of Schemes 4 and 5, we fixed $\gamma = 0.5$ and set $\beta$ as a free parameter (Scheme 4) or set both of them as free parameters (Scheme 5). The obtained rms scatters for Schemes 4 and 5 are slightly smaller compared to those of Schemes 2 and 3 by $\sim$ 0.02$-$0.04 dex.   

In the case of the fiducial mass by $\rm FWHM_{\rm H\beta}$ and $\rm L_{5100}$ (see Table \ref{table:mbh_fwhm}), our calibration is improved with smaller intrinsic scatter (0.14$-$0.25 dex) and rms scatter (0.21$-$0.30 dex) than those in our previous studies \citep{Woo+18}. Nevertheless, the calibration based on the fiducial mass from $\rm FWHM_{\rm H\beta}$ and $\rm L_{5100}$ is less reliable with a larger scatter, compared to the \mbh\ estimators based on the fiducial mass from $\sigma_{\rm H\beta}$ and $\rm L_{5100}$. 

Based on our results, we found that the best \mbh\ estimator based on \MgII\ is achieved when using $\sigma_{\rm MgII}$ and $\rm L_{3000}$, with smallest intrinsic and rms scatters ($\rm \sigma_{inst}$ = 0.09-0.12 dex and $\rm \sigma_{rms}$ = 0.17-0.20 dex). Among the five Schemes, we found that Schemes 2, 3, 4 and 5 give small $\rm \sigma_{inst}$ and $\rm \sigma_{rms}$, 0.09 and 0.17 dex, respectively. However, in those Schemes 4 and 5, $\beta \sim 1.5$ breaks the virial relation ($\beta = 2.0$). The two other cases (Schemes 2 and 3) show similar $\rm \sigma_{inst}$ and $\rm \sigma_{rms}$, 0.11 and 0.19 dex, respectively.
However, we recommend Scheme 2 as the best \MgII\ \mbh\ estimator since it follows the virial relation and the expected size-luminosity relation ($\beta = 2.0$ and $\gamma = 0.5$).  

In short, by enlarging the sample over a large luminosity range, we improve the calibration of \MgII\ based mass estimators. 
For the best pair of L$_{3000}$ and line dispersion of \MgII\ ($\sigma_{\rm MgII}$), we found an intrinsic scatter of $\sim$0.1 dex and a rms scatter
of $\sim$0.2 dex, indicating that the \mbh\ estimated based on the \MgII\ line and UV continuum luminosity is only slightly less reliable
compared to the \mbh\ based on the \Hb\ line and L$_{5100}$. 
\begin{turnpage}
\begin{figure*}
\centering
   \includegraphics[width=0.43\textwidth]{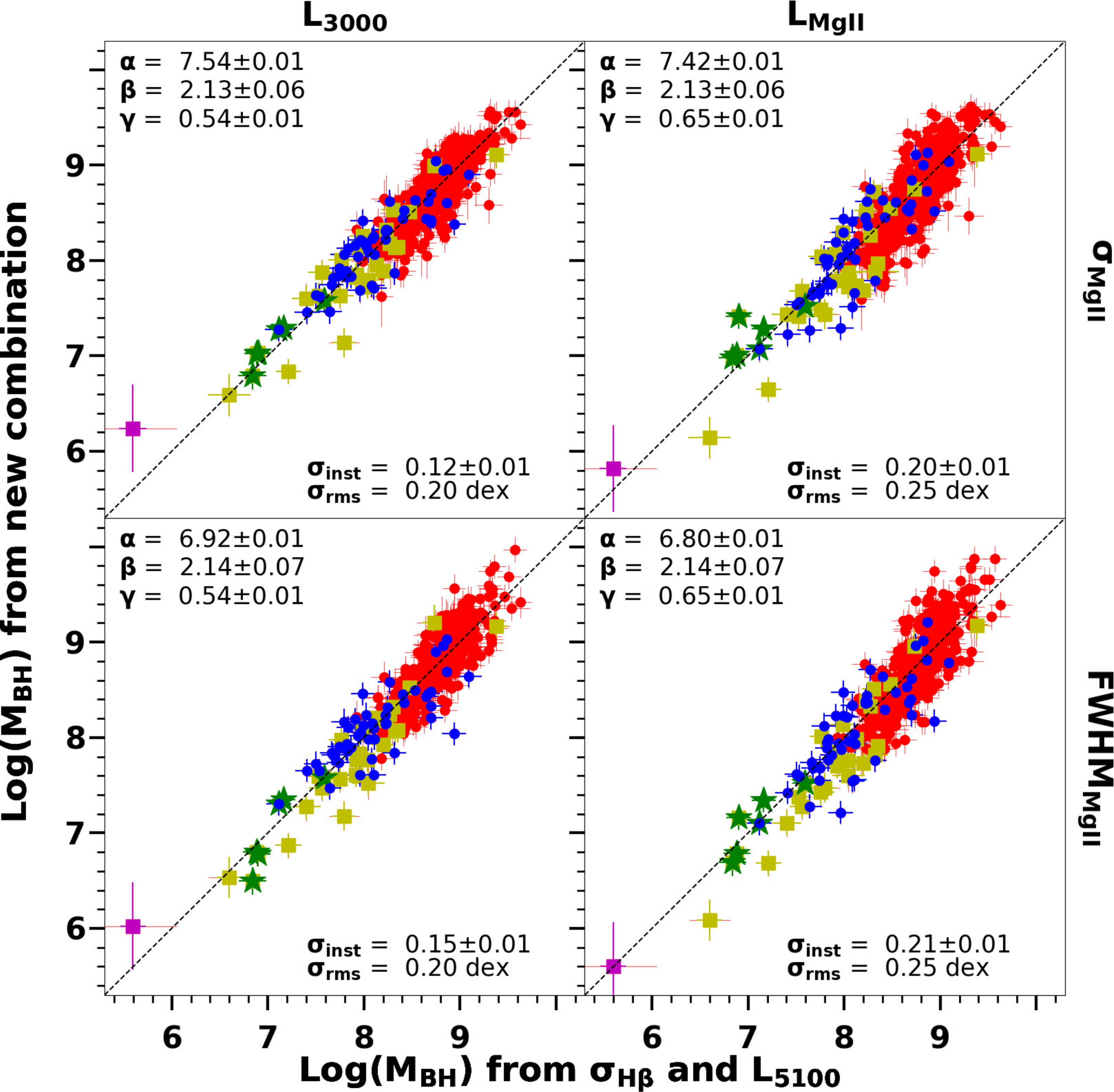} 
   \includegraphics[width=0.43\textwidth]{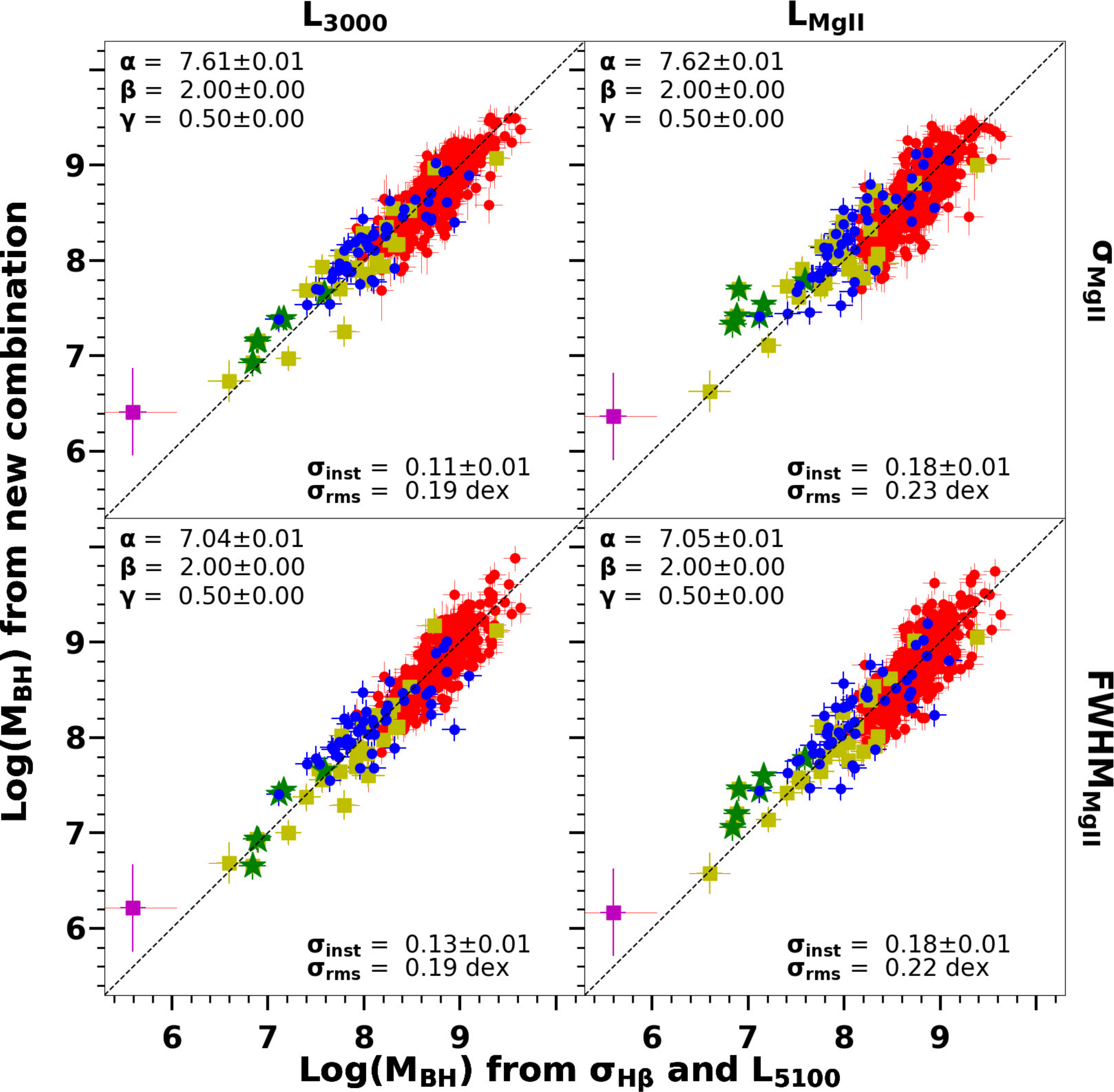}
   \includegraphics[width=0.43\textwidth]{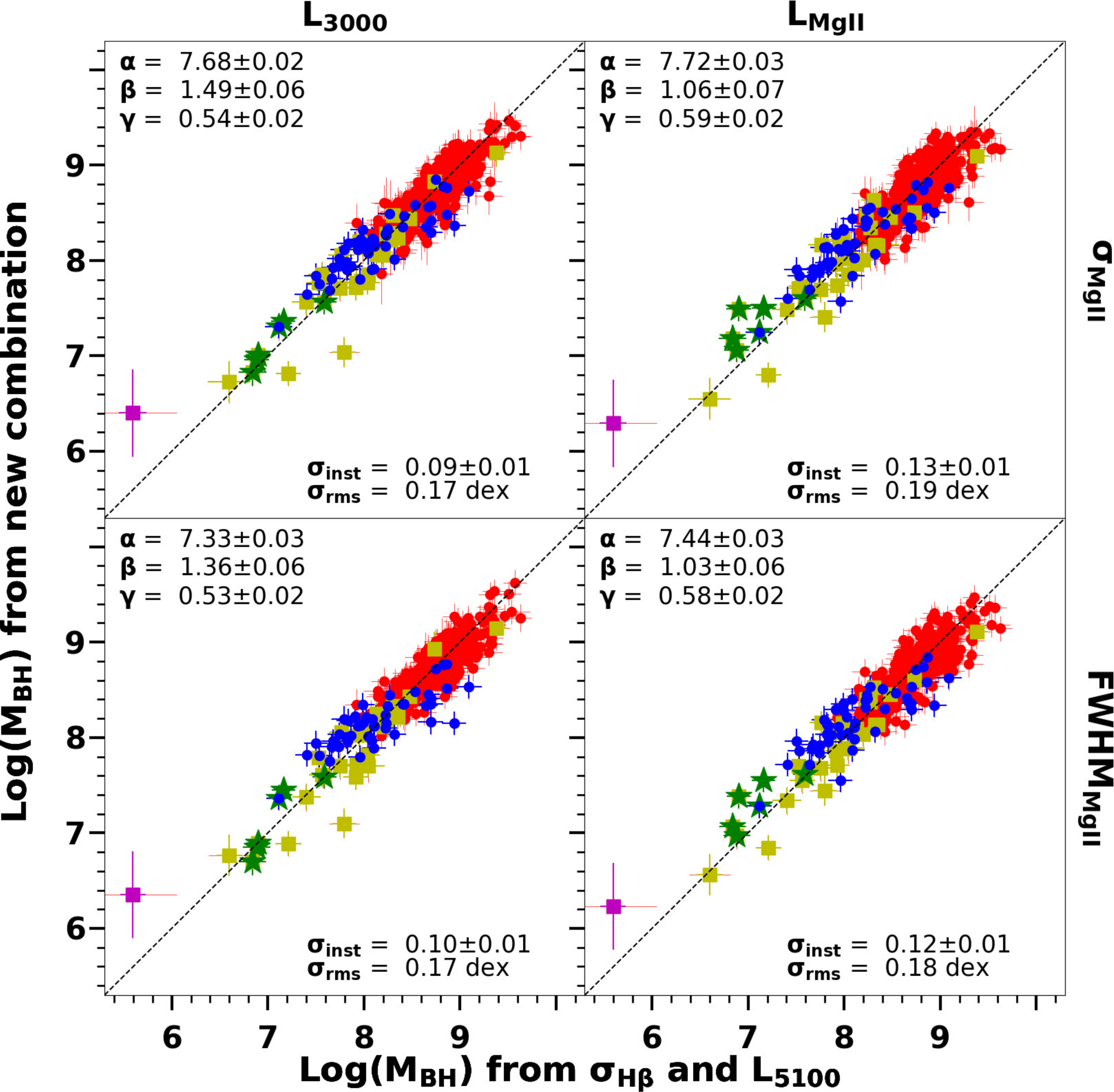}\\ 
   \includegraphics[width=0.43\textwidth]{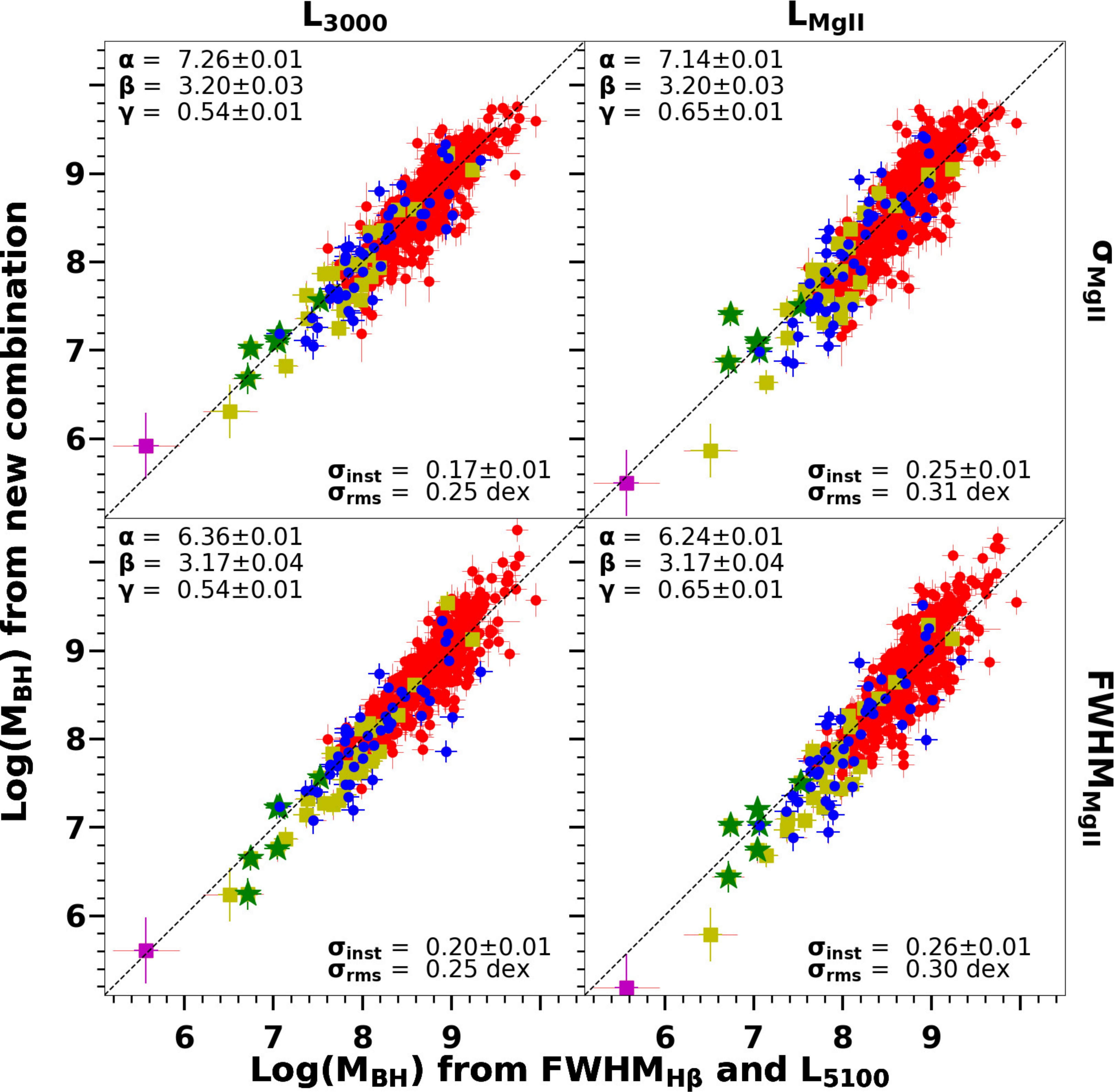}
   \includegraphics[width=0.43\textwidth]{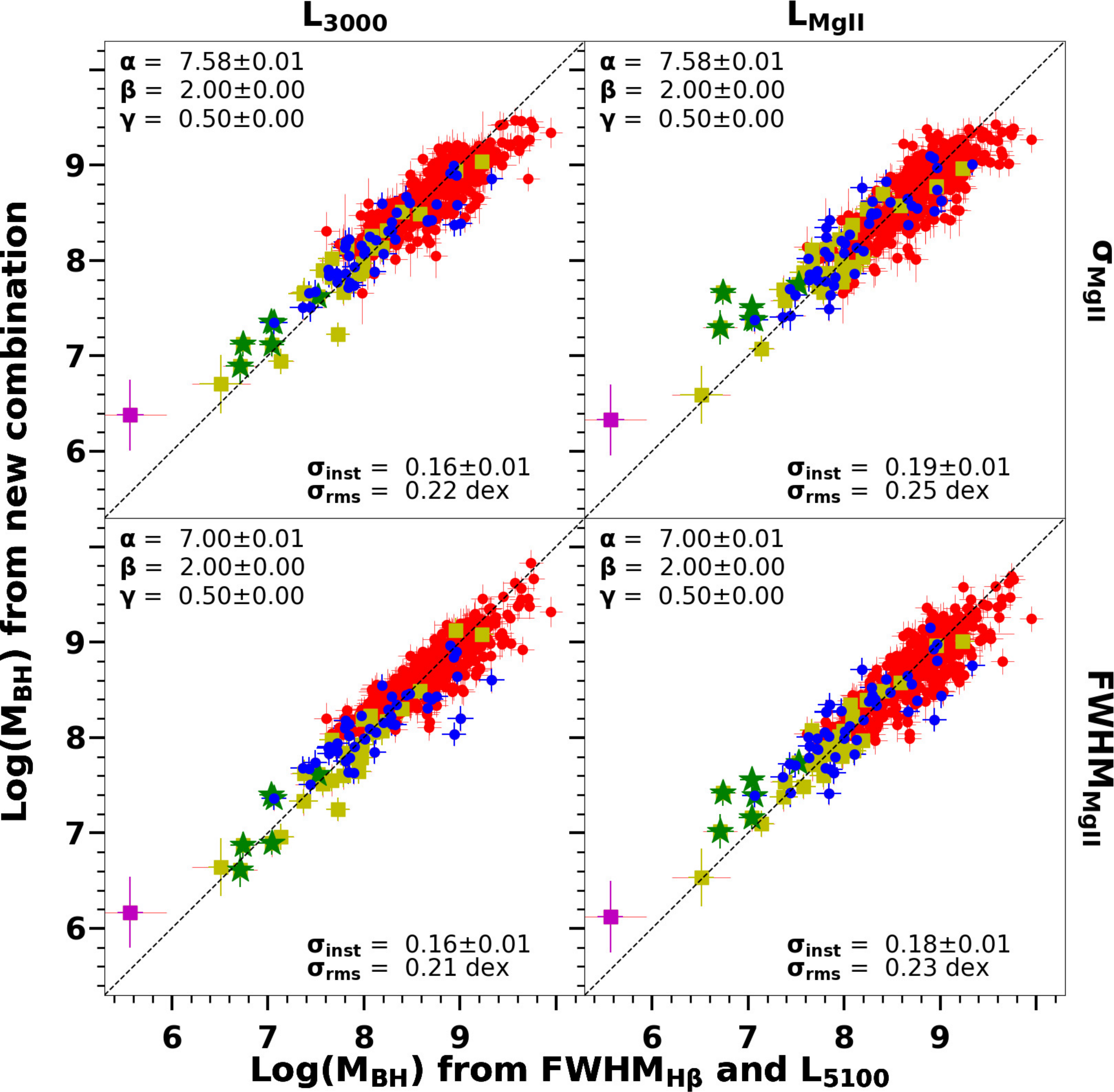}
   \includegraphics[width=0.43\textwidth]{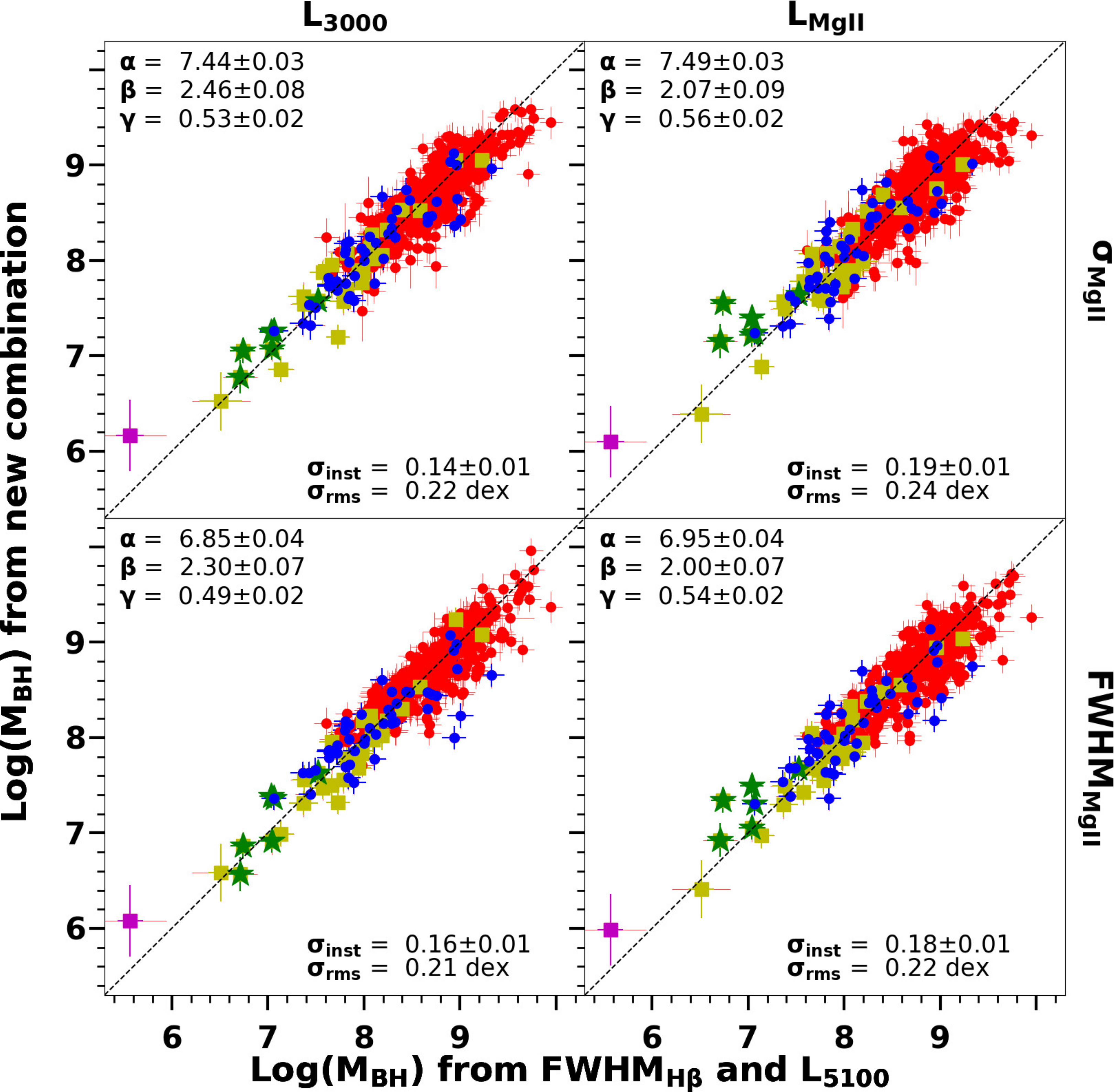}
  \caption{Same as Figure \ref{fig:hbcross}, but for \mbh\ estimator from \ion{Mg}{2}. Top panels: cross-calibration fitting between newly derived \mbh\ and fiducial mass with estimator based on $\sigma_{\rm H\beta}$ and $\rm L_{5100}$. $\beta$ and $\gamma$ is obtained from scaling relations of line width and luminosity in Sections \ref{section:wcompare} and \ref{section:lcompare} (top left panel). Top middle panel: $\beta$ is 2.0, based on the virial assumption and $\gamma$ is 0.5, from expected size-luminosity relation. Top right panel: $\beta$ and $\gamma$ is obtained from the best-fit results based on Scheme 5 in Section \ref{section:mbh:MgII}. Bottom panels: same as the top panels, but for the fiducial mass with estimator based on $\rm FWHM_ {\rm H\beta}$ and $\rm L_{5100}$. The symbols are shown for the moderate-luminosity AGNs (blue), the SDSS sample (red), the 25 RM sources (yellow), the six HST targets (green), and NGC~4051 (pink). The dash-lines represent one to one relation.}
\label{fig:mgcross_2}
\end{figure*}
\end{turnpage}

\section{Discussion}\label{section:discuss}

\subsection{Uncertainties of the \MgII-based mass}

In this section we discuss the systematic uncertainties of the calibrated \MgII\ \mbh\ estimators. For simplicity, we present the results from Scheme 2 for this comparison. Note that we also investigated the systematic uncertainties using the other 4 Schemes and obtained the consistent results. Figure \ref{fig:property} shows the systematic difference between UV and optical \mbh\ estimators as a function of AGN properties. We found no significant correlation of the \mbh\ difference with Eddington ratio, $\rm FWHM_{H\beta}$, F$_{\rm OIII}$/F$_{\rm FeII}$, and F$_{\rm OIII}$/F$_{\rm H\beta, narrow}$. However, we found that the difference between \MgII\ and \Hb-based masses anti-correlates with the systematic difference of the line profiles ($\Delta \rm P$) between \MgII\ and \Hb, which is parameterized by the ratio between FWHM and line dispersion. By performing a regression analysis, we obtained the best-fit result:
\begin{equation}
	\Delta \rm P  = -0.83 \times \log \left (\frac{\rm FWHM_{MgII}/\sigma _{MgII}}{\rm FWHM_{H\beta}/\sigma _{H\beta}} \right) - 0.01. 
	\label{eq:cor1}
\end{equation}

We also found a positive correlation between the mass difference and the UV-to-optical continuum luminosity ratio as similarly reported
by \cite{Woo+18}. We obtained the best-fit slope of 0.73 $\pm$ 0.05 and the intercept of 0.21 $\pm$ 0.02 when comparing the systematic differences of the UV and optical \mbh\ estimates with the UV-to-optical luminosity ratios. Similar to our previous work, in order to minimize the systematic effect
to the different slope of the UV-to-optical spectral slope, we propose to add this correction term to the Equation \ref{eq:mbh2}:
 \begin{equation}
	\Delta \rm C  = -0.73 \times \log (\rm L_{3000}/L_{5100}) + 0.21,
	\label{eq:cor2}
\end{equation}
As similarly suggested by \citet{Woo+18} based on the modeling of the local UV/optical AGN continuum using a power law function, 
we derived the correction factor as a function of the power law coefficient of the AGN continuum $\alpha_{\lambda}$:
\begin{equation}
	\Delta \rm C = 0.17 (1+\alpha_{\lambda}) + 0.20.
	\label{eq:cor3}
\end{equation}
where, the mean $\alpha_{\lambda}$ of our sample is -2.69 $\pm$ 0.87 (i.e., $\alpha_{\nu}$ = 0.69 $\pm$ 0.87). We measured this $\alpha_{\lambda}$ in the wavelength range of 2800-5200 \AA.
Note that this correction for \mbh\ is relative small, $\sim$0.1 dex. In practice the observed spectrum is likely to be limited in the rest-frame UV for high-z AGNs, and if so, the spectral slope cannot be measured from the 2800-5200 \AA\ range. Thus, we present the effect of the different spectral slope as a bias in the \mbh\ estimation.

The difference in the line profiles between \MgII\ and \Hb\ has a significant effect on the \mbh\ estimation. The nonlinear relationship between $\rm FWHM_{MgII}$ and $\rm FWHM_{H\beta}$ will have a significant effect in the UV and optical \mbh\ estimators. Particularly, as in Figure \ref{fig:scale_w}, the different slopes between the narrow and broad $\rm FWHM_{MgII}$ sample shows significant changes in the line profiles of \Hb\ and \MgII\, hence will raise systematic uncertainty between the \MgII\ and \Hb\ \mbh\ estimators. We found that the discrepancy between \MgII\ and \Hb-based masses shows a negative correlation with the line profile of \MgII\ (Figure \ref{fig:correct_mgii}). Since the line profile of \MgII\ has significant effect on the difference of the \MgII\ and \Hb-based masses, we also suggest a correction factor, $\Delta \rm M$ based on the best-fit result as follow:
\begin{equation}
	\Delta \rm M  = -1.14 \times \log \left (\frac{\rm FWHM_{MgII}}{\sigma _{MgII}} \right) + 0.33. 
	\label{eq:cor1_2}
\end{equation}

To demonstrate the effect of the color-correction term and the correction term of \MgII\ line profile, we compared the
\MgII\ and \Hb-based \mbh\ with the corrections of $\rm \Delta C$, $\rm \Delta M$ and the combination of them in Figure \ref{fig:correct_L}. We found that the systematic uncertainty, i.e., the rms scatter between the UV and optical \mbh\ can be reduced from 0.19 dex to 0.15 dex. In practice, when only the rest-frame UV spectrum is available for estimating \mbh, the correction term $\Delta \rm M$ would be useful since $\Delta \rm C$ or $\Delta \rm P$ cannot be obtained without the rest-frame optical spectrum. 

 
We have presented that the change of the UV-to-optical continuum luminosity and the difference of the line profile between \Hb\ and \MgII\ cause a large systematic difference between the \MgII\ and \Hb\ \mbh\ estimators. In addition, we list here sources of systematic uncertainties, which could bias the \MgII\ \mbh\ estimator. First, as we mentioned in Section \ref{section:intro}, there is no available size-luminosity relation based on \MgII\ emission line. In this study, we calibrated the \MgII\ \mbh\ estimator based on the fiducial mass as the single-epoch \Hb\ based mass based on the size-luminosity relation which has an uncertainty, $\sim$0.19 dex \citet{Bentz+13}. Second, the uncertainty of the virial factor f is $\sim$0.12-0.40 dex (e.g., \citealp{Woo15}; \citealp{Pancoast+14}). Third, the variability between the line width and luminosity could introduce a bias with scatter of $\sim$0.1 dex \citep{Park12}. Therefore, we keep in mind that we estimated the \MgII\ \mbh\ estimator that is calibrated based on the fiducial mass which has also various uncertainties of itself, $\sim$0.40-0.70 dex.

We note that there are also significant uncertainties of the \mbh\ estimator based on the analysis of \MgII\ line. First, luminosity and velocity in AGNs show variability. This variability could bias the \MgII\ \mbh\ estimator based on the \Hb\ based mass if the rest-frame spectra of the UV and optical are not observed at the same time. Second, there are reports that the line width FWHMs of \Hb\ and \MgII\ are not comparable, and $\rm FWHM_{H\beta}$ is larger than that of \MgII\ by $\sim20\%$ for the broad \Hb\ line AGNs (e.g., \citealp{Marziani+13}, also see Figure \ref{fig:scale_w} in Section \ref{section:scaling}). We adopted the virial relation, and apply the same $\beta = 2.0$ for the \Hb\ and \MgII\ based masses, respectively. This could bias the \MgII\ \mbh\ estimator since $\rm FWHM_{H\beta}$ is larger than that of \MgII, and $\beta$ value needs to be higher than 2.0. Third, the measurements of the \MgII\ line could introduce a bias in the \MgII\ \mbh\ estimator. For example, a careful analysis of fitting and subtracting for Balmer continuum in the UV spectra may be required for an accurate determination of $\rm L_{3000}$ \citep{Kovacevic+17}. Fourth, in our analysis, we did not subtract the narrow component of \MgII\ since there is no clear narrow \MgII\ component in our spectra. We note that subtracting the narrow component of \MgII\ should be performed with caution since it is difficult to determine how much the narrow component contributes to the line profile.
 

 \subsection{Comparison with previous \MgII-based \mbh\ estimators}
 
There have been various \mbh\ estimators based on the \MgII\ emission line in the literature (e.g., \citealp{McGill+08}; \citealp{Wang+09}; \citealp{Shen+11}; \citealp{Shen+12}; \citealp{Tilton+13}; \citealp{Woo+18}; \citealp{Bahk+19}). As detailed by \citet{Woo+18}, the difference among these \mbh\ estimators is originated by various factors, i.e., \FeII\ templates, the narrow component of \MgII, and the virial factor f. 
For example, we adopted the \FeII\ template from \citet{Tsuzuki06} while other studies such as \citet{Shen+11} and \citet{Shen+12} used the \FeII\ template from \citet{VW01} and \citet{Salviander07}, respectively. The line dispersion of \MgII\ becomes smaller if the \FeII\ template from \citet{Tsuzuki06} is adopted in the fitting process (see Figure 2 in \citealp{Woo+18}). Also, \citet{Shin+19} pointed out that the flux ratio of \FeII/\MgII\ could be different up to $\sim$0.2 dex between the \FeII\ modeled by \citet{Tsuzuki06} and \citet{VW01}. Regarding the measurement of FWHM, the subtraction of the potential narrow component in \MgII\ significantly changes the result as we mentioned in Section \ref{section:measure}. For example, as \citet{Wang+09} subtracted the narrow component in their \MgII\ model, their FWHM measurements could be systematically different from our measurement. In addition, using a different scaling factor f also causes discrepancies between \mbh\ mass estimators. 
In this section, we determined \mbh\ using various mass estimators and compared them to previous studies to investigate systematic differences, using the measurements of the FWHM and line dispersion of \MgII\ line as well as continuum luminosity at 3000\AA\ and \MgII\ line luminosity. For this comparison we chose the results based on Scheme 2 as the best \MgII\ \mbh\ estimator from our calibration. 

 
First, we note that our new calibration is very close to that presented by \citet{Woo+18}, who used only intermediate luminosity AGNs. 
In the calibration with the fiducial mass from the pair of $\sigma_{\rm H\beta}$ and $\rm L_{5100}$, both intrinsic scatter (0.09$-$0.21 dex) and rms scatter (0.17$-$0.25 dex) become smaller than those reported by \citet{Woo+18}, i.e., $\rm \sigma_{inst}$ = 0.13$-$0.23 dex and $\rm \sigma_{rms}$ = 0.18$-$0.36 dex. 
When we used the fiducial mass based on $\rm FWHM_{\rm H\beta}$ and $\rm L_{5100}$, we also obtained consistent mass estimators 
compared to those of \citet{Woo+18}, with slightly improved scatters by $\rm \sigma_{inst}$ = 0.14$-$0.26 dex and $\rm \sigma_{rms}$ = 0.21$-$0.31 dex.

We compared our results with that of \citet{Bahk+19}, the 31 \Hb\ RM AGNs are applied by the same method as we did in our Scheme 2. However, there is systematic difference between our estimator and that of \citet{Bahk+19} with an offset of $\sim$0.27 dex. In \citet{Bahk+19}, the authors discussed that the large systematic difference between the two estimators is from the difference of $\sigma_{\rm H\beta}$/$\sigma_{\rm MgII}$ and $\rm L_{5100}$/$\rm L_{3000}$. This result indicates that there is a systematic uncertainty in the \MgII-based \mbh\ estimator which is raised from the different line profiles between the \MgII\ and \Hb\ emission lines. The relation between the UV and optical mass estimator ratio and the systematic difference of the line profiles between \MgII\ and \Hb, shown in Figure \ref{fig:property}, support this argument.

Second, we compared our mass estimator with the previously reported estimators by \citet{Wang+09}, \citet{Shen+11}, \citet{Shen+12} and \citet{Tilton+13}
in Figure \ref{fig:mbh}. In this comparison, we adjusted the $\alpha$ values of other works by -0.09 dex since they adopted $\log$f = 0.74 from \citep{Onken+04}}. Since these studies only provided the measurements of FWHM of \MgII, we compared \mbh\ based on $\rm FWHM_{MgII}$
and L$_{3000}$.    

We found our \mbh\ is systematically larger by $\sim$0.25 dex than the \mbh\ calculated with the recipes from \citet{Wang+09}, \citet{Shen+11} and \citet{Shen+12}. 
For the case of using $\rm FWHM_{MgII}$ and $\rm L_{MgII}$, our estimator has a large offset of 0.33 dex compared to that of \citet{Tilton+13}. 

The \mbh\ based on our best estimator is higher than that based on the recipe of \citet{Wang+09} by 0.25 dex. As we mentioned, \citet{Wang+09}
subtracted the narrow component in the modeling of the \MgII\ line profile. Thus, $\rm FWHM_{MgII}$ is systematically higher in their analysis, 
leading to a smaller $\alpha$ for comparison with given fiducial \mbh. 
Compared to the mass estimators presented by \citet{Shen+11} and \citet{Shen+12}, our mass estimator provides higher \mbh\
by 0.22$-$0.25 dex. Similar to \citet{Wang+09}, \citet{Shen+11} and \citet{Shen+12} subtracted the narrow component of \MgII, which lead to a smaller $\alpha$ in Equation \ref{eq:mbh2} compared to our \mbh\ estimator. In contrast, \mbh\ is more consistent between our estimator and the estimators of \citet{Shen+11} and \citet{Shen+12}. This is due to the limited luminosity range of their calibrations. \citet{Shen+11} used a high luminosity sample from SDSS ($\rm L_{5100} > 10^{44}\ erg s^{-1}$) at z = 0.4$-$0.8, while \citet{Shen+12} utilized higher luminosity sample ($\rm L_{5100} > 10^{45.4}\ erg s^{-1}$) at z = 1.5$-$2.2. Therefore, these \mbh\ estimators are not properly calibrated for low-luminosity AGNs. 


Finally, we compared our \mbh\ estimator with that of \citet{Tilton+13}, who used $\rm FWHM_{MgII}$ and $\rm L_{MgII}$ for \mbh\ estimation. 
We found a systematic offset of 0.33 dex. \citet{Tilton+13} used the 44 single-epoch \mbh\ sample to calibrate the \mbh\ estimator. \citet{Tilton+13} fixed the $\beta$ = 2.0 and obtained $\gamma$ = 0.53 which are close to those of our values. However, the mass used in \citet{Tilton+13} is based on single-epoch \Hb\ mass, for which the equations from \citet{VP06} were used. The systematic difference of the reference mass is responsible for the systematic offset between our and their mass estimators. 

Our study is performed for the first time based on a large dynamic range covering low-to-high luminosity AGNs, in order to minimize any uncertainty due to the limited luminosity range. Our \mbh\ estimators are different with a systematic offset of $\sim$0.22$-$0.33 dex compared to those of other \mbh\ calibrations performed with limited luminosity range samples. We should be careful when choosing the \mbh\ estimator since systematic discrepancy will affect our understanding of BH mass function and its evolution. 


\begin{figure*}
    \centering
    \includegraphics[width=0.33\textwidth]{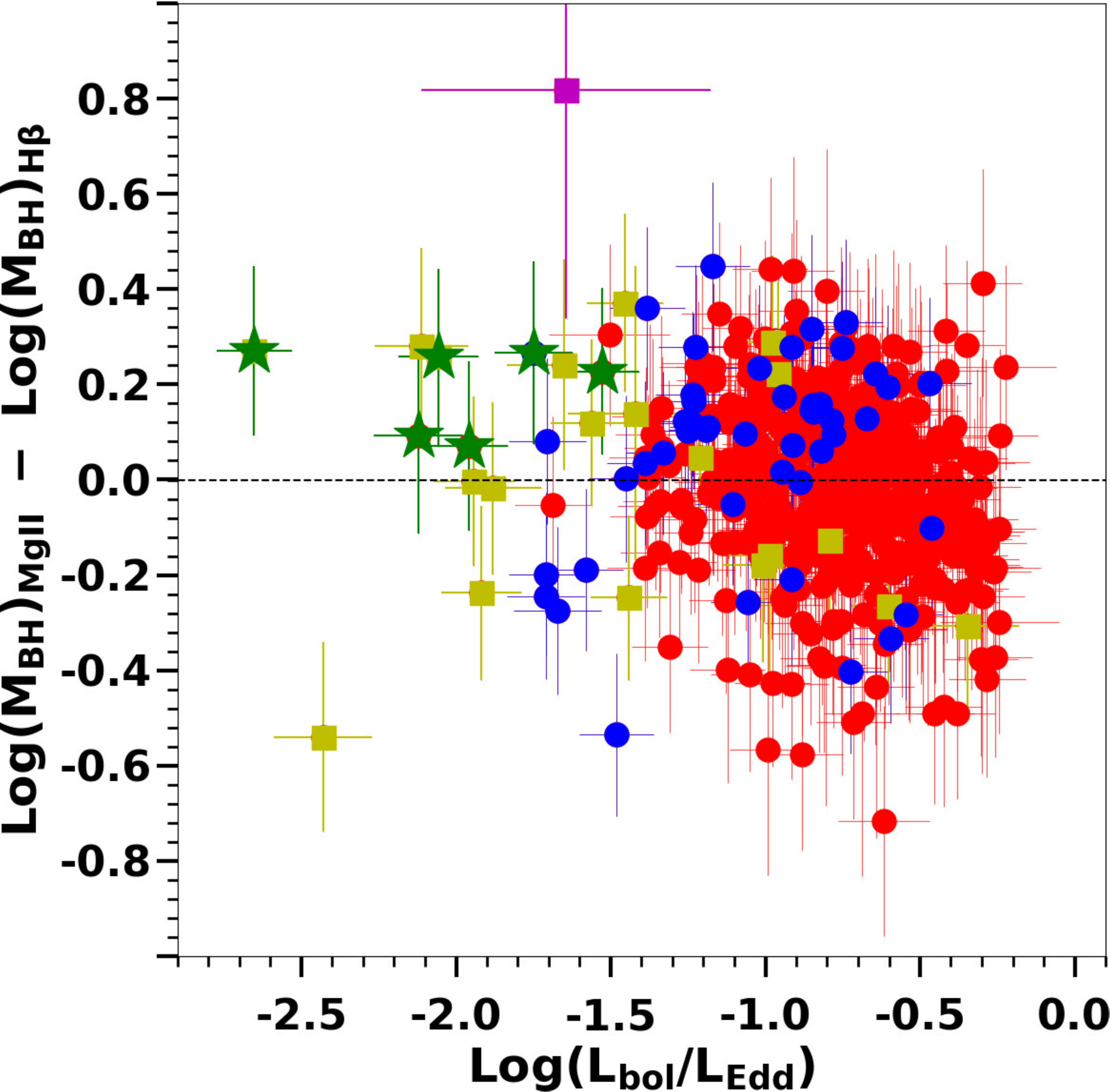} 
    \includegraphics[width=0.33\textwidth]{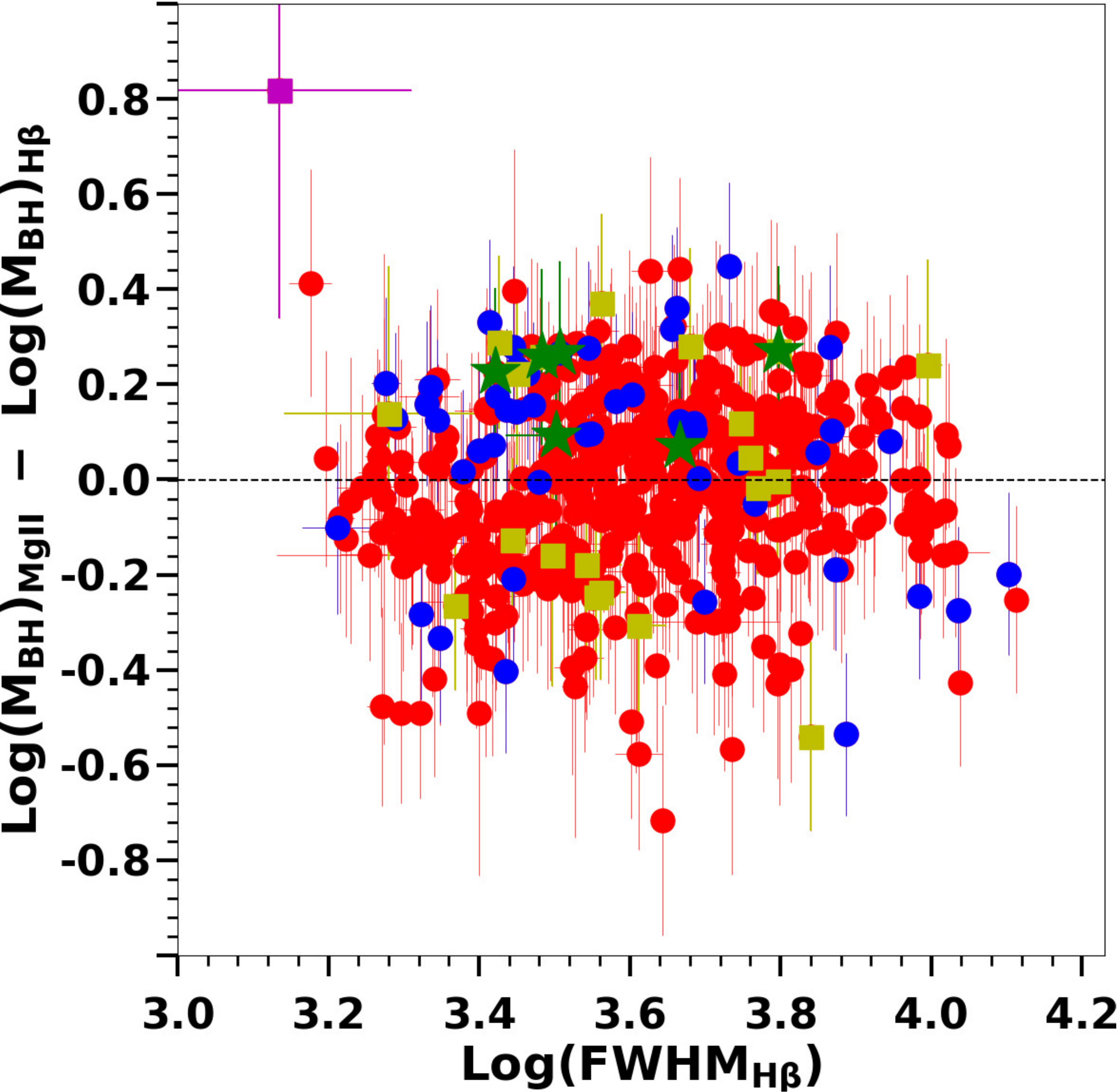}
    \includegraphics[width=0.33\textwidth]{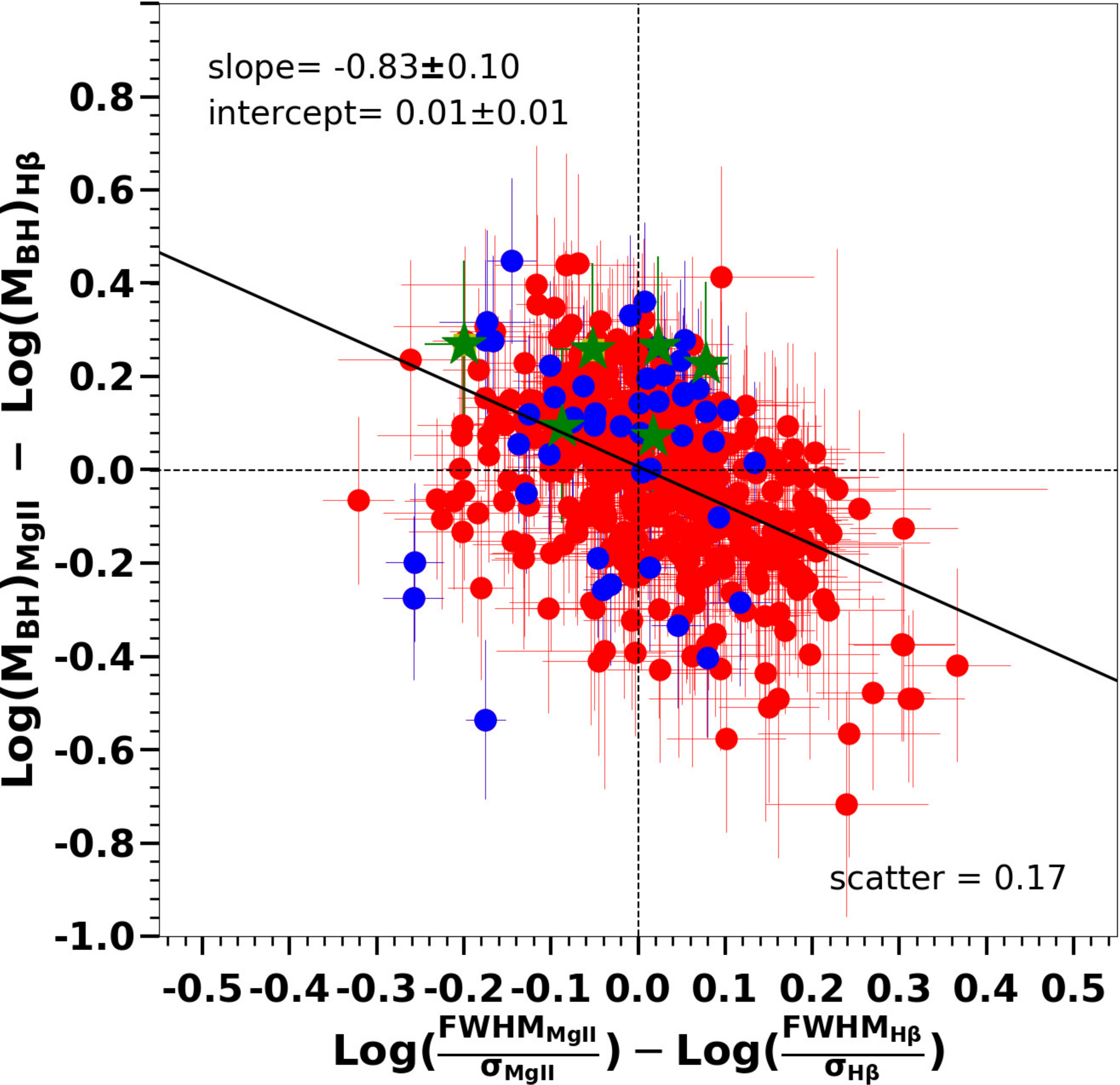} \\
    \includegraphics[width=0.33\textwidth]{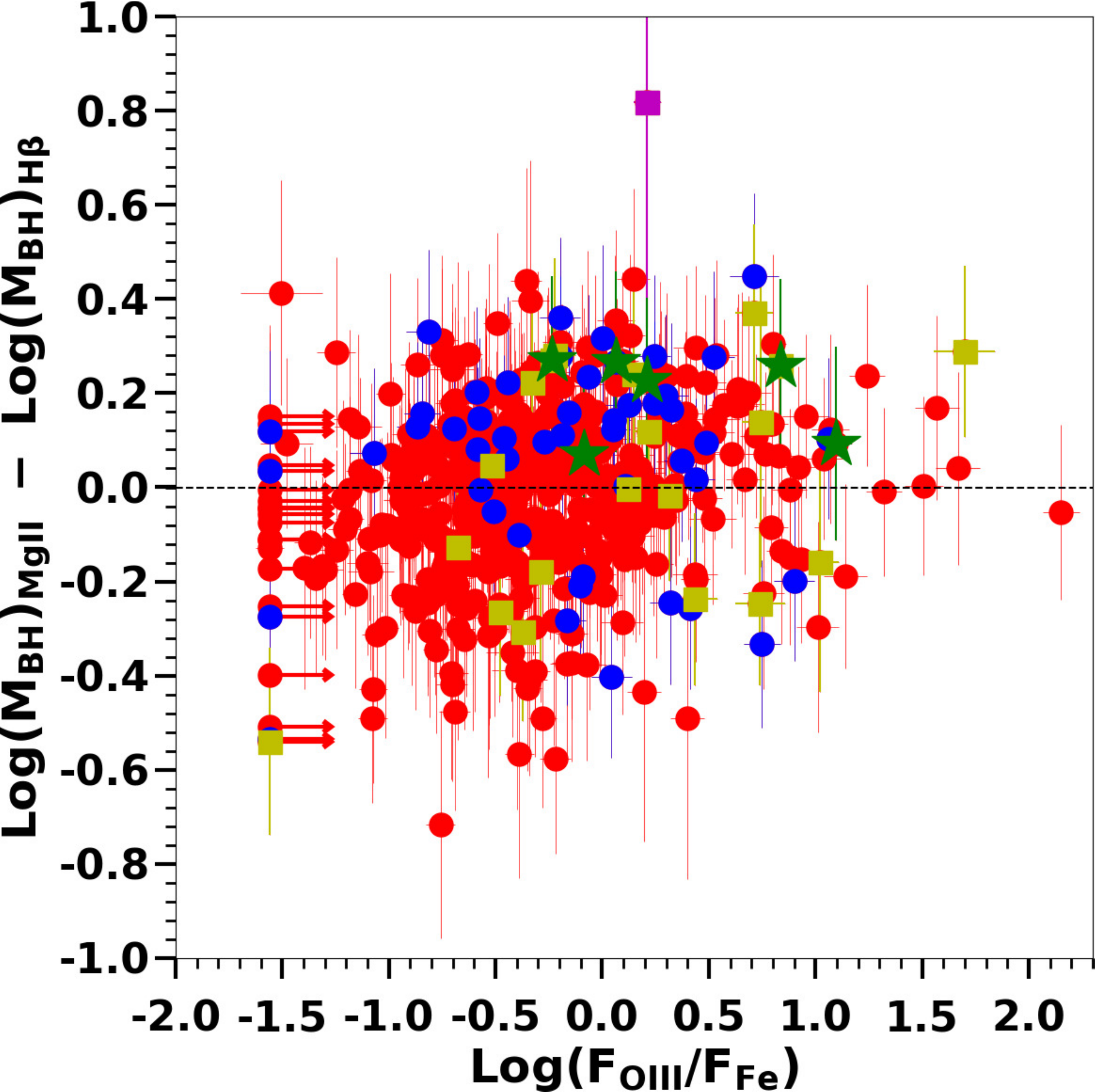}
    \includegraphics[width=0.33\textwidth]{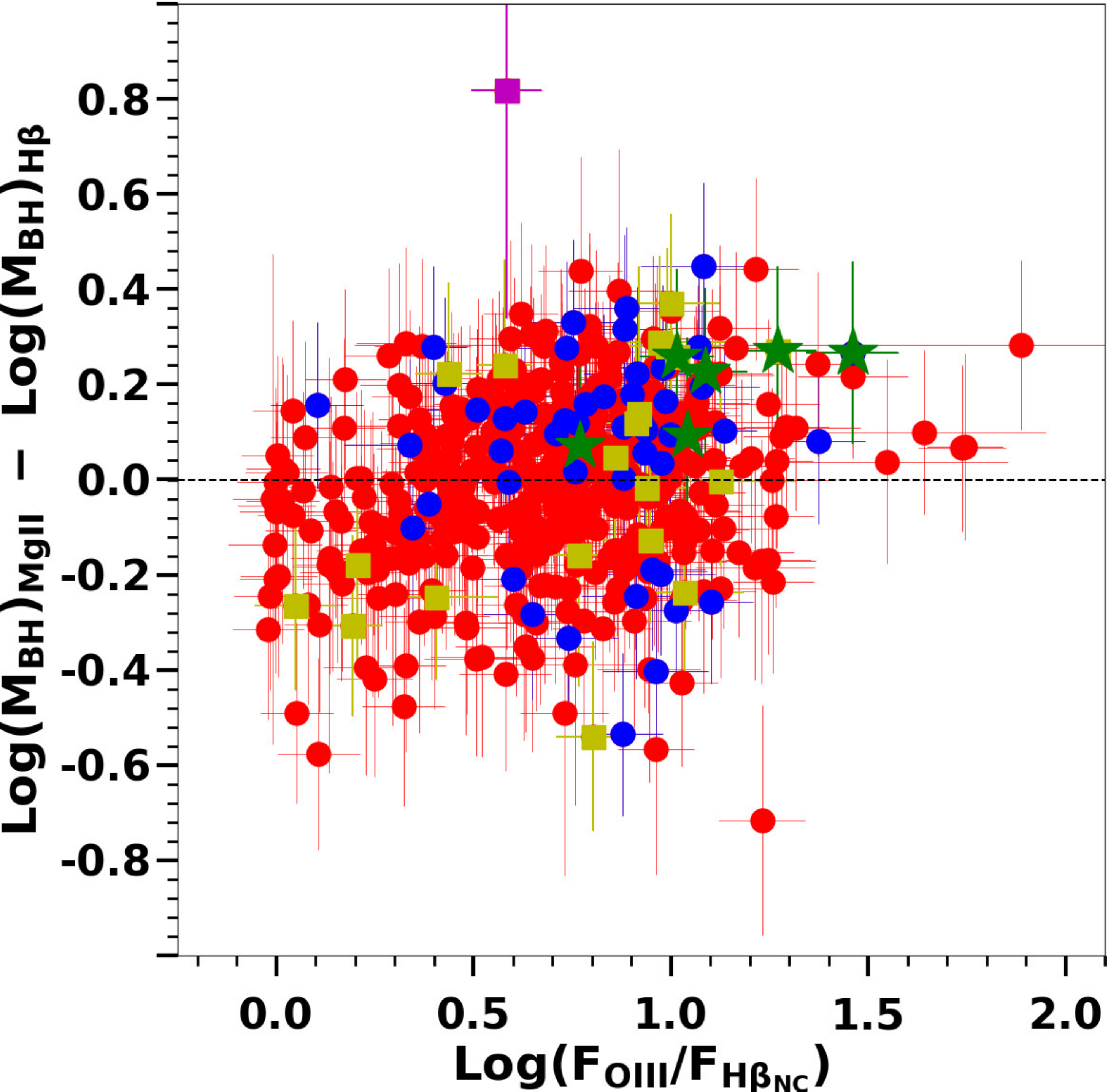}
    \includegraphics[width=0.33\textwidth]{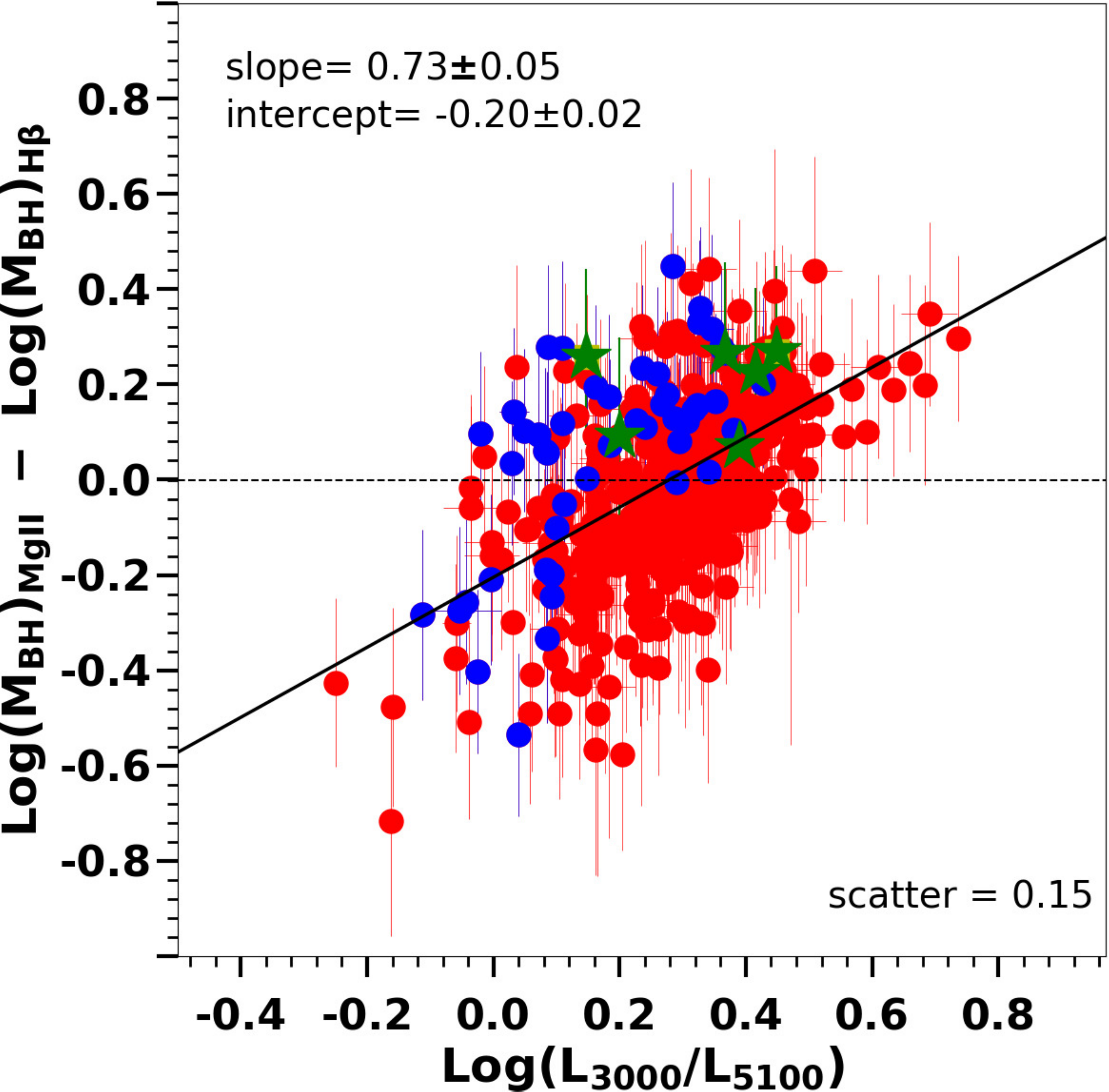}
    \caption{Comparison of systematic difference between \mbh\ estimators from \ion{Mg}{2} (Scheme 2) and \Hb\ as a function of AGN parameters, i.e., Eddington ratio (top left), $\rm FWHM_{H\beta}$ (top middle), the difference of line profiles between \Hb\ and \MgII\ (top right), F$_{\rm OIII}$/F$_{\rm FeII}$ (bottom left), F$_{\rm OIII}$/F$_{\rm H\beta, narrow}$ (bottom middle), L$_{3000}$/L$_{5100}$ (bottom right). The symbols are shown for the moderate-luminosity AGNs (blue), the SDSS sample (red), the 25 RM sources (yellow), the six HST targets (green), and NGC~4051 (pink). The best-fit slope is shown in thick black solid line. The dash-lines show different values of 0.} 
\label{fig:property}     
\end{figure*}

\begin{figure}
	\includegraphics[width=0.40\textwidth]{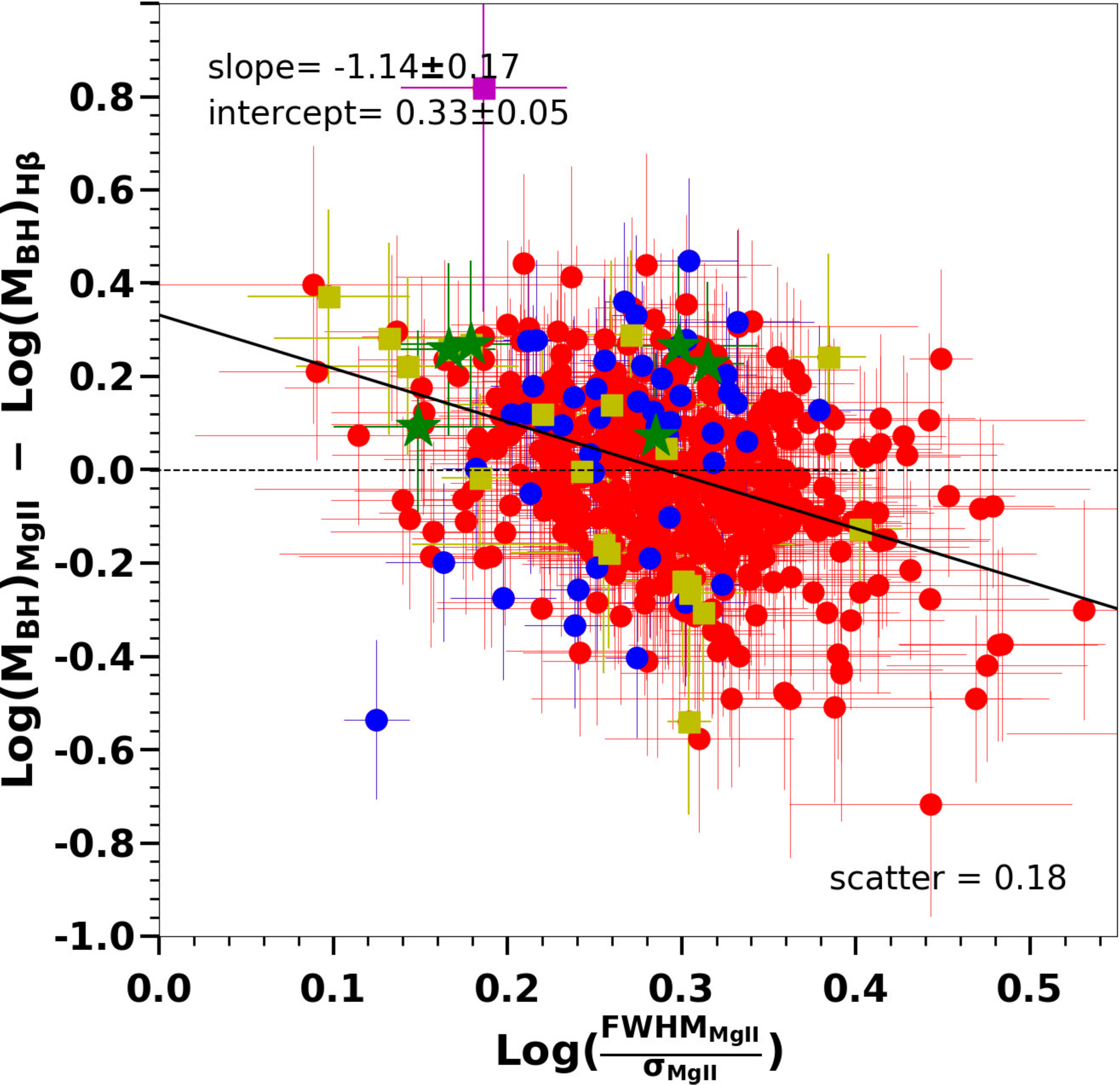}
	\centering
	\caption{Comparison of systematic difference between \mbh\ estimators from \ion{Mg}{2} (Scheme 2) and \Hb\ as a function of \MgII\ profile. The symbols are shown for the moderate-luminosity AGNs (blue), the SDSS sample (red), the 25 RM sources (yellow), the six HST targets (green), and NGC~4051 (pink). The best-fit slope is shown in thick black solid line. The dash-lines show different values of 0.
\label{fig:correct_mgii}}
\end{figure}

\begin{figure*}
	\includegraphics[width=0.33\textwidth]{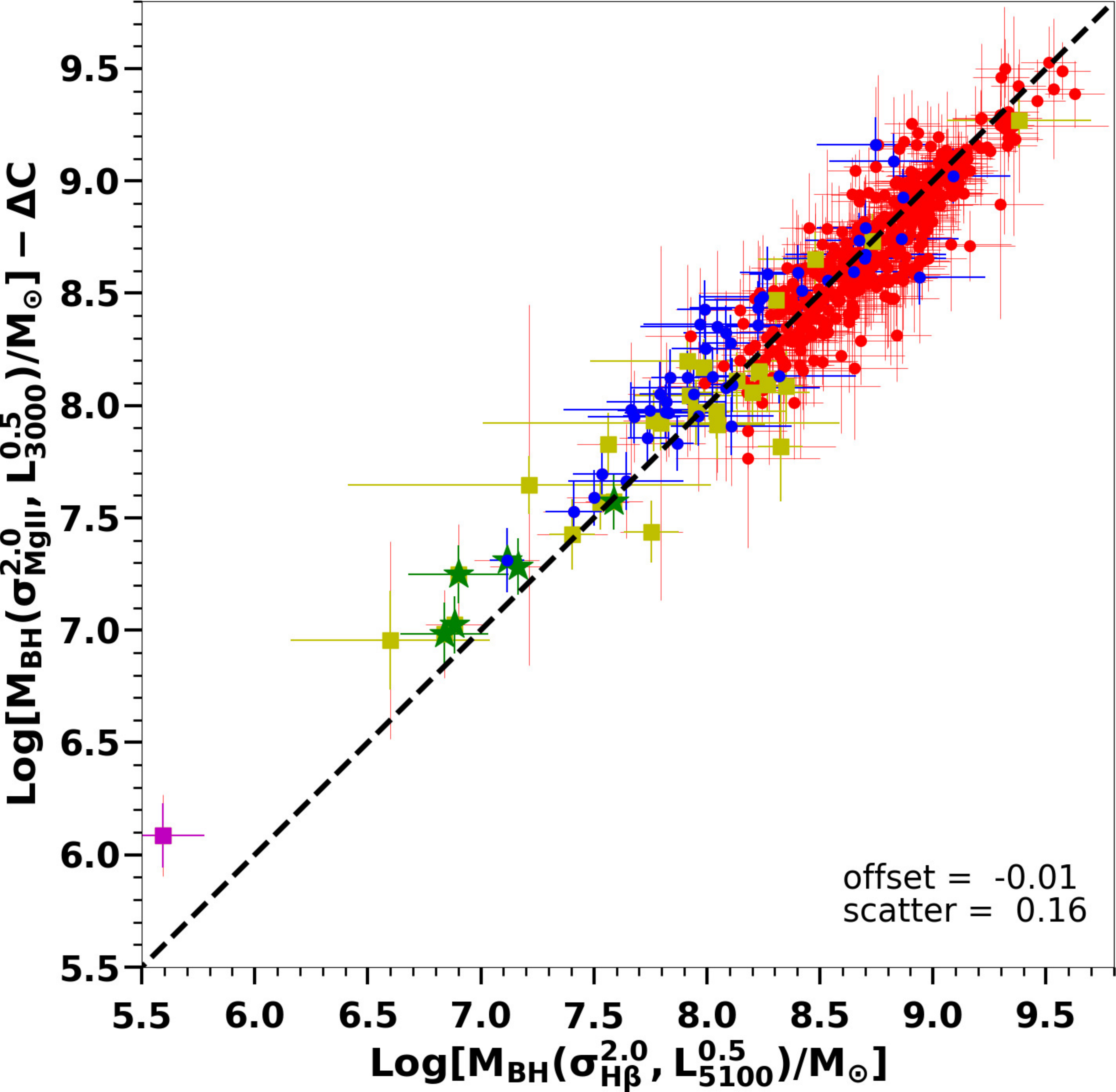}
	\includegraphics[width=0.33\textwidth]{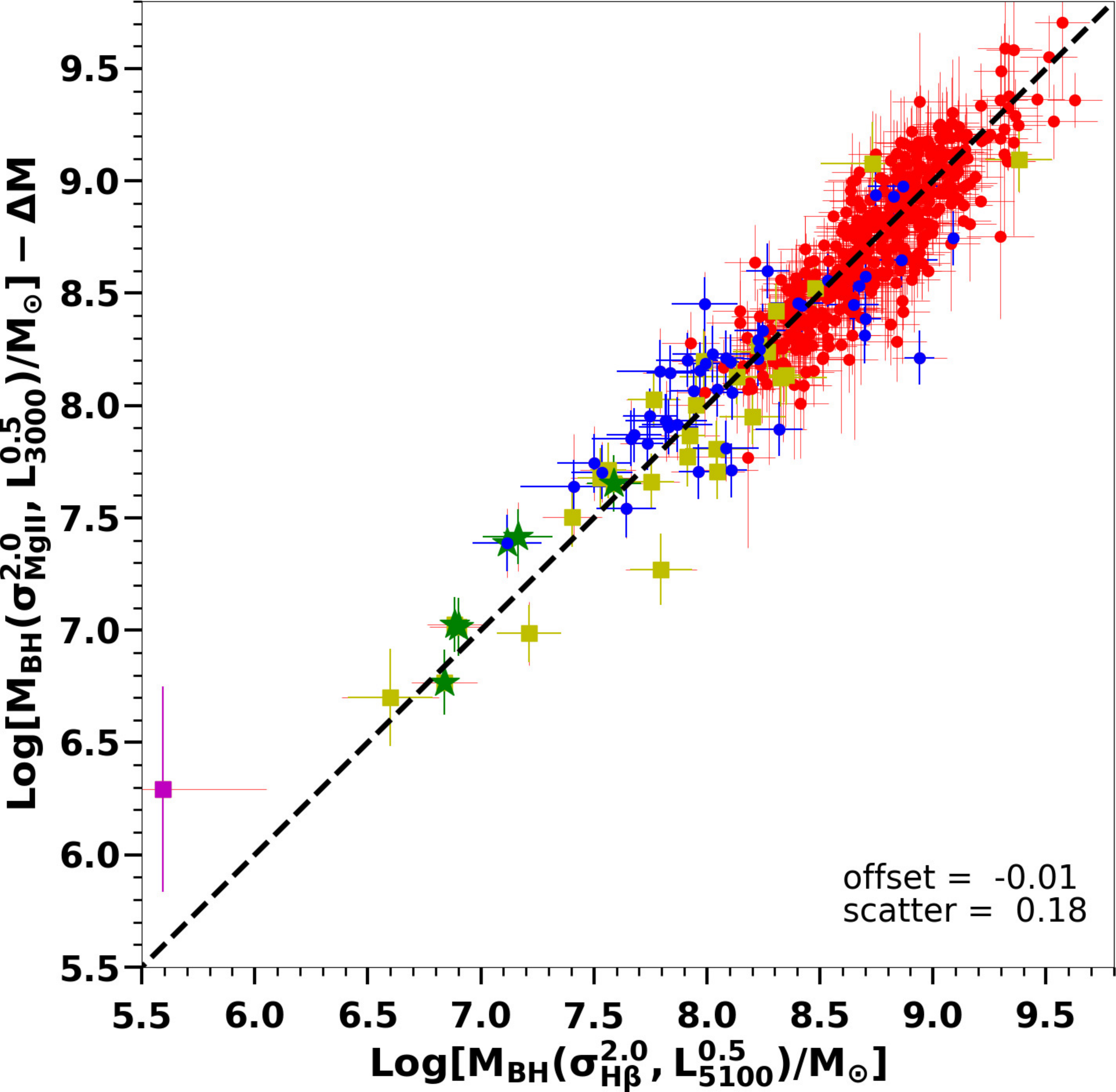}
	\includegraphics[width=0.33\textwidth]{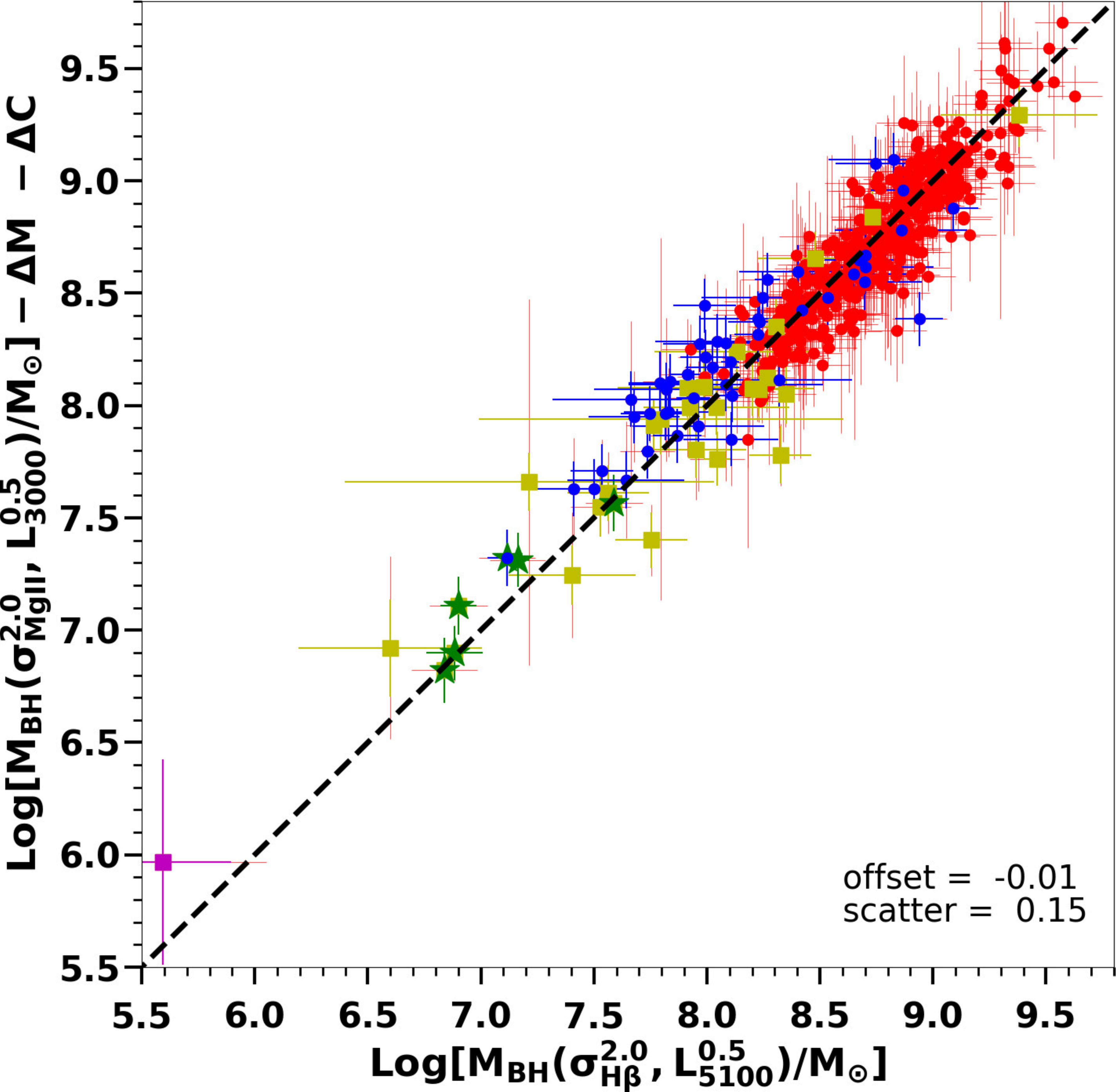}	
	\centering
	\caption{Comparison of \ion{Mg}{2} and \Hb-based based mass estimators after applying the correction term, $\rm \Delta C$ in Equation \ref{eq:cor3} (left panel), $\rm \Delta M$ in Equation \ref{eq:cor1_2} (middle panel) and the combination of the two terms (right panel). The symbols are shown for the moderate-luminosity AGNs (blue), the SDSS sample (red), the 25 RM sources (yellow), the six HST targets (green), and NGC~4051 (pink). The dash-lines show different values of 0.
\label{fig:correct_L}}
\end{figure*}

\begin{figure*}
	\includegraphics[width=0.4\textwidth]{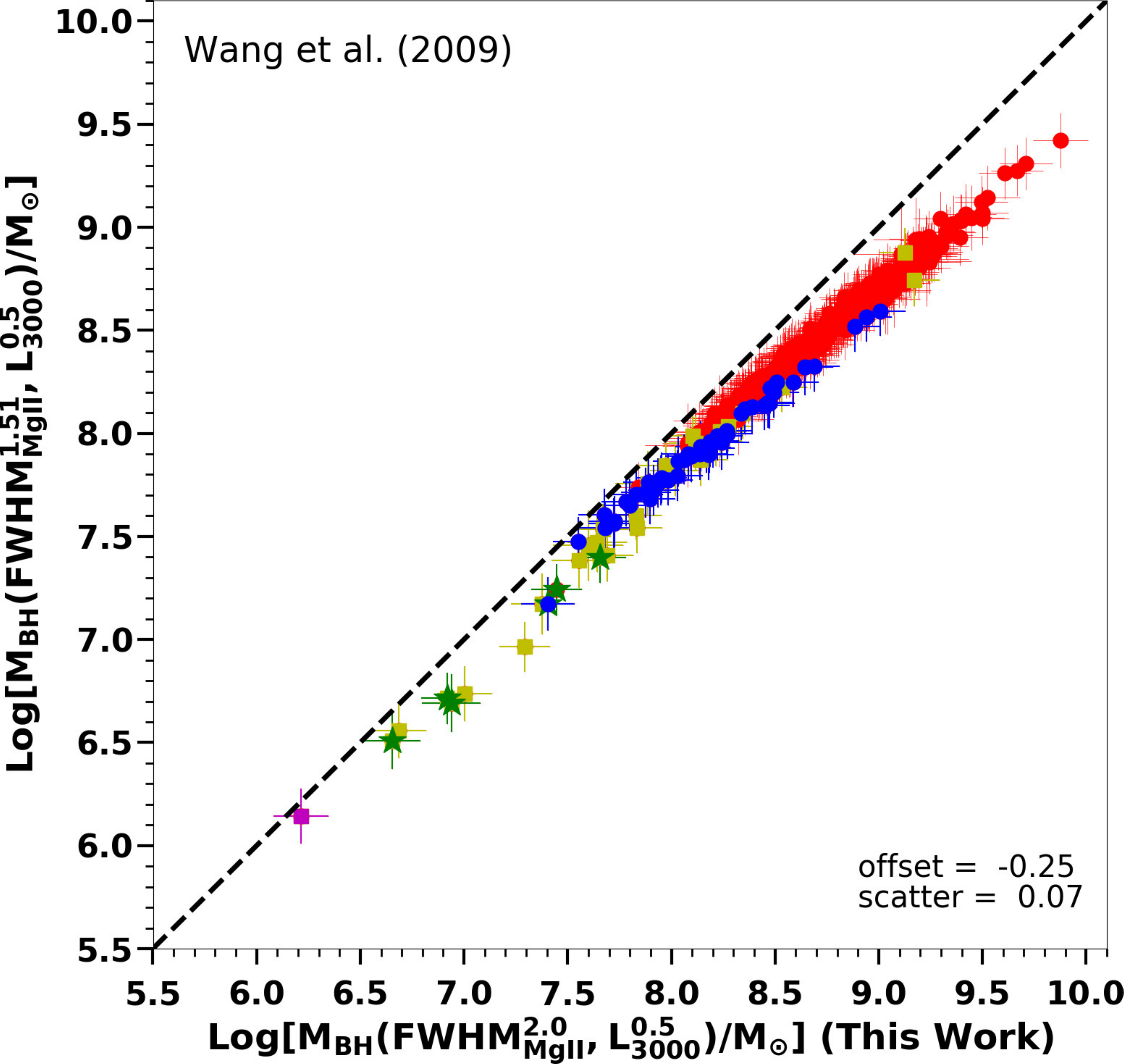}
	\includegraphics[width=0.4\textwidth]{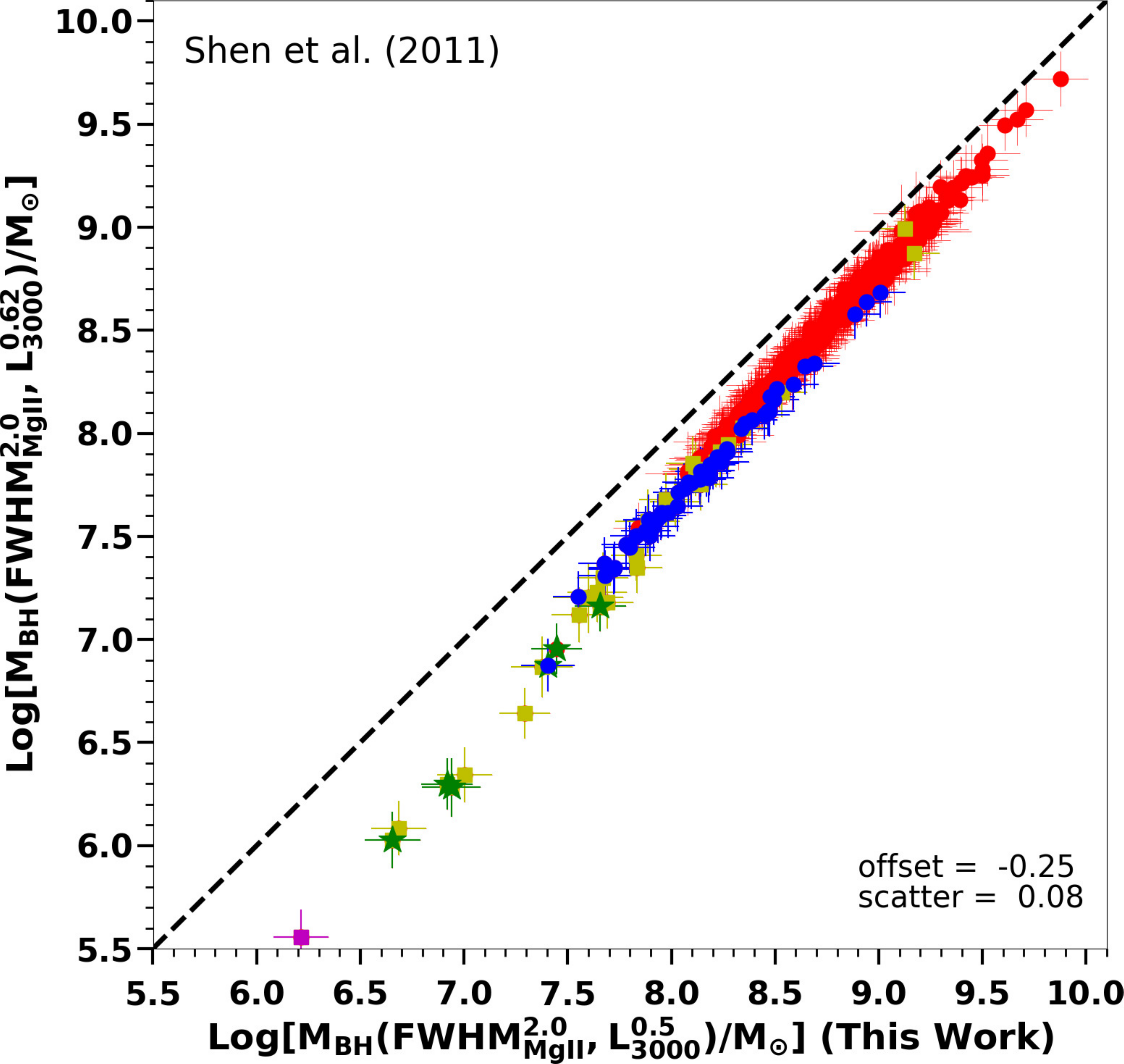}
	\includegraphics[width=0.4\textwidth]{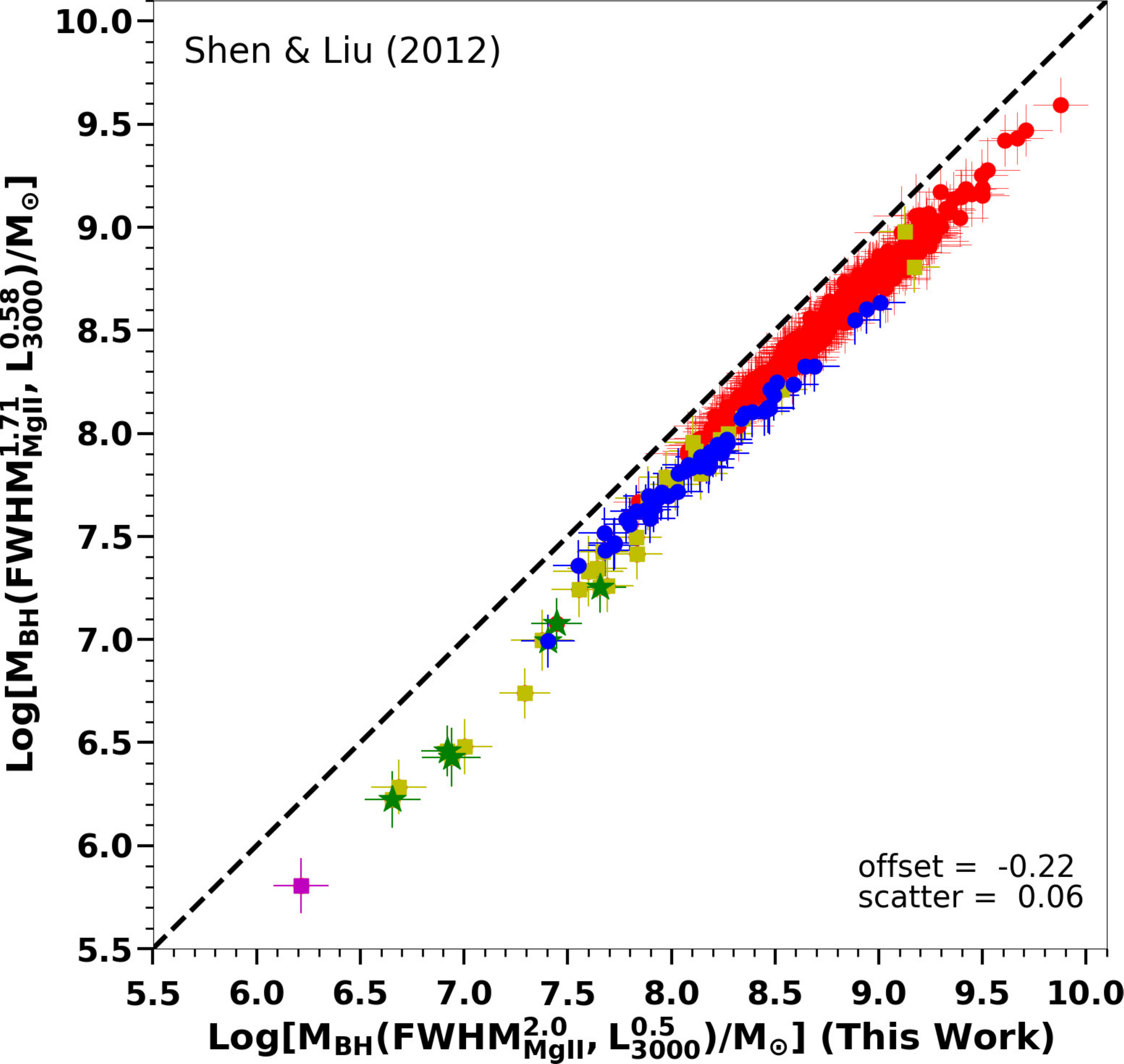}
	\includegraphics[width=0.4\textwidth]{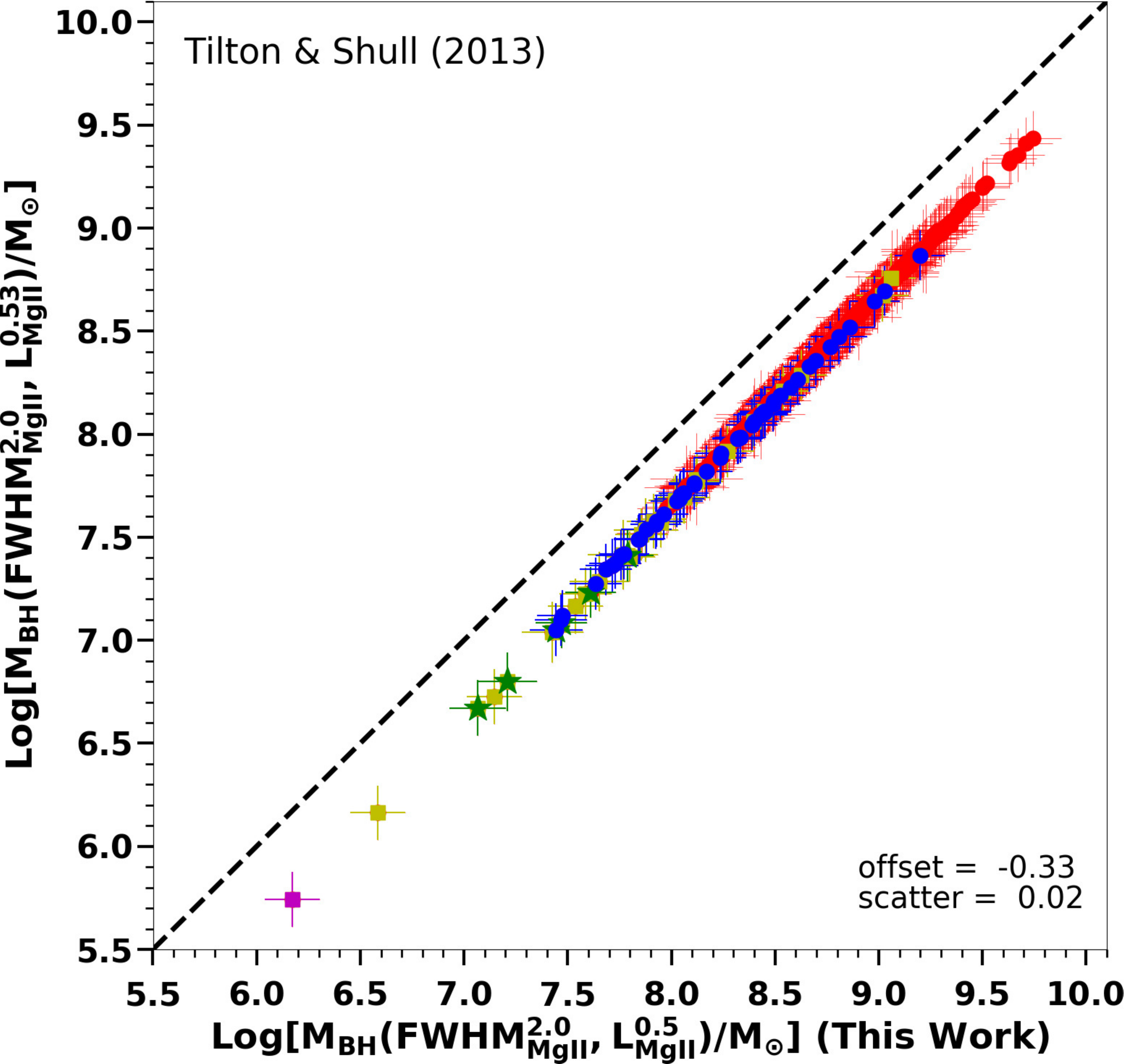}
	\centering
	\caption{Comparison of our \mbh\ calibration with various \ion{Mg}{2}-based mass estimators e.g., \citet{Wang+09} (top left panel), \citet{Shen+11} (top right panel), \citet{Shen+12} (bottom left panel), and \citet{Tilton+13} (bottom right panel). The color schemes of sample are similar to those in Figure \ref{fig:compare}. The symbols are shown for the moderate-luminosity AGNs (blue), the SDSS sample (red), the 25 RM sources (yellow), the six HST targets (green), and NGC~4051 (pink). The dash-lines show different values of 0. 
\label{fig:mbh}}
\end{figure*}


\section{Conclusions}\label{section:sum}

In this study, we present the calibration of \mbh\ estimators, using the combined sample of low, intermediate, and high luminosity AGNs with high quality 
spectra that have \MgII\ and \Hb\ lines observed simultaneously. The dynamic range of $\mathrm{\lambda L_{5100} \sim10^{41.3} - 10^{46.5}}$ \ergs, and 5.5 $<$ $\log$\mbh\ $<$ 9.5
provides reliable mass estimators. We summarize the main results as follows: 

(1) From the comparison of line width between the \MgII\ and \Hb\ emission lines, $\sigma_{\rm MgII}$ and $\sigma_{\rm H\beta}$ show a linear relationship, while, FWHMs of both emission lines show a somewhat sub-linear relationship. 

(2) Similar to our previous work, we found the linear relationship between the continuum luminosities of 3000\AA\ and 5100\AA. In addition, line luminosity of \MgII\ shows somewhat sub-linear relationship with that of \Hb, indicating the Baldwin effect in the UV range. 

(3) In the case of the optical mass estimator, by using the fiducial mass from $\sigma_{\rm H\beta}$ and $\rm L_{5100}$, we obtained the \Hb-based \mbh\ estimator with a small intrinsic scatter $<$ 0.01 dex and rms scatter $<$ 0.13 dex. By using the reference mass from $\rm FWHM_{\rm H\beta}$ and $\rm L_{5100}$, we obtained an intrinsic scatter $<$ 0.12 dex and rms scatter $<$ 0.19 dex.


(4) Using the reference mass from $\sigma_{\rm H\beta}$ and $\rm L_{5100}$, we presented the \MgII\ \mbh\ estimator with the intrinsic scatter of 0.09$-$0.21 dex and rms scatter of 0.17$-$0.25 dex. With the reference mass from $\rm FWHM_{\rm H\beta}$ and $\rm L_{5100}$, we obtained an intrinsic scatter of 0.14$-$0.26 dex and rms scatter of 0.21$-$0.31 dex. In general, depending on the choice of line width ($\sigma$ or FWHM) and luminosity (emission line or continuum), we obtained different systematic uncertainties, i.e., rms scatter larger than $\sim$0.15 dex. Based on our calibrated estimators, $\sigma_{\rm MgII}$ and $\rm L_{3000}$ provide the best UV \mbh\ estimator with an intrinsic scatter of 0.09 dex and rms scatter of 0.17 dex.

(5) From the comparison of the systematic difference between the \MgII\ and \Hb-based \mbh\ estimators as a function of AGN properties, we found strong correlations between the UV and optical \mbh\ ratio and the ratio of line profiles of the \MgII\ and \Hb\ lines and $\rm L_{5100}$/$\rm L_{3000}$. The discrepancy between \MgII\ and \Hb-based \mbh\ estimators is strongly dependent on the difference of line profiles between \MgII\ and \Hb. In addition, we suggested to add additional $\Delta$M correction factor (Equation \ref{eq:cor1_2}) to reduce the systematic uncertainty between the UV and optical \mbh\ estimators.  \\

\acknowledgements

This work has been supported by the Basic Science Research Program through the National Research Foundation of Korea government (2016R1A2B3011457 and 2017R1A5A1070354). H.A.N.L. and Y.Q.X. acknowledge support from the Chinese Academy of Sciences President's International Fellowship Initiative (Grant No. 2019PM0020), the National Natural Science Foundation of China (NSFC-11890693, NSFC-11421303), the CAS Frontier Science Key Research Program (QYZDJ-SSW-SLH006), and K.C. Wong Education Foundation. H. A. N. L. thanks Dr. Jia Li for reading this manuscript.


\begin{turnpage}
\begin{table*}
\begin{center}
\tablewidth{0.1\textwidth}
\fontsize{7}{5}\selectfont
\caption{\label{table:mbh_sigma} \mbh\ estimators based on \ion{Mg}{2}, using the fiducial mass from $\sigma_{\rm H\beta}$ and $\rm L_{5100}$}
\begin{tabular}{cccccccccccc}

\tableline\tableline
Case  & $\alpha$  & $\beta$  & $\gamma$  & $\rm \sigma_{inst}$  & $\rm \sigma_{rms}$   &  & $\alpha$  & $\beta$  & $\gamma$  & $\sigma_{inst}$  & $\rm \sigma_{rms}$ \\
(1)&(2)&(3)&(4)&(5)& (6)& &(7) & (8)&(9)&(10)&(11) \\
\tableline
\tableline

~ \\
  &   &  & {\bf L$_{3000}$ \& $\sigma_{\rm MgII}$ }&  &  & & &   & {\bf L$_{3000}$ \& FWHM$_{\rm MgII}$ }\\
\cline{1-6} \cline{8-12} \\
1) $\beta$ \& $\gamma$ from scaling & 7.54 $\pm$ 0.01 & 2.13 $\pm$ 0.06 & 0.54 $\pm$ 0.01 & 0.12 $\pm$ 0.01 & 0.20
& & 6.92 $\pm$ 0.01 & 2.14 $\pm$ 0.07 & 0.54 $\pm$ 0.01 & 0.15 $\pm$ 0.01 & 0.20 \\
\cline{1-6} \cline{8-12} \\
{\bf 2) $\beta$=2 \& $\gamma$=0.5} & {\bf 7.61 $\pm$ 0.01} & {\bf 2.00} & {\bf 0.50} & {\bf 0.11 $\pm$ 0.01} & {\bf 0.19}
& & {\bf 7.04 $\pm$ 0.01} & {\bf 2.00} & {\bf 0.50} & {\bf 0.13 $\pm$ 0.01} & {\bf 0.19} \\
\cline{1-6} \cline{8-12} \\
3) $\beta$ = 2 & 7.57 $\pm$ 0.02 & 2.00 & 0.53 $\pm$ 0.02 & 0.11 $\pm$ 0.01 & 0.19
& & 7.03 $\pm$ 0.02 & 2.00 & 0.51 $\pm$ 0.02 & 0.13 $\pm$ 0.01 & 0.19 \\
\cline{1-6} \cline{8-12} \\
4) $\gamma$=0.5 & 7.72 $\pm$ 0.02 & 1.51 $\pm$ 0.06 & 0.50 & 0.09 $\pm$ 0.01 & 0.17
& & 7.35 $\pm$ 0.03 & 1.38 $\pm$ 0.05 & 0.50 & 0.10 $\pm$ 0.01 & 0.17 \\
\cline{1-6} \cline{8-12} \\
5) Free $\beta$ \& $\gamma$ & 7.68 $\pm$ 0.02 & 1.49 $\pm$ 0.06 & 0.54 & 0.09 $\pm$ 0.01 & 0.17
& & 7.33 $\pm$ 0.03 & 1.36 $\pm$ 0.06 & 0.53 & 0.10 $\pm$ 0.01 & 0.17 \\

\tableline 
\tableline \\
~\\

  &  &  &  {\bf L$_{\rm MgII}$ \& $\sigma_{\rm MgII}$ } &  &  &  & &  &  {\bf L$_{\rm MgII}$ \& FWHM$_{\rm MgII}$ } \\
\cline{1-6} \cline{8-12} \\

1) $\beta$ \& $\gamma$ from scaling & 7.42 $\pm$ 0.01 & 2.13 $\pm$ 0.06 & 0.65 $\pm$ 0.01 & 0.20 $\pm$ 0.01 & 0.25
& & 6.80 $\pm$ 0.01 & 2.14 $\pm$ 0.07 & 0.65 $\pm$ 0.01 & 0.21 $\pm$ 0.01 & 0.25 \\
\cline{1-6} \cline{8-12} \\
{\bf 2) $\beta$=2 \& $\gamma$=0.5} & {\bf 7.62 $\pm$ 0.01} & {\bf 2.00} & {\bf 0.50} & {\bf 0.18 $\pm$ 0.01} & {\bf 0.23}
& & {\bf 7.05 $\pm$ 0.01} & {\bf 2.00} & {\bf 0.50} & {\bf 0.18 $\pm$ 0.01} & {\bf 0.22} \\
\cline{1-6} \cline{8-12} \\
3) $\beta$ = 2 & 7.57 $\pm$ 0.03 & 2.00 & 0.54 $\pm$ 0.02 & 0.17 $\pm$ 0.01 & 0.23
& & 7.03 $\pm$ 0.03 & 2.00 & 0.51 $\pm$ 0.02 & 0.18 $\pm$ 0.01 & 0.22 \\
\cline{1-6} \cline{8-12} \\
4) $\gamma$=0.5 & 7.80 $\pm$ 0.02 & 1.16 $\pm$ 0.07 & 0.50 & 0.14 $\pm$ 0.01 & 0.20
& & 7.49 $\pm$ 0.03 & 1.12 $\pm$ 0.06 & 0.50 & 0.13 $\pm$ 0.01 & 0.19 \\
\cline{1-6} \cline{8-12} \\
5) Free $\beta$ \& $\gamma$ & 7.72 $\pm$ 0.03 & 1.06 $\pm$ 0.07 & 0.59 & 0.13 $\pm$ 0.01 & 0.19
& & 7.44 $\pm$ 0.03 & 1.03 $\pm$ 0.06 & 0.58 & 0.12 $\pm$ 0.01 & 0.18 \\

\tableline
\tableline
\end{tabular}
\end{center}
\tablecomments{Col. (1): Method of calibration. Col. (2): Number of data use in the calibration.  Col. (3) \& (8): $\alpha$ values. Col. (4) \& (9): 
$\beta$ values. Col. (5) \& (10): $\gamma$ values. Col. (6) \& (11): intrinsic scatter. Col. (7) \& (12): rms scatter. We recommend Scheme 2 (bold font) as the best estimator among various Schemes.
}
\end{table*}
\end{turnpage}

%
%
\begin{turnpage}
\begin{table*}
\begin{center}
\tablewidth{1\textwidth}
\fontsize{7}{5}\selectfont
\caption{\label{table:mbh_fwhm} \mbh\ estimators based on \ion{Mg}{2}, using the fiducial mass from $\rm FWHM_{H\beta}$ and $\rm L_{5100}$}
\begin{tabular}{cccccccccccc}

\tableline\tableline
Case  & $\alpha$  & $\beta$  & $\gamma$  & $\rm \sigma_{inst}$  & $\rm \sigma_{rms}$   &  & $\alpha$  & $\beta$  & $\gamma$  & $\rm \sigma_{inst}$  & $\rm \sigma_{rms}$ \\
(1)&(2)&(3)&(4)&(5)& (6)& &(7) & (8)&(9)&(10)&(11) \\
\tableline
\tableline

~ \\
  &   &  & {\bf L$_{3000}$ \& $\sigma_{\rm MgII}$ }&  &  & & &   & {\bf L$_{3000}$ \& FWHM$_{\rm MgII}$ }\\
\cline{1-6} \cline{8-12} \\
1) $\beta$ \& $\gamma$ from scaling & 7.26 $\pm$ 0.01 & 3.20 $\pm$ 0.03 & 0.54 $\pm$ 0.01 & 0.17 $\pm$ 0.01 & 0.25
& & 6.36 $\pm$ 0.01 & 3.17 $\pm$ 0.04 & 0.54 $\pm$ 0.01 & 0.20 $\pm$ 0.01 & 0.25 \\
\cline{1-6} \cline{8-12} \\
2) $\beta$=2 \& $\gamma$=0.5 & 7.58 $\pm$ 0.01 & 2.00 & 0.50 & 0.16 $\pm$ 0.01 & 0.22
& & 7.00 $\pm$ 0.01 & 2.00 & 0.50 & 0.16 $\pm$ 0.01 & 0.21 \\
\cline{1-6} \cline{8-12} \\
3) $\beta$ = 2 & 7.54 $\pm$ 0.03 & 2.00 & 0.53 $\pm$ 0.02 & 0.16 $\pm$ 0.01 & 0.22
& & 6.99 $\pm$ 0.02 & 2.00 & 0.50 $\pm$ 0.02 & 0.16 $\pm$ 0.01 & 0.21 \\
\cline{1-6} \cline{8-12} \\
4) $\gamma$=0.5 & 7.47 $\pm$ 0.02 & 2.47 $\pm$ 0.08 & 0.50 & 0.15 $\pm$ 0.01 & 0.22
& & 6.85 $\pm$ 0.04 & 2.30 $\pm$ 0.07 & 0.50 & 0.16 $\pm$ 0.01 & 0.21 \\
\cline{1-6} \cline{8-12} \\
5) Free $\beta$ \& $\gamma$ & 7.44 $\pm$ 0.03 & 2.46 $\pm$ 0.08 & 0.53 & 0.14 $\pm$ 0.01 & 0.22
& & 6.85 $\pm$ 0.04 & 2.30 $\pm$ 0.07 & 0.49 & 0.16 $\pm$ 0.01 & 0.21 \\

\tableline 
\tableline \\
~\\

  &  &  &  {\bf L$_{\rm MgII}$ \& $\sigma_{\rm MgII}$ } &  &  &  & &  &  {\bf L$_{\rm MgII}$ \& FWHM$_{\rm MgII}$ } \\
\cline{1-6} \cline{8-12} \\
1) $\beta$ \& $\gamma$ from scaling & 7.14 $\pm$ 0.01 & 3.20 $\pm$ 0.03 & 0.65 $\pm$ 0.01 & 0.25 $\pm$ 0.01 & 0.31
& & 6.24 $\pm$ 0.01 & 3.17 $\pm$ 0.04 & 0.65 $\pm$ 0.01 & 0.26 $\pm$ 0.01 & 0.30 \\
\cline{1-6} \cline{8-12} \\
2) $\beta$=2 \& $\gamma$=0.5 & 7.58 $\pm$ 0.01 & 2.00 & 0.50 & 0.19 $\pm$ 0.01 & 0.25
& & 7.00 $\pm$ 0.01 & 2.00 & 0.50 & 0.18 $\pm$ 0.01 & 0.23 \\
\cline{1-6} \cline{8-12} \\
3) $\beta$ = 2 & 7.51 $\pm$ 0.03 & 2.00 & 0.56 $\pm$ 0.02 & 0.19 $\pm$ 0.01 & 0.24
& & 6.96 $\pm$ 0.03 & 2.00 & 0.54 $\pm$ 0.02 & 0.18 $\pm$ 0.01 & 0.22 \\
\cline{1-6} \cline{8-12} \\
4) $\gamma$=0.5 & 7.55 $\pm$ 0.02 & 2.13 $\pm$ 0.09 & 0.50 & 0.19 $\pm$ 0.01 & 0.25
& & 6.98 $\pm$ 0.04 & 2.04 $\pm$ 0.07 & 0.50 & 0.18 $\pm$ 0.01 & 0.23 \\
\cline{1-6} \cline{8-12} \\
5) Free $\beta$ \& $\gamma$ & 7.49 $\pm$ 0.03 & 2.07 $\pm$ 0.09 & 0.56 & 0.19 $\pm$ 0.01 & 0.24
& & 6.95 $\pm$ 0.04 & 2.00 $\pm$ 0.07 & 0.54 & 0.18 $\pm$ 0.01 & 0.22 \\

\tableline
\tableline
\end{tabular}
\end{center}
\tablecomments{Col. (1): Method of calibration. Col. (2): Number of data use in the calibration.  Col. (3) \& (8): $\alpha$ values. Col. (4) \& (9): 
$\beta$ values. Col. (5) \& (10): $\gamma$ values. Col. (6) \& (11): intrinsic scatter. Col. (7) \& (12): rms scatter. 
}
\end{table*}
\end{turnpage}

\end{document}